\documentclass[preprint,12pt,letterpaper]{elsarticle}

\usepackage{graphicx}
\usepackage{amsmath}
\usepackage{bbm}
\usepackage{bm}
\usepackage{amssymb}
\usepackage{amsthm}
\usepackage[toc,page]{appendix}

\usepackage{lineno}
\usepackage{subfig}
\usepackage{fullpage}
\usepackage{algorithm}
\usepackage{algpseudocode}
\usepackage{array}
\usepackage{pbox}
\usepackage{tikz}
\usepackage[]{changes}
\usepackage[colorinlistoftodos]{todonotes}

\def\onedot{$\mathsurround0pt\ldotp$}
\def\cddots{
  \mathbin{\vcenter{\baselineskip.67ex
    \hbox{\onedot}\hbox{\onedot}}%
  }}%

\newcolumntype{L}[1]{>{\raggedright\let\newline\\\arraybackslash\hspace{0pt}}m{#1}}
\newcolumntype{C}[1]{>{\centering\let\newline\\\arraybackslash\hspace{0pt}}m{#1}}
\newcolumntype{R}[1]{>{\raggedleft\let\newline\\\arraybackslash\hspace{0pt}}m{#1}}

\usepackage{url}
\usepackage{listings}
\usepackage[colorlinks=true]{hyperref}

\usepackage{color,soul}
\usepackage{mathtools}
\usepackage{booktabs}

\usepackage[colorlinks=true]{hyperref}

\usepackage{forest}	
\usetikzlibrary{arrows.meta, shapes.geometric, calc, shadows}

\usepackage{multicol}
\usepackage{imakeidx}
\usepackage{nomencl}
\makenomenclature
\usepackage{etoolbox}
\renewcommand\nomgroup[1]{%
  \item[\itshape
  \ifstrequal{#1}{A}{Symbols}{%
  \ifstrequal{#1}{B}{Roman Letters}{%
  \ifstrequal{#1}{C}{Greek Letters}{%
  \ifstrequal{#1}{D}{Abbreviations}{}}}}%
]}
 
\renewcommand{\nompreamble}{\begin{multicols}{2}}
\renewcommand{\nompostamble}{\end{multicols}}

\definecolor{lightblue}{rgb}{.90,.95,1}
\definecolor{lightgreen}{rgb}{.90,1,.95}
\definecolor{darkgreen}{rgb}{0,.5,0.5}

\colorlet{col1in}{red!30}
\colorlet{col1out}{red!50}
\colorlet{col5in}{blue!10}
\colorlet{col5out}{blue!20}
\colorlet{col6in}{blue!20}
\colorlet{col6out}{blue!30}
\colorlet{col9in}{gray!80}
\colorlet{col9out}{gray!40}
\colorlet{linecolk}{black!80}
\colorlet{linecol}{blue!80}

\newcommand*\patchAmsMathEnvironmentForLineno[1]{%
  \expandafter\let\csname old#1\expandafter\endcsname\csname #1\endcsname
  \expandafter\let\csname oldend#1\expandafter\endcsname\csname end#1\endcsname
  \renewenvironment{#1}%
     {\linenomath\csname old#1\endcsname}%
     {\csname oldend#1\endcsname\endlinenomath}}%
\newcommand*\patchBothAmsMathEnvironmentsForLineno[1]{%
  \patchAmsMathEnvironmentForLineno{#1}%
  \patchAmsMathEnvironmentForLineno{#1*}}%
\AtBeginDocument{%
\patchBothAmsMathEnvironmentsForLineno{equation}%
\patchBothAmsMathEnvironmentsForLineno{align}%
\patchBothAmsMathEnvironmentsForLineno{flalign}%
\patchBothAmsMathEnvironmentsForLineno{alignat}%
\patchBothAmsMathEnvironmentsForLineno{gather}%
\patchBothAmsMathEnvironmentsForLineno{multline}%
}

\usepackage{etoolbox}
\appto\appendix{\addtocontents{toc}{\protect\setcounter{tocdepth}{0}}}

\biboptions{numbers,sort&compress}

\newcommand{\cov}{\mathbb{C}\text{ov}}
\newcommand{\var}{\mathbb{V}\text{ar}}

\graphicspath{ {./figures/}}

\journal{Progress in Aerospace Sciences}

\makeindex

\makeatletter
\def\@author#1{\g@addto@macro\elsauthors{\normalsize%
    \def\baselinestretch{1}%
    \upshape\authorsep#1\unskip\textsuperscript{%
      \ifx\@fnmark\@empty\else\unskip\sep\@fnmark\let\sep=,\fi
      \ifx\@corref\@empty\else\unskip\sep\@corref\let\sep=,\fi
      }%
    \def\authorsep{\unskip,\space}%
    \global\let\@fnmark\@empty
    \global\let\@corref\@empty  
    \global\let\sep\@empty}%
    \@eadauthor={#1}
}
\makeatother

\begin{document}

\begin{frontmatter}


\title{Quantification of Model Uncertainty in RANS Simulations:\\ A Review}




\author{Heng Xiao\fnref{cofirst}\corref{corhx}}
\cortext[corhx]{Corresponding author. Tel: +1(540)2310926}
\ead{hengxiao@vt.edu}
\address{Kevin T. Crofton Department of Aerospace and Ocean Engineering, Virginia Tech, Blacksburg, VA 24060, USA}

\author{Paola Cinnella\fnref{cofirst}}
\ead{paola.cinnella@ensam.eu}
\address{Laboratoire DynFluid, Arts et M\'{e}tiers ParisTech, 151 Boulevard de l'Hopital, 75013 Paris, France}

\fntext[cofirst]{Contributed equally.}

\begin{abstract}
In computational fluid dynamics simulations of industrial flows, models based on the Reynolds-averaged Navier--Stokes (RANS) equations are expected to play an important role in decades to come. However, model uncertainties are still a major obstacle for the predictive capability of RANS simulations. This review examines both the parametric and structural uncertainties in turbulence models. We review recent literature on data-free (uncertainty propagation) and data-driven (statistical inference) approaches for quantifying and reducing model uncertainties in RANS simulations. Moreover, the fundamentals of uncertainty propagation and Bayesian inference are introduced in the context of RANS model uncertainty quantification. Finally, the literature on uncertainties in scale-resolving simulations is briefly reviewed with particular emphasis on large eddy simulations.
\end{abstract}

\end{frontmatter}



\tableofcontents

\mbox{}
\hrule
\begin{thenomenclature} 

 \nomgroup{A}

  \item [{$\overline{\boxed{\cdot}}$}]\begingroup ensemble averaging or spatial filtering\nomeqref {0}
		\nompageref{2}
  \item [{$\lvert\lvert \cdot \rvert\rvert_{L^2}$ }]\begingroup L2 norm\nomeqref {0}
		\nompageref{2}
  \item [{$\lvert\lvert A \rvert\rvert_{\mathsf{P}}$}]\begingroup norm of $A$ weighted by covariance matrix $\mathsf{P}^{-1}$, i.e., $\lvert\lvert A^\top \mathsf{P}^{-1} A\rvert\rvert$\nomeqref {0}
		\nompageref{2}
  \item [{$\star$}]\begingroup perturbed quantities\nomeqref {0}
		\nompageref{2}
  \item [{$\boxed{\cdot}^\top$}]\begingroup transpose of vectors and matrices\nomeqref {0}
		\nompageref{2}
  \item [{$:$}]\begingroup double dot of tensors $\tau_{ij} \frac{\partial U_i}{\partial x_j} \equiv \bm{\tau} \cddots \nabla \mathbf{U}$\nomeqref {0}
		\nompageref{2}
  \item [{$\circ$}]\begingroup Hadamard (element-wise) multiplication\nomeqref {0}
		\nompageref{2}
  \item [{$\frac{D \cdot}{D t}$}]\begingroup material derivative\nomeqref {0}
		\nompageref{2}

 \nomgroup{B}

  \item [{$\mathbf{a}$}]\begingroup anisotropy tensor\nomeqref {0}
		\nompageref{2}
  \item [{$c_1$, $c_2$, $c_3$}]\begingroup barycentric coordinates\nomeqref {0}
		\nompageref{2}
  \item [{$C_{\varepsilon1}$,$C_{\varepsilon2}$,$C_{\mu}$}]\begingroup RANS model coefficients\nomeqref {0}
		\nompageref{2}
  \item [{$C_{s}$}]\begingroup Smagorinsky constant\nomeqref {0}
		\nompageref{2}
  \item [{$\cov$}]\begingroup covariance of random variables\nomeqref {0}
		\nompageref{2}
  \item [{$\bm{d}$}]\begingroup discrepancy of observation and truth\nomeqref {0}
		\nompageref{2}
  \item [{$\frac{D \cdot}{D t}$}]\begingroup material derivative\nomeqref {0}
		\nompageref{2}
  \item [{$\mathcal{D}$}]\begingroup data used for inference\nomeqref {0}
		\nompageref{2}
  \item [{$D_\omega$}]\begingroup dissipation of turbulent frequency\nomeqref {0}
		\nompageref{2}
  \item [{$\mathbb{E}[Z]$}]\begingroup expectation of random variable $Z$\nomeqref {0}
		\nompageref{2}
  \item [{$f$}]\begingroup functional mapping\nomeqref {0}
		\nompageref{2}
  \item [{$\mathcal{GP}(\cdot, \cdot)$}]\begingroup Gaussian process\nomeqref {0}
		\nompageref{2}
  \item [{$\mathbf{h}$}]\begingroup unit quaternion\nomeqref {0}
		\nompageref{2}
  \item [{$\mathsf{H}$}]\begingroup observation matrix\nomeqref {0}
		\nompageref{2}
  \item [{$i, j, k$}]\begingroup indices\nomeqref {0}\nompageref{2}
  \item [{$\mathbf{I}$}]\begingroup second-order identity tensor\nomeqref {0}
		\nompageref{2}
  \item [{$I$}]\begingroup number of scenarios\nomeqref {0}
		\nompageref{2}
  \item [{$J$}]\begingroup objective function in optimization\nomeqref {0}
		\nompageref{2}
  \item [{$k$}]\begingroup turbulent kinetic energy\nomeqref {0}
		\nompageref{2}
  \item [{$K(\cdot, \cdot)$}]\begingroup kernel for Gaussian processes\nomeqref {0}
		\nompageref{2}
  \item [{$\mathsf{K}$}]\begingroup Kalman gain matrix (in EnKF)\nomeqref {0}
		\nompageref{2}
  \item [{$K$}]\begingroup number of models\nomeqref {0}\nompageref{2}
  \item [{$l$}]\begingroup length scale in covariance kernel\nomeqref {0}
		\nompageref{2}
  \item [{$\mathcal{L}$}]\begingroup linear differential operator\nomeqref {0}
		\nompageref{2}
  \item [{$\mathcal{M}$, $M_i$}]\begingroup set of models; model\nomeqref {0}
		\nompageref{2}
  \item [{$\mathbf{n}$}]\begingroup axis of rotation\nomeqref {0}
		\nompageref{2}
  \item [{$\mathsf{N}$}]\begingroup normal distribution\nomeqref {0}
		\nompageref{2}
  \item [{$\mathcal{N}$}]\begingroup nonlinear differential operator\nomeqref {0}
		\nompageref{2}
  \item [{$\mathcal{O}(\cdot)$}]\begingroup of the order of\nomeqref {0}
		\nompageref{2}
  \item [{$p$}]\begingroup instantaneous pressure\nomeqref {0}
		\nompageref{2}
  \item [{$p'$ }]\begingroup pressure fluctuation\nomeqref {0}
		\nompageref{2}
  \item [{$p(z)$}]\begingroup probability distribution of $Z$\nomeqref {0}
		\nompageref{2}
  \item [{$P_1$, $P_2$}]\begingroup two locations in wing--body juncture flow\nomeqref {0}
		\nompageref{2}
  \item [{$\mathbb{P}$}]\begingroup (discrete) probability mass function\nomeqref {0}
		\nompageref{2}
  \item [{$\mathcal{P}$}]\begingroup production (of TKE, Reynolds stresses, or turbulent frequency) \nomeqref {0}
		\nompageref{2}
  \item [{$\mathsf{P}$}]\begingroup covariance matrix of state vector\nomeqref {0}
		\nompageref{2}
  \item [{$P$}]\begingroup mean pressure\nomeqref {0}\nompageref{2}
  \item [{$\bm{q}$}]\begingroup mean flow features\nomeqref {0}
		\nompageref{2}
  \item [{$Q^{\delta}$}]\begingroup rotation matrix\nomeqref {0}
		\nompageref{2}
  \item [{$\mathbb{R}$}]\begingroup real number space\nomeqref {0}
		\nompageref{2}
  \item [{$\mathsf{R}$}]\begingroup covariance matrix of observation error\nomeqref {0}
		\nompageref{2}
  \item [{$\mathbf{S}$}]\begingroup strain rate tensor\nomeqref {0}
		\nompageref{2}
  \item [{$\mathcal{S}$}]\begingroup source terms\nomeqref {0}
		\nompageref{2}
  \item [{$\tilde{S}_i$}]\begingroup scenario (in BMSA)\nomeqref {0}
		\nompageref{2}
  \item [{$t$}]\begingroup time\nomeqref {0}\nompageref{2}
  \item [{$T_\omega$}]\begingroup transport of turbulent frequency\nomeqref {0}
		\nompageref{2}
  \item [{$u_i$, $\bm{u}$}]\begingroup instantaneous velocity\nomeqref {0}
		\nompageref{2}
  \item [{$u'_i$, $\bm{u}'$}]\begingroup velocity fluctuation\nomeqref {0}
		\nompageref{2}
  \item [{$U_i$, $\bm{U}$}]\begingroup mean velocity\nomeqref {0}
		\nompageref{2}
  \item [{$\mathbb{V}\text{ar}[Z]$}]\begingroup variance of random variable $Z$\nomeqref {0}
		\nompageref{2}
  \item [{$\mathbf{V}$}]\begingroup eigenvectors of second order tensor\nomeqref {0}
		\nompageref{2}
  \item [{$w_\alpha$ }]\begingroup coefficients in expansion of random field\nomeqref {0}
		\nompageref{2}
  \item [{$W$ }]\begingroup Wiener process (in SDEs) \nomeqref {0}
		\nompageref{2}
  \item [{$x_i$, $\bm{x}$ }]\begingroup spatial coordinates\nomeqref {0}
		\nompageref{2}
  \item [{$y$ }]\begingroup model output\nomeqref {0}\nompageref{2}
  \item [{$\bm{z}$}]\begingroup augmented state vector\nomeqref {0}
		\nompageref{2}
  \item [{$Z$, $z$}]\begingroup random variable and its realization\nomeqref {0}
		\nompageref{2}

 \nomgroup{C}

  \item [{$\alpha$}]\begingroup index for basis functions\nomeqref {0}
		\nompageref{2}
  \item [{$\beta$}]\begingroup multiplicative discrepancy field\nomeqref {0}
		\nompageref{2}
  \item [{$\gamma$}]\begingroup parameter in regularization term\nomeqref {0}
		\nompageref{2}
  \item [{$\Delta_g$}]\begingroup grid spacing/filter width in LES\nomeqref {0}
		\nompageref{2}
  \item [{$\delta$}]\begingroup discrepancies\nomeqref {0}
		\nompageref{2}
  \item [{$\delta_{ij}$}]\begingroup Kronecker delta, second-order identity tensor\nomeqref {0}
		\nompageref{2}
  \item [{$\epsilon$}]\begingroup noise in experimental data\nomeqref {0}
		\nompageref{2}
  \item [{$\varepsilon$}]\begingroup dissipation rate\nomeqref {0}
		\nompageref{2}
  \item [{$\zeta$}]\begingroup truth in the context of model uncertainty\nomeqref {0}
		\nompageref{2}
  \item [{$\theta$, $\bm{\theta}$}]\begingroup model parameter(s)\nomeqref {0}
		\nompageref{2}
  \item [{$\vartheta$}]\begingroup angle of rotation\nomeqref {0}
		\nompageref{2}
  \item [{$\kappa$}]\begingroup von Karman constant\nomeqref {0}
		\nompageref{2}
  \item [{$\lambda_i$}]\begingroup eigenvalues for anisotropy tensor\nomeqref {0}
		\nompageref{2}
  \item [{$\mathbf{\Lambda}$}]\begingroup diagonal matrix of eigenvalues for anisotropy tensor\nomeqref {0}
		\nompageref{2}
  \item [{$\mu$}]\begingroup dynamic viscosity of fluids\nomeqref {0}
		\nompageref{2}
  \item [{$\nu$}]\begingroup kinematic viscosity\nomeqref {0}
		\nompageref{2}
  \item [{$\nu_t$}]\begingroup turbulent eddy viscosity\nomeqref {0}
		\nompageref{2}
  \item [{$\xi$}]\begingroup physical state of the system\nomeqref {0}
		\nompageref{2}
  \item [{$\rho$}]\begingroup fluid density\nomeqref {0}\nompageref{2}
  \item [{$\bm{\varsigma}$}]\begingroup latent variables  (e.g., geometry, boundary conditions in CFD model)\nomeqref {0}
		\nompageref{2}
  \item [{$\sigma$}]\begingroup variance (field) of random fields\nomeqref {0}
		\nompageref{2}
  \item [{$\sigma_k$, $\sigma_{\varepsilon}$}]\begingroup coefficients in turbulence models\nomeqref {0}
		\nompageref{2}
  \item [{$\Sigma$}]\begingroup covariance matrix\nomeqref {0}
		\nompageref{2}
  \item [{$\bm{\tau}$}]\begingroup Reynolds stress\nomeqref {0}
		\nompageref{2}
  \item [{$\nu_t$}]\begingroup turbulent viscosity\nomeqref {0}
		\nompageref{2}
  \item [{$\phi_i(\bm{x})$}]\begingroup basis functions (e.g., from Karhunen--Loeve expansion)\nomeqref {0}
		\nompageref{2}
  \item [{$\varphi_i$}]\begingroup Euler angles\nomeqref {0}
		\nompageref{2}
  \item [{$\Psi$}]\begingroup quantities to be predicted\nomeqref {0}
		\nompageref{2}
  \item [{$\omega$ }]\begingroup turbulent frequency\nomeqref {0}
		\nompageref{2}
  \item [{$\mathbf{\Omega}$}]\begingroup rotation-rate tensor\nomeqref {0}
		\nompageref{2}

 \nomgroup{D}

  \item [{BMSA}]\begingroup Bayesian model--scenario averaging\nomeqref {0}
		\nompageref{2}
  \item [{CFD}]\begingroup computational fluid dynamics\nomeqref {0}
		\nompageref{2}
  \item [{DNS}]\begingroup direct numerical simulation\nomeqref {0}
		\nompageref{2}
  \item [{EARSM}]\begingroup explicit algebraic Reynolds stress model \nomeqref {0}
		\nompageref{2}
  \item [{EnKF}]\begingroup ensemble Kalman filtering\nomeqref {0}
		\nompageref{2}
  \item [{gPC}]\begingroup generalized  polynomial chaos \nomeqref {0}
		\nompageref{2}
  \item [{LES}]\begingroup large eddy simulation\nomeqref {0}
		\nompageref{2}
  \item [{LHS}]\begingroup Latin hypercube sampling\nomeqref {0}
		\nompageref{2}
  \item [{PCE}]\begingroup polynomial chaos expansion\nomeqref {0}
		\nompageref{2}
  \item [{PDE}]\begingroup partial differential equation\nomeqref {0}
		\nompageref{2}
  \item [{pdf}]\begingroup  probability density function\nomeqref {0}
		\nompageref{2}
  \item [{pmf}]\begingroup probability mass function\nomeqref {0}
		\nompageref{2}
  \item [{MAP}]\begingroup maximum a posteriori\nomeqref {0}
		\nompageref{2}
  \item [{QoI}]\begingroup quantity of interest\nomeqref {0}
		\nompageref{2}
  \item [{MLMC}]\begingroup multilevel Monte Carlo\nomeqref {0}
		\nompageref{2}
  \item [{MCMC}]\begingroup Markov chain Monte Carlo\nomeqref {0}
		\nompageref{2}
  \item [{NS}]\begingroup Navier--Stokes\nomeqref {0}\nompageref{2}
  \item [{RANS}]\begingroup Reynolds-averaged Navier--Stokes\nomeqref {0}
		\nompageref{2}
  \item [{RSTE}]\begingroup Reynolds stress transport equation\nomeqref {0}
		\nompageref{2}
  \item [{RSTM}]\begingroup Reynolds stress transport model\nomeqref {0}
		\nompageref{2}
  \item [{SA}]\begingroup Spalart--Allmaras (turbulence model)\nomeqref {0}
		\nompageref{2}
  \item [{SDE}]\begingroup stochastic differential equation\nomeqref {0}
		\nompageref{2}
  \item [{SGS}]\begingroup sub-grid scale\nomeqref {0}\nompageref{2}
  \item [{TKE}]\begingroup turbulent kinetic energy\nomeqref {0}
		\nompageref{2}
  \item [{UQ}]\begingroup uncertainty quantification\nomeqref {0}
		\nompageref{2}

\end{thenomenclature}

\hrule
\mbox{}


\section{Introduction}
\label{sec:introduction}

Turbulence affects natural and engineered systems from sub-meter to planetary scales yet it is among the last unsolved problems in classical physics.   Accurate predictions of turbulent flows are of vital importance for the design, analysis, and operation of many critical systems in aerospace engineering such as aircraft, spacecraft, and gas turbine engines.  The dynamics of  fluid flows are described by the Navier--Stokes (NS) equations.  While  many applications in aerospace engineering involve compressible flows, reacting flows or two-phase flows, for illustration purposes we restrict our attention to the NS equations for incompressible flows of constant-property, Newtonian fluids are shown below:
  \begin{subequations} \label{eq:ns}
  \begin{align}
    \quad \frac{\partial u_i}{\partial x_i} & = 0 \label{eq:ns-mass} \\
    \frac{\partial u_i}{\partial t}+\frac{\partial \left( u_i u_j \right)}{\partial x_j} & =
    -\frac{\partial {p}}{\partial x_i} +\frac{1}{Re}\frac{\partial^2 u_i}{\partial x_j \partial
      x_j} ,
\label{eq:ns-momentum}
\end{align}
\end{subequations}
where $u_i$, $p$, $x_i$ and $t$ are, respectively, the flow velocity, pressure, and spatial and temporal coordinates. Although simpler in form than the partial differential equations governing the above-mentioned problems, incompressible NS equations cover a very wide variety of flow configurations and bear the key difficulty that leads to the turbulence modeling dilemma, i.e., the nonlinear convective  term in Equation~(\ref{eq:ns-momentum}). Equation~(\ref{eq:ns}) is normalized with respect to a reference length $L_{\text{ref}}$, a reference velocity $U_{\text{ref}}$, and the density $\rho$ and viscosity $\mu$ of the fluid.  The parameter $Re=\rho U_{\text{ref}} L_{\text{ref}}/\mu$ is the Reynolds number, a measure of the relative importance of inertia to viscous forces.  Because of the nonlinearity of the convection terms ${\partial \left( u_i u_j \right)}/{\partial x_j}$, the NS equations admit chaotic solutions when the Reynolds number is beyond some flow-dependent critical value. As the Reynolds number increases, eventually the flow reaches a state of motion characterized by strong three-dimensional and unsteady chaotic fluctuations of the velocity and pressure fields, which is referred to as the turbulent regime.

\subsection{Landscape of turbulence modeling}
\label{sec:landscape}

\begin{figure}[!htbp]
  \centering
    \includegraphics[width=0.55\textwidth]{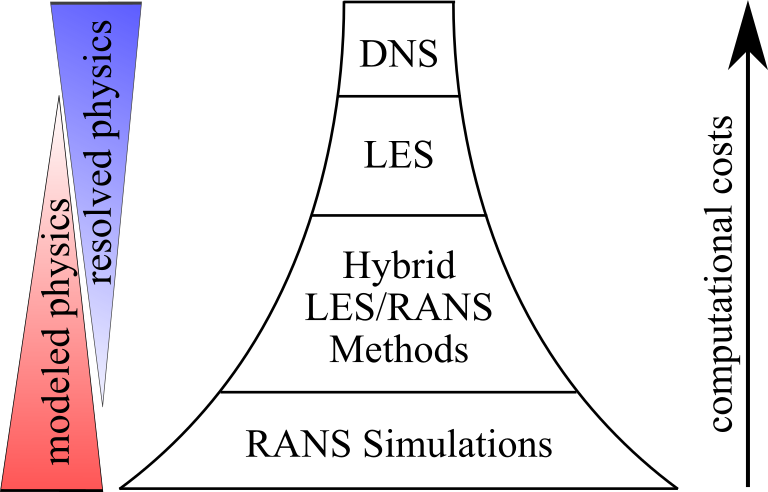}
   \caption{A schematic representation of the hierarchy of turbulence modeling approaches based on computational costs and the amounts of resolved versus modeled physics. Figure inspired by \citet{sagaut2006multiscale}. Abbreviations: DNS, direct numerical simulations; LES, large eddy simulations; RANS, Reynolds-Averaged Navier--Stokes.}
  \label{fig:classification}
\end{figure}

Turbulent flows are characterized by a wide range of spatial and temporal scales.  Consequently, performing direct numerical simulations (DNS) by solving the NS equations and resolving all the turbulence scales are prohibitively expensive, particularly for high Reynolds number flows.  Practically used turbulence modeling strategies range from DNS with the highest fidelity, where all physics of spatial and temporal scales are \emph{resolved} and no modeling is involved, to Reynolds averaged Navier--Stokes (RANS) simulations with the lowest fidelity, where the entire range of turbulent flow scales is \emph{modeled}. This model hierarchy is illustrated in Figure~\ref{fig:classification}, with the top represented by the most physics-resolving and computationally expensive approach (DNS) and the bottom by the most empirical and computationally affordable approach (RANS).  Lower fidelity models toward the bottom of the hierarchy involve more flow-dependent, uncertain closures than the higher-fidelity, scale-resolving approaches towards the top of the hierarchy. On the other hand, high-fidelity, scale-resolving models are more susceptible to influences from numerical uncertainties as well as initial and boundary conditions.

A compromise between DNS and RANS simulations at two ends of the spectrum is large eddy simulation (LES), in which only the larger, more energetic scales are resolved, while scales below a cutoff threshold are filtered out. The filtered Navier--Stokes equations contain a subgrid-scale (SGS) stress that is unclosed and needs to be modeled. The SGS stress term represents the interactions between the filtered and resolved scales, which result from the nonlinear, convection term~\cite{sagaut2006large}.  Large eddy simulations have significantly reduced computational costs compared to DNS for shear flows far removed from wall boundaries. Unfortunately, they remain prohibitively expensive for wall bounded flows at high Reynolds number due to the small yet energetic scales dominating the dynamics in the near-wall regions~\cite{spalart2009detached-eddy}. This challenge has led to the development of methods combining LES in free shear regions with RANS models or other simplified models (e.g., boundary layer equation or law of the wall) in the under-resolved near-wall regions. Such approaches include hybrid RANS/LES models~\citep{frohlich2008hybrid,chaouat2017state} and wall-modeled LES~\cite{cabot2000approximate,piomelli2002wall,kawai2012wall,yang2015integral}, among others.

While scale-resolving simulations such as DNS, LES, and hybrid RANS/LES provide more insights of fluid flow physics, in many simulations of engineering turbulent flows such as those for aerodynamic design and optimization, the quantities of interest depend on the mean flow only, and the instantaneous flow fields are not of concern.  In these cases it is desirable to solve for the mean flow more efficiently.  For that purpose, the instantaneous velocity $u_i$ and pressure $p$ are decomposed into the sum of the mean\footnote{Note that several definitions exist for the mean or average quantities~\citep[see, e.g.,][]{wilcox2006turbulence}. The most general one is the statistical ensemble average, which however is rarely used in current practice due to the large number of independent flow realizations required for convergence. For statistically steady flow, time average is used instead based on an ergodicity hypothesis. The same is also used for unsteady flows, although its validity is still controversial.} components $U_i$ and $P$ and the fluctuations~$u'_i$ and~$p'$, respectively.  Substituting the decomposition into the Navier--Stokes equations and taking the ensemble-average leads to the RANS equations:
\begin{subequations}
  \label{eq:rans}
  \begin{align}
    \quad \frac{\partial U_i}{\partial x_i} & = 0  \\
    \frac{\partial U_i}{\partial t}+\frac{\partial \left(U_i U_j \right)}{\partial x_j}
                                            & =  -\frac{\partial {P}}{\partial x_i}
                                              +\frac{1}{Re}\frac{\partial^2 U_i}{\partial x_j \partial  x_j}
                                              - \frac{\partial \overline{  u_i'  u_j'}}{\partial x_j}  \; .
 \label{eq:rans-momentum}
  \end{align}
  \end{subequations}
 The RANS equations are  similar in form to the Navier--Stokes equations except for the term involving the tensor $-\overline{u_i' u_j'}$. As with the SGS stress term in the filtered NS equations for LES, this term stems from the nonlinear convection term in the NS equation and represents the cross-component covariance among the velocity fluctuations. It is often referred to as Reynolds stress due to its formal similarity to the viscous stresses and is denoted as 
\begin{equation}
\tau_{ij} = -\overline{u_i' u_j'}  \;  .  
\end{equation}
Since the velocity fluctuations are not available in RANS simulations, one must resort to closure models to supply Reynolds stresses, which lies at the root of most efforts of turbulence modeling.

The choice of the appropriate modeling level remains a matter of expert judgment. In particular, it inevitably involves a compromise between computational cost and predictive accuracy.  Even after a given fidelity level is selected (e.g., RANS or LES), several possible closure models may be designed for relating the unclosed terms to the resolved variables. These closure models differ both by their mathematical structure and by the associated model parameters.  The common practice in turbulence modeling is to leave the choice of a specific closure model to user judgment and to treat model parameters as adjustable coefficients that are generally calibrated to reproduce simple, canonical flows.  Both of the preceding aspects, however, represent sources of uncertainty in the prediction of new flows.  Recent development of turbulence modeling in RANS, LES, and hybrid approaches has been reviewed by~\citet{durbin2018some}.  Despite considerable progress recently made in LES and hybrid RANS/LES models (e.g., \citep{sagaut2006large,girimaji2006partially-averaged,frohlich2008hybrid,spalart2009detached-eddy,xiao2012consistent}), RANS models are expected to remain the workhorse in engineering practice for decades to come, due to their much lower computational costs and superior robustness.  For this reason,  this review mainly focuses on the quantification and reduction of uncertainties in RANS models.

\begin{figure}[!htbp]
  \centering
    \includegraphics[width=0.8\textwidth]{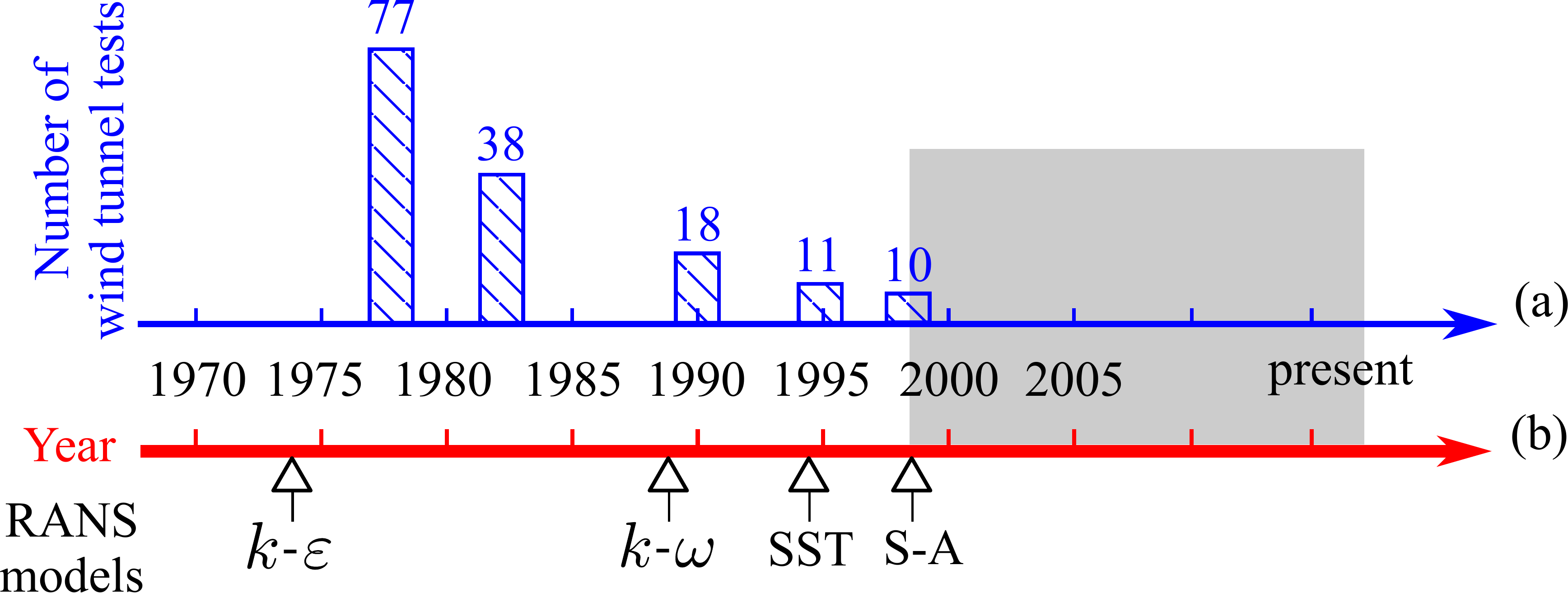}
   \caption{Stagnation of turbulence modeling in the past few decades (shaded regions), showing (a) the number of wind tunnel tests required in the design cycle of commercial aircraft in the past five decades~\citep{johnson2005thirty}  and (b) the time at which commonly used models were developed. }
  \label{fig:wind-tunnel}
\end{figure}

The landscape of RANS-based turbulence modeling has not changed for decades. The stagnation is evident from two observations as illustrated in Figure~\ref{fig:wind-tunnel}. First, the number of wind tunnel tests performed in a typical design cycle of a commercial airplane was reduced from 75 in the 1970s to 10 in the 1990s, but this number has been stagnant since then, with turbulence models being the major bottleneck in predictive accuracies~\citep{johnson2005thirty}. Second, most of the currently used turbulence models were developed decades ago  and provide unsatisfactory performance for many flows.  Generations of researchers have labored for many decades on dozens of turbulence models, yet none of them achieved predictive generality.  Flow-specific tuning and fudge functions are still an indispensable part of RANS simulations~\citep{spalart2015philosophies}.  Current development of improved turbulence models faces the dilemma of conserving the low computational costs and high robustness of RANS approaches while incorporating as much physics as possible.

\subsection{Origin of  uncertainties in RANS models}
\label{sec:origin-rans-uq}

\begin{figure}[!htbp]
  \centering
  \includegraphics[width=0.93\textwidth]{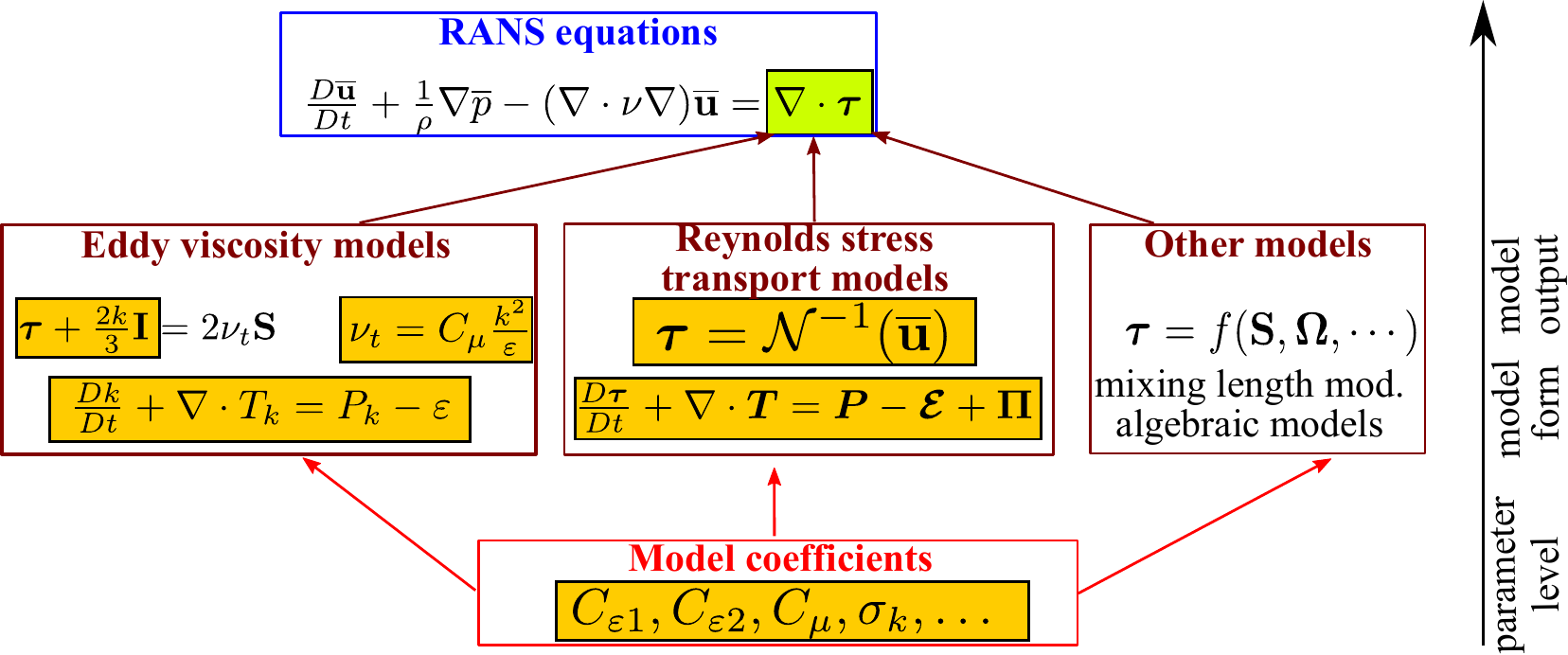}
  \caption{Stages of turbulence modeling in commonly used models with Reynolds stress transport models and linear eddy viscosity models as examples. Such a hierarchy provides a clear map on where model uncertainties can be introduced and inferred (shown as shaded items).  $\frac{D \cdot}{D t}$ denotes material derivative.}
  \label{fig:hierarchy}
\end{figure}

A recent review on data-driven turbulence modeling strategies~\citep{duraisamy2019turbulence} classified the model uncertainties in RANS simulations into four levels, including uncertainties due to information loss in the Reynolds-averaging process, uncertainties in representing the Reynolds stress as a functional form of the mean fields,  uncertainties in the choice of the specific function, and uncertainties in the parameters of a given model.   In this review, we will  
focus on the uncertainties due to the choice of functional forms and parameters in the turbulence models.  Figure~\ref{fig:hierarchy} 
shows a graphical representation of different sources of model uncertainties in typical RANS models.

The following observations about the Reynolds stress tensor have profound implications for turbulence modeling and RANS model uncertainty quantification. First, it is a covariance tensor of velocity fluctuations as pointed out above, and mathematically any covariance tensor must be symmetric positive semi-definite. This is referred to as \emph{realizability requirement}. Second, it appears in the RANS momentum equation in its divergence $\nabla \cdot \boldsymbol{\tau}$. While the Reynolds stress as a symmetric rank-two tensor has six independent components, the divergence $\nabla \cdot \boldsymbol{\tau}$ as a forcing term only has three components.  The majority of existing turbulence models use the Reynolds stress as the target of modeling (Figure~\ref{fig:hierarchy}). The rationale behind this choice is that the divergence form makes it easier to ensure conservation of momentum. That is, in this form the momentum is introduced into the system by the modeled Reynolds stress only through the boundaries and not within the volume.  In contrast, directly constructing such a conservative forcing term is not straightforward~\cite{perot1996new}.  Therefore, in the remainder of this paper we discuss only turbulence models based on the Reynolds stress $\boldsymbol{\tau}$.

Reynolds stress based turbulence models require prescribing a constitutive relation for~$\boldsymbol{\tau}$ as a function of the mean flow fields.  The most widely used class of models, generally known as linear eddy viscosity models, relies on the Boussinesq analogy (see, e.g., \cite{wilcox2006turbulence}).  This assumption states that the anisotropic part of~$\boldsymbol{\tau}$  behaves similarly to the viscous stress tensor of a Newtonian fluid, i.e. it is a linear function of the local mean flow rate-of-strain~$S_{ij}$:
\begin{subequations}
\label{eq:evm}
\begin{align}
 \tau_{ij} + \frac{2k}{3} \delta_{ij}& =  2 \nu_t S_{ij}     \label{eq:evm-tau} \\
\textrm{with}  \quad S_{ij} & =\frac{1}{2}\left(\frac{\partial U_i}{\partial x_j}+\frac{\partial U_j}{\partial x_i}\right) ,
 \label{eq:evm-s}
\end{align}
\end{subequations}
where $\tau_{ij} + \frac{2k}{3} \delta_{ij}$ is the Reynolds stress anisotropy, $k = \frac{1}{2} \overline{u'_i u'_i} = - \frac{1}{2}\tau_{ii}$ is the turbulent kinetic energy with a summation over index $i$ implied, $\delta_{ij}$ is the Kronecker delta (or the second order identity tensor in its vector form $\mathbf{I}$), and the eddy viscosity $\nu_t$ is the proportionality scalar.

\begin{figure}[!htbp]
  \centering
    \subfloat[Effects of turbulence model]{\includegraphics[width=0.45\textwidth]{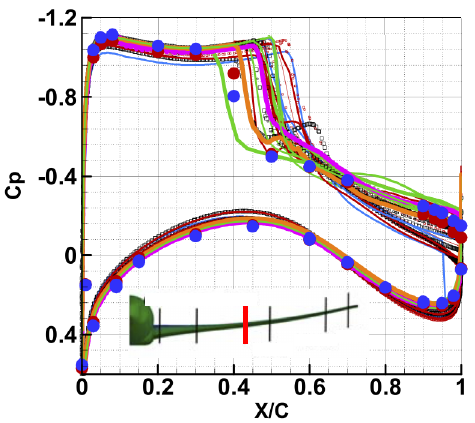}}
    \subfloat[Effects of model coefficients]{\includegraphics[width=0.5\textwidth]{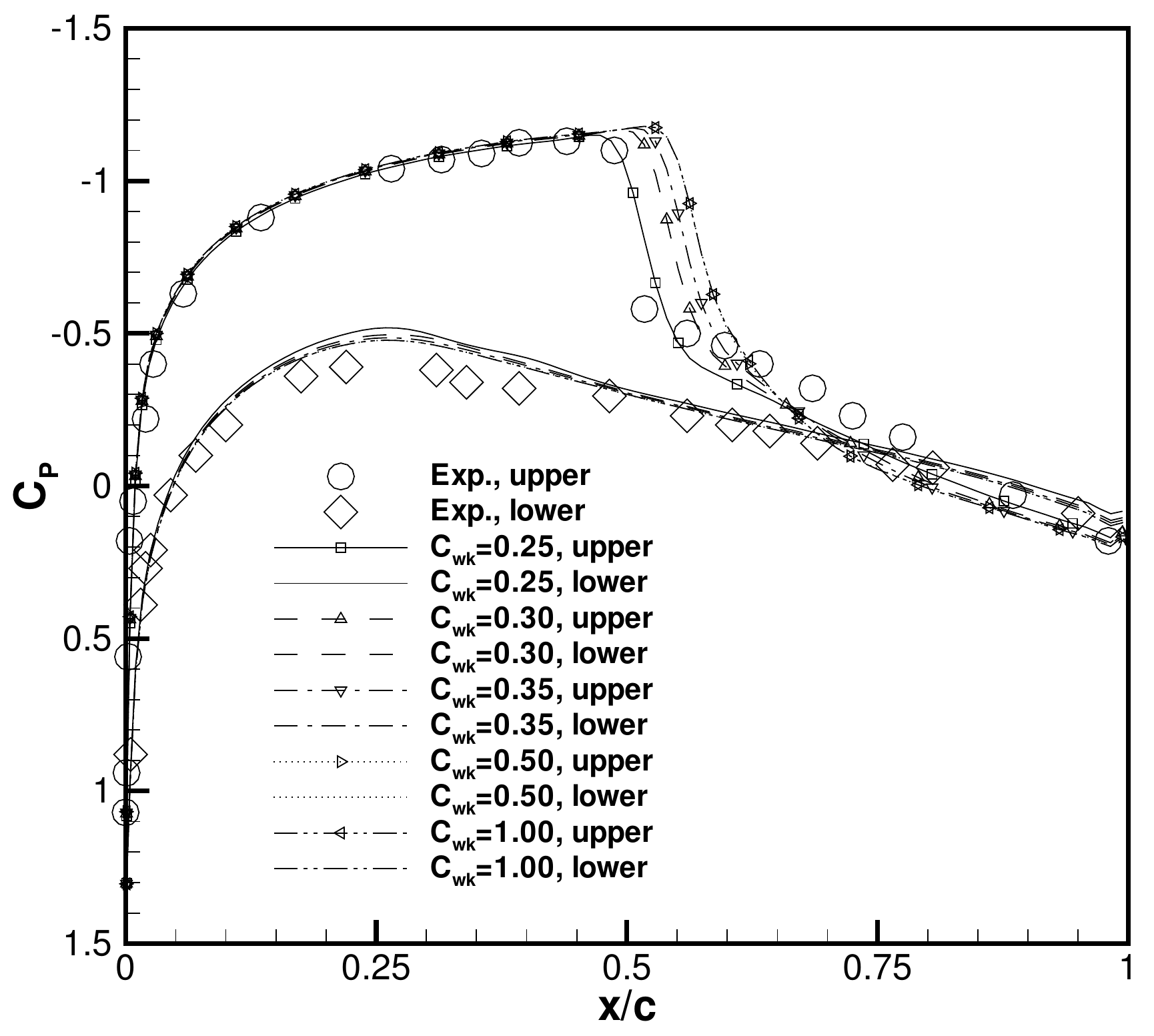}}
    \caption{Examples of uncertainties in RANS predictions of pressure coefficient $C_p$ distribution on wings and airfoils due to (a) model form and (b) model coefficients. Panel (a) shows the $C_p$ profile on a CRM wing-body configuration at $4.0^\circ$ angle of attack.  Results are from the $6^{\text{th}}$ AIAA CFD drag prediction workshop based on different RANS models, including $k$--$\varepsilon$ model, $k$--$\omega$ model, SA model, SA with quadratic constitutive relation (QCR), and EARSM. The location of the presented pressure distribution is indicated by the red/solid line on the wing (see inset; showing the port half of the fuselage and the wing only).  Figure reprinted with permission from~\citet{tinoco2018summary}. Panel (b) shows the $C_p$ profile on a NACA0012 airfoil in a transonic flow with freestream Mach number 0.8 and Reynolds number $9\times 10^6$, obtained from RANS simulations with the algebraic model of Baldwin and Lomax~\cite{baldwin1978thin}.  The figure shows the effect of varying $C_{wk}$, one of the seven model parameters, from 0.25 to 1, adopted from an unpublished report of the second author~\cite{cinnella2016review}.
      \label{fig:CRM-RANS}
    }
\end{figure}

The limitations of the Boussinesq assumption have been widely recognized in the literature, particularly for flows with separation, streamline curvature, or strong pressure gradients (see, e.g., \citep{wilcox2006turbulence} for a review).  Since it is often not possible to know beforehand if one or more of such flow features will be present in a new flow configuration, predictions based on the RANS equations are flawed by a structural (i.e. model-form) uncertainty~\citep{draper1995assessment,kennedy2001bayesian}.  Several attempts have been made to overcome the weaknesses of linear eddy-viscosity models, e.g., by developing nonlinear eddy viscosity models~\citep{speziale1987on}, explicit algebraic Reynolds stress models (EARSM)~\citep{gatski1993on}, and Reynolds stress transport models (RSTM)~\cite{launder1975progress,wilcox2006turbulence}.  All such models rely on more sophisticated constitutive relations than Equation~(\ref{eq:evm}). Nevertheless, such sophisticated models lack the robustness of the simple linear eddy viscosity models. For example, cubic eddy viscosity models involve many more parameters, which are difficult to calibrate with available data~\cite{ray2018learning}.  As another example, the Reynolds stress transport equations have a pressure--strain-rate that needs to be modeled, and the predictive performance of RSTM are highly sensitive to its modeling. Consequently, the lack of robustness restricts these advanced models to a small fraction of practical turbulent flows despite their theoretical superiority~\cite{spalart2015philosophies}, and no turbulence models are able to accurately predict the flow physics in all circumstances.  The importance of model uncertainty is clearly illustrated in Figure~\ref{fig:CRM-RANS}a, which shows the predicted pressure distribution on the wing section of a Common Research Model (CRM) predicted by a number of turbulence models. A large scattering of the predictions is observed, particularly downstream of the shock wave generated at the upper wing surface.

In addition to the structural uncertainties, parametric uncertainties arising from the coefficients closure models also have to be accounted for.  Such coefficients are usually calibrated against experimental data for a set of simple flows (e.g., the decay of homogeneous and isotropic turbulence, flat plate boundary layers, and simple shear flows), which are generally far from practical applications.  Moreover, the calibration data suffer from measurement errors, which inevitably impair the credibility of the calibrated parameters. Finally, many of the nominal coefficients found in the RANS modeling literature may not correspond to best-fit of calibration dataset, but were chosen based on numerical considerations.  In practice, the closure coefficients are often empirically re-tuned by using heuristic and trial-and-error approaches in order to fit available data for a target class of flows.  Figure~\ref{fig:CRM-RANS}b illustrates the effect of varying only one of the seven parameters in the algebraic model of \citet{baldwin1978thin}. In particular, the location of the shock wave at the airfoil upper surface and the post-shock pressure are very sensitive to the varied coefficient~\cite{cinnella2016review}.

Both the parametric and the structural uncertainties mentioned above are of \emph{epistemic} nature, i.e. theoretically they could be reduced when better knowledge of turbulent flow physics and/or more abundant or more accurate data become available.  This is in contrast to \emph{aleatory} uncertainties, which arise from intrinsic variability of a process, e.g., uncertainties in manufactured geometries~\citep{parussini2007fictitious,liu2017quantification}, operation conditions of turbines or aircraft~\citep{avdonin2018quantification} or inflow conditions~\citep{gorle2015quantifying,mariotti2016freestream}.  In practice, reducing epistemic uncertainties by leveraging additional knowledge (e.g., by developing more advanced models to incorporate such knowledge) is far from straightforward.  Additionally, sophisticated models may lack numerical robustness or incur excessive computational costs.  Except for a few canonical examples, it is challenging, if not impossible, to identify the dominant source of uncertainty with definitive evidence, even for a given flow and a specific turbulence model. For instance, in many cases it is possible to improve the results of a model flawed by structural inadequacy by over-tuning its closure parameters. However, such over-tuning typically leads to poor predictions when applying the model to different flows from the calibration flows.  Such a phenomenon is referred to as over-fitting in statistics and machine learning~\citep{james2013introduction}.

\subsection{Approaches for quantifying uncertainties in turbulence models}
\label{sec:uq-turbulence-model}

Empirical assessment of uncertainties in turbulence models dates back to the early days of turbulence modeling, but rigorous treatments of such uncertainties in a statistical framework is only a recent development.  While it is a consensus that aleatory uncertainties are best represented in a probabilistic framework, different approaches have been pursued for epistemic uncertainties. Because epistemic uncertainties come from lack of knowledge, it is a philosophical question whether to treat such uncertainties in probabilistic framework.  In the Bayesian framework, all sources of uncertainty are represented as subjective beliefs and assigned a measure of probability. This review primarily focuses on Bayesian approaches. However, many other non-Bayesian or non-probabilistic approaches for treating epistemic uncertainties exist. Examples include imprecise probability theory~\citep{klir2006uncertainty}, probability bounds analysis~\citep{ferson1996what,ferson1996different,liu2004arithmetic}, Dempster--Shafer evidence theory~\citep{shafer1976mathematical}, fuzzy sets~\citep{zadeh1996fuzzy}, and credal sets~\citep{bernardini2010bounding}.  For an overview and applications of some of these approaches, see refs.~\cite{oberkampf2010verification,roy2011comprehensive}.

Current approaches for quantifying the model-form uncertainties associated with RANS simulations can be classified into parametric and non-parametric approaches\footnote{Here we have used the terminology (``parametric'' and ``non-parametric'') rather liberally, which is closely related to, but not strictly consistent with, the standard terminology in the statistics literature. In statistics, parametric models refer to those parameterized by a finite set of parameters, while non-parametric models refer to those with infinite degrees of freedom (e.g., spatial random fields).} depending on where the uncertainties are introduced.  In parametric approaches, uncertainties are introduced to the closure coefficients of chosen turbulence models, based on which the overall prediction uncertainties are assessed.  Although neglecting uncertainties in the model forms and constrained by the baseline models, the parametric approach has the advantage of being non-intrusive and thus readily available to CFD practitioners.  On the other hand, non-parametric approaches directly investigate the uncertainties on modeled terms (fields in RANS solvers), e.g., the eddy viscosity~\citep{dow2011quantification}, source terms in the turbulent transport equations~\citep{singh2016using}, or the Reynolds stress itself~\citep{xiao2016quantifying,ling2016reynolds}. An advantage of these approaches is that the uncertainties of modeled terms reveal more physical insights than the uncertainties of the model coefficients, e.g., allowing the flow regions more prone to model inaccuracies to be identified. However, non-parametric approaches also introduce new challenges, since the uncertainties are now quantified for spatial fields, which theoretically have infinite degrees of freedom.  The dimensionality (and thus the cost of the uncertainty quantification) increases with the size of mesh used to discretize the RANS equations. Additionally, such methods are intrusive by nature and thus are less friendly to industrial practitioners who are limited to black-box CFD solvers. 

\begin{figure}[!htbp]
  \centering
   \subfloat[Uncertainty propagation (forward propagation)]{\includegraphics[width=0.5\textwidth]{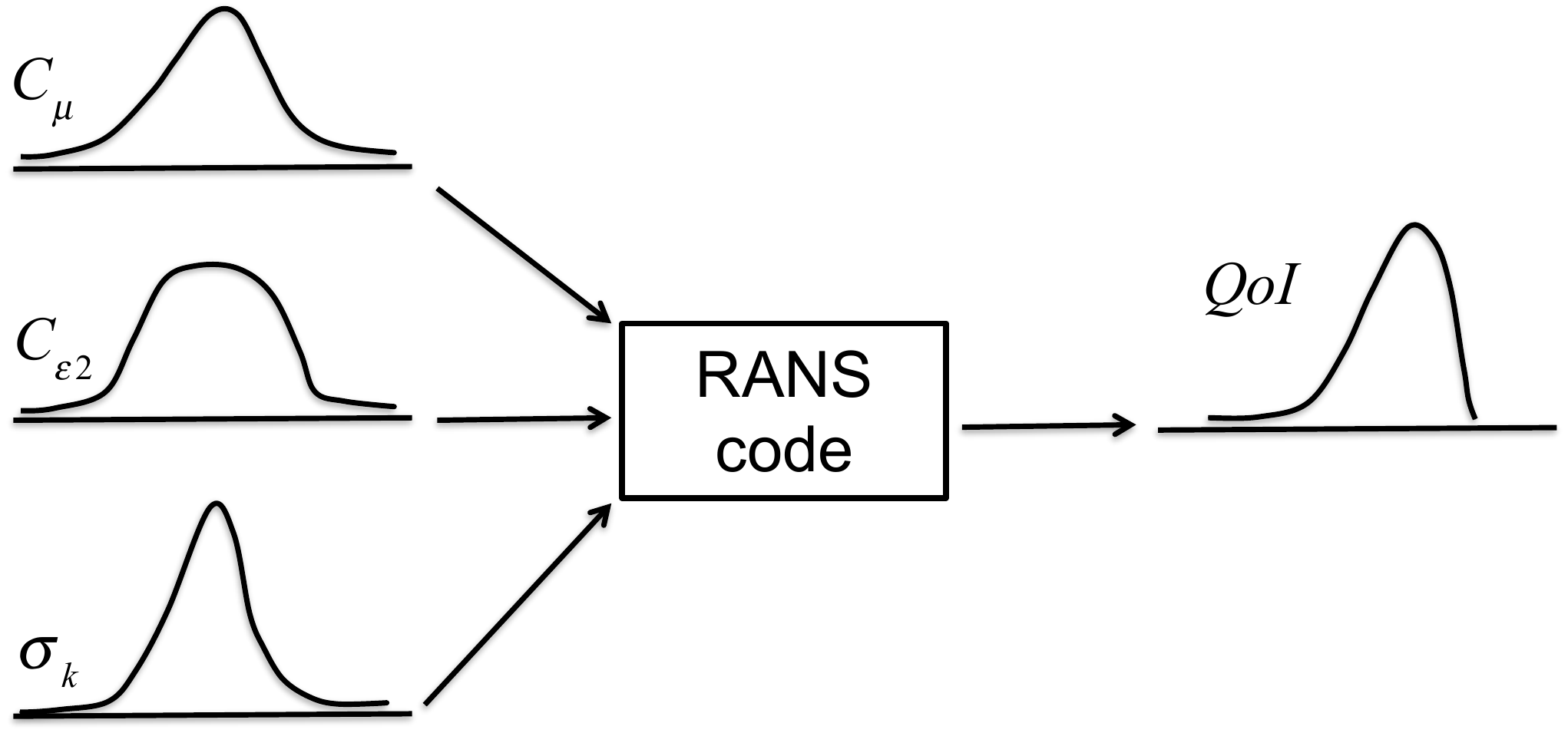}\label{fig:forward-backward_forward}} \\
   \subfloat[Statistical inference (backward analysis)]{\includegraphics[width=0.6\textwidth]{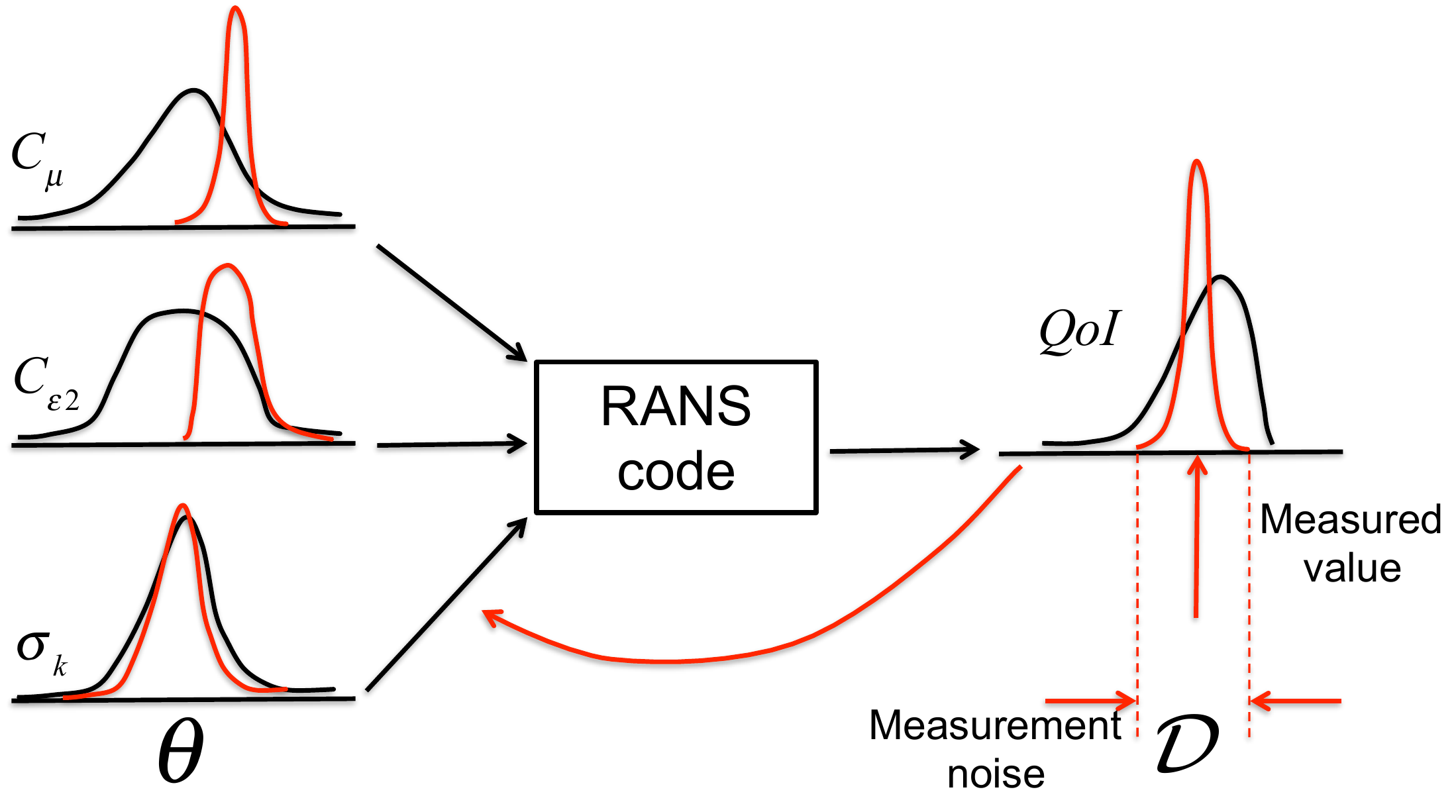}\label{fig:forward-backward_backward}}
   \caption{Illustration of uncertainty propagation (forward analysis) and statistical inference (backward analysis) in the context of RANS simulations. Uncertainty propagation (forward analysis) involves propagating specified prior distributions on the input $\bm{\theta}$ (e.g., angle of attack/AoA, Reynolds number, model coefficients, or modeled terms such as Reynolds stresses) through a RANS simulation code and investigate the uncertainties in the solutions (quantities of interests/QoIs, e.g., lift and drag coefficients). Statistical inference (backward  analysis) involves  assimilating available measurement data to reduce uncertainties in the aforementioned input (e.g., AoA or Reynolds number). The inferred input distributions can be subsequently propagated to make predictions on the QoIs.}
  \label{fig:forward-backward}
\end{figure}

In addition to the parametric/non-parametric classification, it is possible to distinguish forward and backward methods, also referred to as data-free and data-driven approaches as illustrated in Figure~\ref{fig:forward-backward}. Forward (data-free) methods consist in propagating some pre-specified probability distributions on the closure coefficients (or on the modeled terms) through the RANS equations and investigating the uncertainty distribution of the solution (Figure~\ref{fig:forward-backward_forward}). On the other hand, backward (data-driven) methods consist in assimilating available data to infer the coefficient distributions or model errors (Figure~\ref{fig:forward-backward_backward}). Such inferred distributions then become available for propagation through the RANS equations in a subsequent prediction step as in the forward analysis.  The applicability of the calibrated RANS models to new flows remains as a main concern for both parametric and non-parametric approaches.  Table~\ref{tab:ref-table} shows a classification of the literature based on their parametric/non-parametric and forward/backward characteristics. Note that the classification omitted data-driven methods that primarily focused on developing turbulence models~\citep[e.g.,][]{weatheritt2016novel,weatheritt2017development} rather than quantifying their uncertainties.
A roadmap is provided in Figure~\ref{fig:roadmap} to help the reader navigate through this review.

\pgfkeys{/forest,
  rect/.append style   = {rectangle, rounded corners = 2pt,
                         inner color = col6in, outer color = col6out},
  ellip/.append style  = {ellipse, inner color = col5in,
                          outer color = col5out},
  orect/.append style  = {rect, 
                         text width = 325pt, text centered,
                         minimum height = 10pt, outer color = col7out,
                         inner color=col7in},
  oellip/.append style = {ellip, inner color = col1in,
                          draw = linecolk,
                          outer color = col1out, text centered}
  }

\begin{figure}[!p]
\begin{forest}
for tree={draw, rectangle, 
      align=center,
      line width=1pt,
      ellip,
      align=center,
      child anchor=north,
      parent anchor=south,
      drop shadow,
      l sep+=14pt, 
      edge path={
        \noexpand\path[color=linecol, rounded corners=5pt,
          >={Stealth[length=10pt]}, line width=1pt, ->, \forestoption{edge}]
          (!u.parent anchor) -- +(0,-5pt) -|
          (.child anchor)\forestoption{edge label};
        }
    }
[
Introduction: turbulence and its modeling \S \ref{sec:introduction} \\
Fundamentals: probability and statistics of UQ
\S \ref{sec:prob-theory},
calign=first, inner color = col9in, outer color = col9out
[RANS model \\uncertainties  \S \ref{sec:parametric} \S \ref{sec:nonparametric}, oellip, tier=rans
	[Parametric \\ approaches \S \ref{sec:parametric}, 
	oellip, tier=para,
	for descendants={rect}
		[Data-free: \\probe \\sensitivity to \\parameters \S \ref{sec:parametric-forward}
		]
		[Data-driven: \\infer \\parameters \\from data 
		\S \ref{sec:parametric-backward},
		minimum width = 7em,
		 for tree={l sep+=3pt}
	        [Statistical \\inference of
	        \\ parameters \\ \S \ref{sec:para-backward-plain},
	        tier=terminal,
	    	   minimum width = 6em
	        ]
	        [Considering 
	        \\structural \\uncertainty \\
	        \S \ref{sec:para-stochastic},
	        tier=terminal,
	   	   minimum width = 6em
	        ]
	        [Considering \\ multiple \\ models \\
	        \S \ref{sec:bsma},
	        tier=terminal,
	        minimum width = 6em
	        ]
	    ]
	]
	[Non-parametric \\ approaches 
	  \S \ref{sec:nonparametric}, oellip, 
	   tier=para,
	  for descendants={rect}
	  [
	  Data-free \\and \\data-driven \\ approaches,
	  minimum width = 8em, 
		[Targeting \\transport \\equations \\ \S \ref{sec:nonpara-beta},
		tier=terminal,
		minimum width = 5em
		]
		[Targeting \\ turbulent \\viscosity \\ \S \ref{sec:nonpara-nut}, 
		 tier=terminal
		]
		[Targeting \\ Reynolds \\stress \\ \S \ref{sec:nonpara-tau}, 
		tier=terminal,
		minimum width = 5em
		]
	  ]
	]
]
[LES \\uncertainties \S \ref{sec:les},
    tier=rans, align=center,
    oellip, 
    inner color = col9in, outer color = col9out,
    line width=1pt
]
]
\end{forest}
\caption[]{
Roadmap of this review with links to relevant sections. Legend: \tikz\draw[black,fill=red!50] (0,0) ellipse (1.6ex and 0.7ex);  major elements of this review;
\tikz\draw[black,fill=black!40] ellipse (1.6ex and 0.7ex); auxiliary topics of this review;
\tikz\draw[black,fill=col6out] (0,0) rectangle (2ex, 1.5ex);  detailed topics in RANS model-form uncertainty.
\label{fig:roadmap}
}
\end{figure}
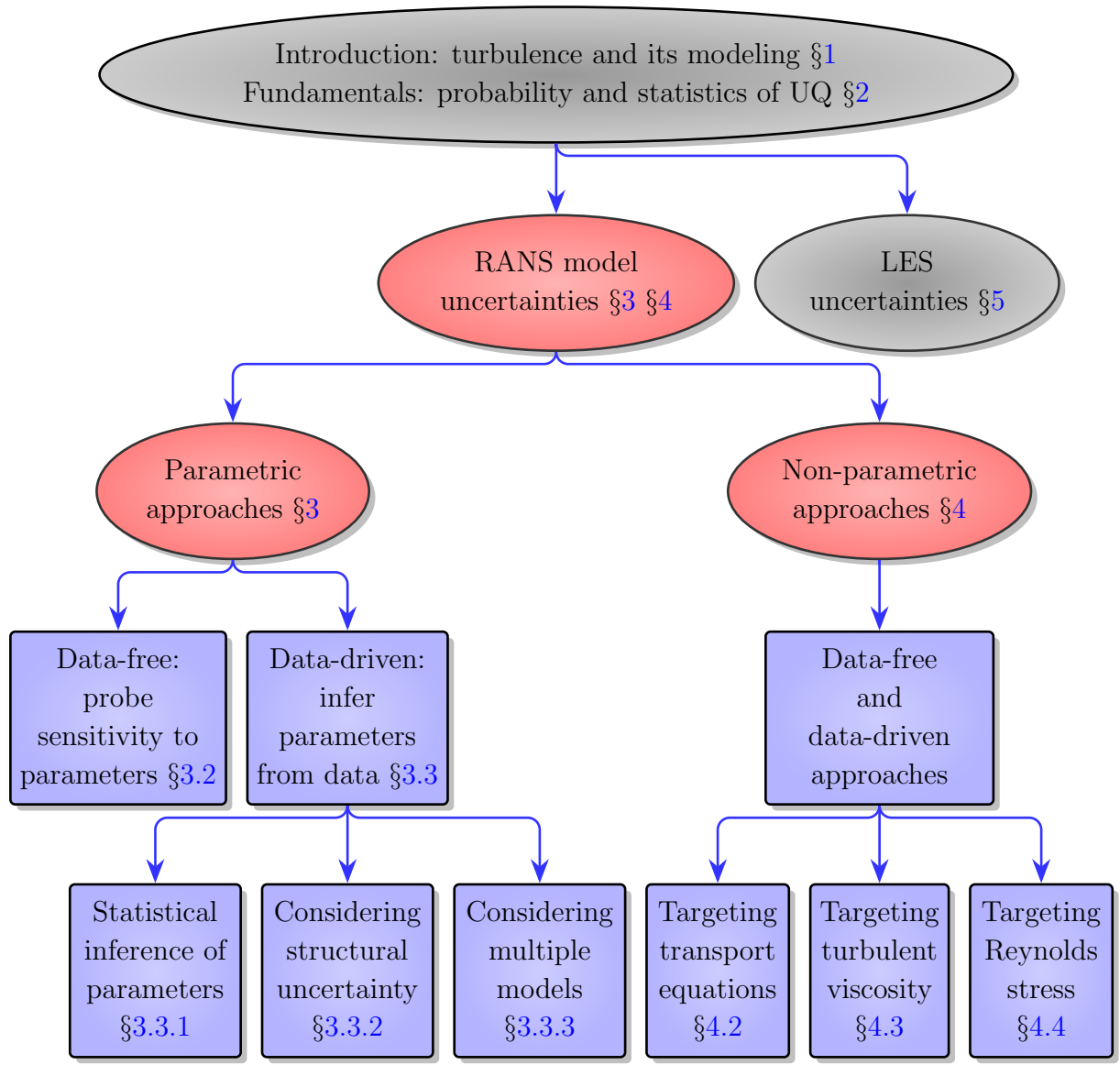

\begin{table}[htbp]
\caption{Classification of literature of RANS model uncertainty quantification based on parametric/non-parametric approaches and data-free (forward) / data-driven (backward) approaches. Works in multi-model approaches are listed along with parametric approaches.}
\begin{center}
\small
  \begin{tabular}{C{2.1cm} | L{6.6cm} | L{5.6cm}  }
\toprule
  & Parametric & Non-parametric \\
  \midrule
  data-free (forward) &
    \pbox{5.9cm}{
(\citet{turgeon2001application}, \citeyear{turgeon2001application})\\
(\citet{dunn2011uncertainty}, \citeyear{dunn2011uncertainty})\\
(\citet{platteeuw2008uncertainty}, \citeyear{platteeuw2008uncertainty})\\
(\citet{margheri2014epistemic}, \citeyear{margheri2014epistemic})\\
(\citet{schaefer2016uncertainty}, \citeyear{schaefer2016uncertainty})
}
&
    \pbox{5.9cm}{
(\citet{emory2011modeling,emory2013modeling}, \citeyear{emory2011modeling,emory2013modeling}) \\
(\citet{iaccarino2017eigenspace}, \citeyear{iaccarino2017eigenspace}) \\
(\citet{mishra2017uncertainty}, \citeyear{mishra2017uncertainty}) \\
(\citet{edeling2017data-free}, \citeyear{edeling2017data-free}) \\
(\citet{xiao2017random}, \citeyear{xiao2017random})
    }
    \\
    \vspace{1em}
    &
    \pbox{5.9cm}{
      \emph{Multi-model}: \\
     (\citet{poroseva2006improving}, \citeyear{poroseva2006improving})
\\
    (\citet{edeling2014predictive,edeling2018bayesian}, \citeyear{edeling2014predictive,edeling2018bayesian}) }
      & {} \\

   \hline
  data-driven
  (backward)
  &
  \pbox{6.8cm}{
    (\citet{cheung2011bayesian}, \citeyear{cheung2011bayesian})\\
     (Kato et al.~\citep{kato2013approach,kato2015data}, \citeyear{kato2013approach,kato2015data})\\
    (\citet{margheri2014epistemic}, \citeyear{margheri2014epistemic})\\
    (\citet{ray2016bayesian,ray2018learning}, \citeyear{ray2016bayesian,ray2018learning})\\
(\citet{edeling2014bayesian,edeling2014predictive,edeling2018bayesian}, 2014--2018) \\
    (\citet{papadimitriou2015bayesian}, \citeyear{papadimitriou2015bayesian})
    }
&
  \pbox{5.5cm}{
        (\citet{dow2011quantification}, \citeyear{dow2011quantification}) \\
    (\citet{singh2016using}, \citeyear{singh2016using}) \\
    (\citet{xiao2016quantifying}, 2016) \\
    (\citet{wu2016bayesian} 2016) \\
    (\citet{wang2016incorporating}, 2016) \\
    (\citet{parish2016paradigm}, \citeyear{parish2016paradigm})\\
    (\citet{edeling2017data-free},
    \citeyear{edeling2017data-free})
}
\\
\bottomrule
\end{tabular}
\end{center}
\label{tab:ref-table}
\end{table}%

The rest of the paper is organized as follows. A brief review of available techniques for uncertainty propagation, data assimilation and statistical inference is presented in Section~\ref{sec:prob-theory}. In Section~\ref{sec:parametric} we review parametric  and multi-model approaches, the latter of which partly accounts for model-form uncertainties.  Section~\ref{sec:nonparametric} is dedicated to non-parametric approaches, which target model-form uncertainties. For completeness, an overview of uncertainties in scale-resolving approaches, and more specifically LES, are briefly reviewed in Sections~\ref{sec:les}.  Finally, conclusions, future research, and perspectives are presented in Section~\ref{sec:conclusion}.

\section{Fundamentals of probability and statistics for uncertainty quantification}
\label{sec:prob-theory}

Probability and statistics lie at the core of most of the work reviewed in this work. Therefore, we provide a brief overview of the relevant methods in this section in the context of quantifying and reducing RANS model uncertainties. Based on these foundations, we briefly introduce the algorithms used for uncertainty propagation (forward analysis) and Bayesian inference (backward analysis). In particular, we discuss some commonly used methods for exact and approximate Bayesian inferences.

\subsection{Representation, sampling, and propagation of model uncertainties}

In the probabilistic approach, the uncertain quantities of concern in the RANS model, such as the model coefficients, can be represented as random variables. A random variable $Z$ is a scalar function that may take a range of possible values $z$, referred to as \emph{realizations}.  A vector of random variables $\bm{Z} = [Z_1, \cdots, Z_n]$, indexed by integers, is a random vector.  An example is the combination of coefficients in a RANS model.  A random field $\bm{Z}(\bm{x})$ is a field of random variables indexed by the spatial coordinate $\bm{x}$. It is also referred to as stochastic process when the index is time coordinate $t$. Random field is a generalization of random vectors to the continuous limit.  The true Reynolds stress field $\bm{\tau}(\bm{x})$ and the discrepancies $\delta_\tau$ in the RANS-modeled Reynolds stress $\bm{\tau}^{\text{rans}}(\bm{x})$ are examples of random fields in RANS model uncertainty quantification.

A continuous random variable can be characterized by its probabilistic distributions such as cumulative distribution function or probability density function $p(z)$.  Common quantities of interest in uncertainty quantification  are statistical moments of the random variables such as expectation $\mathbb{E}[Z]$ and variance $\var[z]$, which can be obtained via integration over all possible outcomes of $Z$, e.g.,
\begin{subequations}
\label{eq:moments}
\begin{align}
\mathbb{E}[Z] & = \int z \; p(z) dz,  \\
\var[Z] & = \int \left(z-\mathbb{E}[Z]\right)^2 \; p(z) dz .
\end{align}
\end{subequations}
The expectations and variances of random vectors and random fields can be obtained by applying Equation~(\ref{eq:moments}) to each component thereof, recalling that random vectors and random fields are collections of random variables indexed by integers and real numbers, respectively.  Moreover, a random vector is further characterized by its covariance matrix $K_{ij} = \cov(Z_i, Z_j)$, which represents the correlation among the components of $\bm{Z}$. A generalization of the covariance matrix of random vectors to random fields leads to covariance kernel $K(\bm{x}, \bm{x}')$, which indicates the pair-wise covariance between the random variables $Z(\bm{x})$ and $Z(\bm{x}')$ corresponding to locations $\bm{x}$ and $\bm{x}'$.  The most commonly used covariance kernel for the random fields representing model discrepancies is the squared exponential kernel:
\begin{equation}
  \label{eq:sq-exp}
    K(\bm{x}, \bm{x}') = \sigma^2 \exp{\left(-\frac{|\bm{x} - \bm{x}'|^2}{2 l^2}\right)}
\end{equation}
with $\sigma$ and $l$ indicating variance and length scale, respectively. Such a kernel implies that the correlation between two random variables depends on their corresponding indexing locations. The farther apart  the two locations $\bm{x}$ and $\bm{x}'$ are, the smaller the correlation between $Z(\bm{x})$ and $Z(\bm{x}')$ is.

In this work, we consider a RANS-based CFD model $M: (\bm{\varsigma}; \bm{\theta}) \mapsto y$, which is parameterized by~$\bm{\theta}$ and maps the latent variables $\bm{\varsigma}$ (e.g., geometry, boundary conditions) to an observable output~$y$. The multidimensional uncertain variable $\bm{\theta}$ can be a vector of model coefficients in parametric approaches or a spatial field in non-parametric approaches, e.g., Reynolds stress field $\bm{\tau}(\bm{x})$ or eddy viscosity field $\nu_t(\bm{x})$.  Two types of analyses can be performed:
\begin{itemize}
\item Uncertainty propagation (forward analysis): When the probability distribution $p(\bm{\theta})$ of the model parameters $\bm{\theta}$ is known, the probability distribution $p(y)$ of the output can be obtained by (i) sampling the specified distribution $p(\bm{\theta})$, e.g., by using a Monte Carlo method, (ii) evaluating the model $M$, and (iii) aggregating the propagated samples. A typical algorithm for plain Monte Carlo sampling is presented in~\ref{app:mc}. The probability distribution $p(\bm{\theta})$ that is known on the parameters is referred to as the \emph{prior} distribution.
\item Bayesian inference (backward analysis): When data $\mathcal{D}$ is available on the output $y$, which may be noisy, biased, or incomplete, the input probability distribution of $\bm{\theta}$ can be inferred. The result is the \emph{posterior} distribution $p(\bm{\theta}| \mathcal{D})$ of $\bm{\theta}$ given data $\mathcal{D}$, representing an update distribution from the prior distribution $p(\bm{\theta})$ after observing the data.
\end{itemize}

\subsection{Uncertainty propagation (Forward analysis)}
\label{sec:prob-theory-uq}

Techniques to propagate uncertainties can be classified into two categories~\citep[see][Chapter 1.4]{le-matre2010spectral}: spectral methods~\citep{ghanem2003stochastic} and Monte Carlo (MC) methods~\citep{glasserman2004monte-carlo}. 
Spectral methods  discretize the uncertainty space of the random variables by using orthogonal basis functions. This is done in a similar way in which orthogonal basis functions (e.g., Fourier functions or orthogonal polynomials) are used for the spatial discretization of deterministic PDEs.  In uncertainty quantification, spectral methods
have faster statistical convergence but they depend on the smoothness of the prior and the function that maps the inputs to outputs.  Another barrier for spectral methods is the ``curse of dimensionality'': the number of function evaluations needed to accurately describe the statistics increases exponentially with the cardinality of the parameter space.  Monte Carlo methods, on the other hand, approximate the solution by using random samples from the input uncertainty space and are not adversely affected by its dimensionality. However, the convergence rate is uniformly slow at a rate of $\mathcal{O}(N^{-1/2})$, where $N$ is the number of samples~\citep{glasserman2004monte-carlo}.

While the Monte Carlo based uncertainty quantification seems straightforward, the slow convergence rate poses a major challenge in applications where the computational cost of propagating each sample is high, as is the case for CFD simulations. Accelerating the statistical convergence of Monte Carlo methods has been a topic of intensive research, and numerous techniques for variance reduction have been proposed. Examples include stratified sampling, Latin hypercube sampling~\citep{helton2003latin}, importance sampling, and control variate~\citep{glasserman2004monte-carlo}. A recent development is multilevel Monte Carlo (MLMC) methods~\citep{giles2008multilevel, barth2011multi-level,mishra2012multi-level}, where simulations on coarser meshes are used as control variate of those on fine meshes to reduce the variances. A generalization of MLMC has led to multi-fidelity Monte Carlos methods~\citep{muller2013multilevel,muller2014solver-based,peherstorfer2018survey}, where a sequence of models with ascending fidelities (e.g., empirical formulas, panel methods, RANS, LES) are combined for input uncertainty propagation, with lower-fidelity models used as control variate of higher fidelity models as in the MLMC methods.  However, so far these methods have been primarily used for propagating input uncertainties and not model uncertainties. One difficulty associated with multi-level and multi-fidelity methods is the possible non-trivial interactions between model uncertainties and numerical discretization uncertainties.

Another approach for overcoming the difficulty of expensive model simulations are surrogate models or response surface methods. In these methods, a surrogate of the original model, e.g., in the form of splines, polynomial chaos, or neural networks, are first constructed based on data obtained by evaluating the original model $M$ at a number of \emph{design points}.  The surrogate models provide an approximate functional mapping $\widetilde{M}: \theta \mapsto y$ that replaces the true mapping $M$ for use in the subsequent sample propagation.  Once constructed, the surrogate models can be evaluated at negligible computational costs. However, as with spectral methods, a main difficulty for the surrogate model approach is the curse of dimensionality, which makes it impractical for high dimensional input space.

\subsection{Statistical inference (Backward analysis)}
\label{sec:prob-theory-da}

Most of the works on inference of model uncertainties (referred to as backward analysis above) are based on  Bayes' theorem:
\begin{equation}
p(\theta | \mathcal{D}) =
\frac{p(\mathcal{D} | \theta ) \,
p(\theta)}{p(\mathcal{D})} ,
   \label{eq:bayes}
\end{equation}
which states that the posterior probability $p(\theta | \mathcal{D})$ is proportional to the prior $p(\theta)$ and the likelihood $p(\mathcal{D} | \theta)$.  The prior $p(\theta)$ summarizes all available knowledge about $\theta$ before observing the data $\mathcal{D}$.  The likelihood function $p(\mathcal{D} | \theta )$ describes the probability of observing the data from a process described by the model $M(\theta)$ parameterized by $\theta$. In the context of RANS uncertainty quantification, evaluating $p(\mathcal{D} | \theta )$ for a given realization of the model parameters $\theta$ involves running the CFD code and is thus a costly operation.  Finally, $p(\mathcal{D})$ is the total probability of observing the data, which normalizes the posterior probability.

\subsubsection{Bayesian inference based on Markov chain Monte Carlo sampling}

\begin{figure}[!htbp]
  \centering
\includegraphics[width=0.6\textwidth]{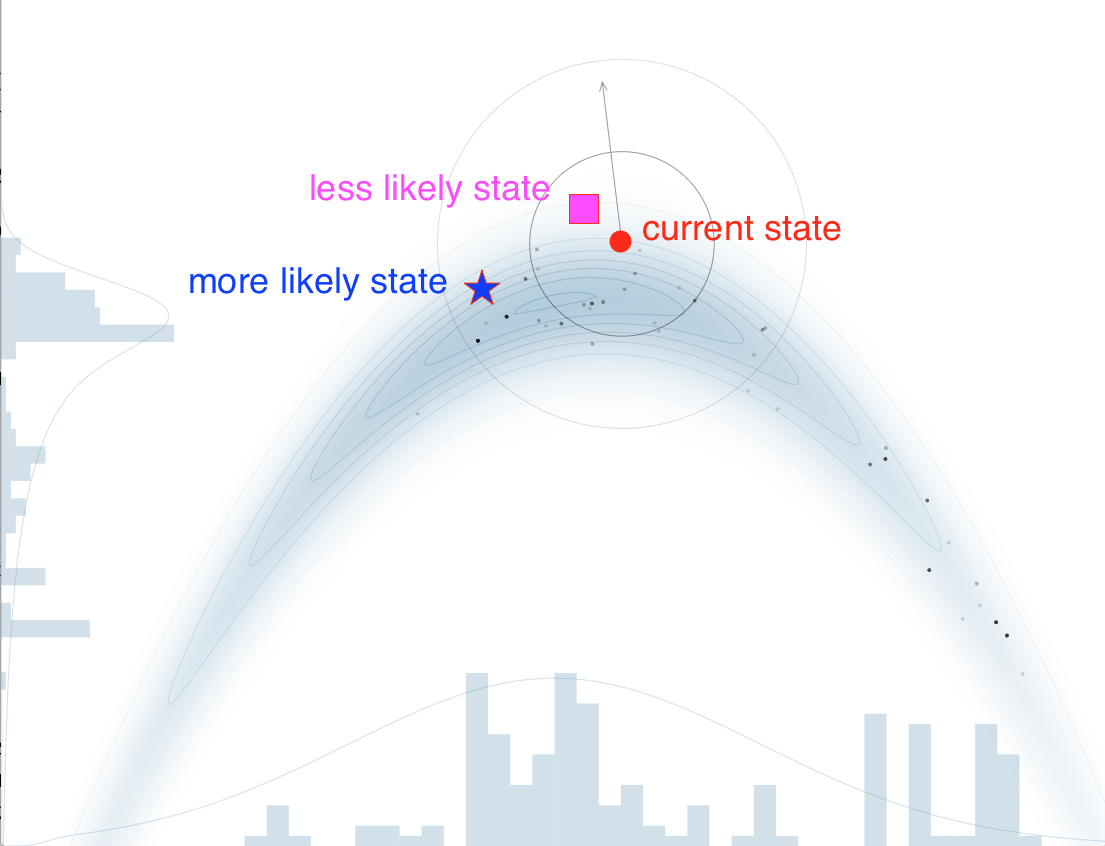}
   \caption{Illustration of Markov chain Monte Carlo sampling of a banana-shaped posterior (shaded contour) in a two-dimensional state space. The sampled distribution is illustrated with the trace of past samples (dots) and the marginal distributions (histograms plotted on the horizontal and vertical axes).   
   Image obtained by using the MCMC demonstration code (\url{https://chi-feng.github.io/mcmc-demo/}) by Chi Feng of MIT.}
  \label{fig:mcmc}
\end{figure}

Theoretically, evaluating the posterior can be straightforward using the following procedure similar to the plain Monte Carlo sampling: (i) draw samples from the prior, (ii) evaluate the likelihood for each sample, and (iii) aggregate the samples to estimate the posterior. However, this is much more challenging than in the forward analysis above. In the forward analysis the probability distribution is known, and thus one can draw more samples from the high probability regions, e.g., by using stratified sampling~\cite{glasserman2004monte-carlo}. In contrast, Bayesian inference involves sampling from the posterior, the high probability regions of which is not known \emph{a priori}. For example, samples drawn from regions with high prior probability may turn out to have very small likelihood after an expensive model evaluation, which may lead to very small posterior probability (see Equation~(\ref{eq:bayes})). Therefore, plain Monte Carlo methods are rarely used due to its difficulty in efficiently targeting the high posterior regions. Instead, Markov chain Monte Carlo (MCMC) methods are commonly used, which are a class of sequential sampling strategies in which the next sampled state only depends on the current state. Such a strategy allows the sampling to focus on high probability regions with occasional excursion to low probability regions (tails).  Given a target distribution, the MCMC algorithm samples from that distribution by constructing a Markov chain whose stationary distribution coincides with the target distribution. A typical MCMC algorithm with Metropolis--Hastings sampling is detailed in \ref{app:mcmc} and illustrated graphically in Figure~\ref{fig:mcmc}.

While the MCMC is the golden standard of Bayesian inference and posterior sampling, a major challenge of its application is that it requires a large number of samples to achieve statistical convergence. Typically the required number of samples range from $\mathcal{O}(10^5)$ to $\mathcal{O}(10^6)$, with the specific number depending on the shape of the posterior distribution and the effectiveness of the sampling.  In CFD applications, each evaluation involves a simulation that takes hours or even weeks to run depending on the complexity of the flow configuration. For example, RANS simulations of a jet in crossflow, which is a geometrically simple yet industrially relevant case, needed $\mathcal{O}(10^7)$ grid points and $\mathcal{O}(10^{4})$ CPU hours to run on a high performance computing cluster~\cite{ray2016bayesian,ray2018learning}. Clearly, it is impractical to perform a full RANS simulation for each evaluation of likelihood in the MCMC sampling.  This is not only due to the large number of required samples but also because of the \emph{sequential}  nature of the MCMC algorithm -- the next proposed sample depends on the evaluated posterior at the current state.

As in the uncertainty propagation discussed above, surrogate models are  commonly used for likelihood evaluation in MCMC-based model uncertainty quantification to alleviate the high computational cost of RANS simulations~\cite{edeling2014bayesian,ray2016bayesian,ray2018learning}. Efficient sampling of high dimensional spaces with MCMC is a topic of active research, with many methods proposed in the past few years, e.g., by adaptively constructing local approximations  during the sampling and by using the likelihood to inform the sampling~\citep[see, e.g.,][]{moselhy2012bayesian,parno2018transport}.

Another difficulty arises from the physical constraints among the state variables (e.g., parameters in closure models or Reynolds stresses at different spatial locations), which is particularly relevant for RANS model uncertainty quantification.  For example, in the parametric approach such constraints on the parameters dictates that points in some regions in the state space may yield nonphysical solutions or fail to converge at all. Consequently, such regions should be excluded when using MCMC to sample the posterior.  Again, this can be done by building surrogate models from simulation data~\citep{guillas2014bayesian,ray2016bayesian,ray2018learning}.  The fraction of excluded regions increases exponentially with the dimension of the sample space. Finally, it is noted that such a surrogate approach is also restricted to state spaces with low dimensions.

\subsubsection{Approximate Bayesian inference based on MAP estimation}

The MCMC method provides the most accurate sampling of the posterior but requires a large number of samples.  When the exact probability is not critical and only the low order moments such as the mean and the variance are important, various approximate Bayesian inference methods can be used~\citep[e.g.,][]{mons2016reconstruction,iglesias2013ensemble-kalman}. These methods use maximum a posteriori (MAP) probability estimate to obtain the mode (peak) of the posterior and not the full posterior distribution.

The MAP estimate can be computed in several ways, among which the most commonly used are variational methods and ensemble methods.  Both methods are used in data assimilation with a wide range of applications ranging from numerical weather forecasting to subsurface flow characterization.  Both variational methods and ensemble methods have been adopted for parameter inferences.  To this end, the system state is first augmented to include both the observable, physical state $\xi_i (t)$ (e.g., velocities, pressure, and/or turbulent kinetic energy) and parameters $\bm{\theta}$ (e.g., model coefficients or Reynolds stress discrepancies, which are not observable and need to be inferred).  Specifically, $\bm{z}$ is written as a vector formed by stacking the unknown parameters and the physical states $\xi_i$:
\begin{equation}
\label{eq:aug-state}
\bm{z} = [\zeta_1, \cdots, \xi_n;  \bm{\theta}]^\top,
\end{equation}
where $\top$ indicates vector transpose, and  $\bm{\theta} = [\theta_1, \theta_2, \cdots, \theta_r] $ is a vector of $r$ parameters.
When computing the MAP estimate, the following objective function is to be minimized:
\begin{equation}
  \label{eq:obj-da}
  J = \| \bm{z}  -  \mathbb{E}[\bm{z}]  \|_{\mathsf{P}} +  \| y -  \mathsf{H}[\bm{z}] \|_{\mathsf{R}}
\end{equation}
where $\mathsf{P}$ and $\mathsf{R}$ are the covariance matrices of the state $\bm{z}$ and the observation errors, respectively, with $\| A \|_{\mathsf{P}} = \|A^\top \mathsf{P}^{-1} A\|$ and $\| \cdot \|_{\mathsf{R}}$ similarly defined; $\mathsf{H}$ is the observation matrix, which maps the state space to the observation space, typically reducing the dimension dramatically. Its interpretation will be further detailed in the context of the ensemble Kalman filtering algorithm (see \ref{app:enkf}).

Obtaining the MAP estimate is equivalent to minimizing the cost function $J$ in Equation~\eqref{eq:obj-da} under the constraint imposed by the models describing the physical system (i.e., RANS equations in case of turbulent flows), during which the set of parameters minimizing the discrepancies between the prediction and the observation data is sought.  In variational methods the minimization problem is often solved by using gradient descent methods, with the gradient obtained with adjoint methods.  In contrast, ensemble methods use samples to estimate the covariance of the state vector, which is further used to solve the optimization problem.  Variational methods have been the standard in data assimilation and still dominate the field, while ensemble methods such as ensemble Kalman filtering have matured in the past decades and are making their way to operational weather forecasting. Hybrid approaches combining both approaches are an area of intense research and have been explored in CFD applications~\cite{mons2016reconstruction}.

Recently, ensemble Kalman filtering (EnKF)~\cite{evensen2003ensemble,evensen2009data} has been widely used in inverse modeling to estimate model uncertainties~\citep{iglesias2013ensemble-kalman,xiao2016quantifying}.  In EnKF-based inverse modeling, one starts with an ensemble of model parameter values drawn from their prior distribution. The filtering algorithm uses a Bayesian approach to assimilate observation data (e.g., data from experiments and high-fidelity simulations) and produces a new ensemble that represents the posterior distribution. In parametric or field inference of concern here, the EnKF method is used in an iterative manner to find the states that optimally fits the model and data with uncertainties of both accounted for, which is essentially a derivative-free optimization. As such, it is referred to as the \emph{iterative ensemble Kalman method}. This is in contrast to the EnKF-based data assimilation as used in numerical weather forecasting, where the observation data arrive sequentially. The algorithm for the iterative ensemble Kalman method is presented in~\ref{app:enkf}.

EnKF has some well known limitations due to its assumptions of linear models and Gaussian distributions, and theoretically they would perform poorly for non-Gaussian priors and highly nonlinear forward models.  However, despite the above-mentioned limitations, EnKF methods have been successfully used in a wide range of applications. Mathematicians have performed analyses to shed light on why they have worked well in view of their theoretical limitations~\cite{ernst2015analysis,schillings2017analysis}.

\section{Parametric and multi-model approaches}
\label{sec:parametric}


In this review we use ``parametric approaches'' to refer to methods that quantify the uncertainty associated with RANS simulations by investigating primarily the sensitivity of the results to the closure coefficients. As mentioned in Section~\ref{sec:introduction},  we will use ``forward approaches'' to refer the methods that consist of perturbing the closure coefficient according to some probability distribution function and quantifying the output uncertainty on the computed solution. This is in contrast to ``backward approaches'', in which observations are used to infer the model coefficients.  In both the forward and backward approaches, the model structure is fixed and only the uncertainty on the coefficients is quantified. This can nevertheless be used to learn about structural inadequacy of the model, as will be shown later.  In some cases, one of the outcomes of the inference process is an estimate of the plausibility of a given model based on the available observations, i.e. the inference may also provide some guidelines for model selection.  Finally, we will discuss multi-model approaches in which the uncertainty on the model choice is tackled by considering a set of alternative model structures.

\subsection{Uncertainties in RANS model parameters}
\label{sec:origin-coefs}

All RANS models have some closure coefficients. A typical example is provided by the well-known $k$--$\varepsilon$  model, initially proposed by \citet{jones1972prediction}. In this model, the Reynolds stress tensor is modeled by using the Boussinesq approximation in Equation~(\ref{eq:evm}), and the turbulent viscosity $\nu_t$ is computed by solving additional transport equations for the turbulent kinetic energy $k$ and the turbulent dissipation $\varepsilon$:
 \begin{subequations} \label{kepseqn}
  \begin{align}
  \nu_t & = C_{\mu}\frac{k^2}{\varepsilon}\\
    \frac{\partial k}{\partial t}+U_i\frac{\partial k}{\partial x_i} & =
         \mathcal{P}_k  
          - \varepsilon  +   \frac{\partial }{\partial x_i} \left[\left(\nu+\frac{\nu_t}{\sigma_k}\right) \frac{\partial k}{\partial x_i}\right]
                                                                       \label{eq:k} \\
    \frac{\partial \varepsilon}{\partial t}+U_i\frac{\partial \varepsilon}{\partial x_i} & = C_{\varepsilon_1} \frac{\varepsilon}{k}
    \mathcal{P}_k
                                                                                           -C_{\varepsilon_2} \frac{\varepsilon^2}{k}  +   \frac{\partial}{\partial x_i}\left[\left(\nu+\frac{\nu_t}{\sigma_{\varepsilon}}\right) \frac{\partial \varepsilon}{\partial x_i}\right]  ,        \label{eq:eps} 
\end{align}
\end{subequations}
where $\mathcal{P}_k$ is the production of turbulent kinetic energy through energy extraction from the mean flow gradient:
\begin{equation}
   \mathcal{P}_k  = \tau_{ij} S_{ij} \equiv \bm{\tau} \cddots  \mathbf{S} \;  ,
   \label{eq:k-prod}
\end{equation}
and $\cddots$ indicates tensor contraction.

The $k$--$\varepsilon$ model above involves coefficients $C_{\mu}$, $C_{\varepsilon_1}$, $C_{\varepsilon_2}$, $\sigma_k$, and $\sigma_{\varepsilon}$.  The nature of these coefficients leads to ambiguity regarding their values, and a set of flow-independent optimal values are unlikely to exist~\cite{edeling2014bayesian}. The above-mentioned coefficients are traditionally calibrated to reproduce results of a few canonical flows.  One of such canonical flows is the decaying homogeneous isotropic turbulence. In this flow the $k$ and $\varepsilon$ equations (Equations~\ref{eq:k}--\ref{eq:eps}) simplify to
\begin{align}
 \frac{dk}{dt} &= -\varepsilon \label{eq:dkdt},\\
\text{and} \quad  \frac{d\varepsilon}{dt} &= -C_{\varepsilon2}\frac{\varepsilon^2}{k} .
\end{align}
These equations can be solved analytically to give
\begin{align}
 k(t) = k_0\left(\frac{t}{t_0}\right)^{-n},
\end{align}
with reference time $t_0 = nk_0/\varepsilon_0$ and the exponent $n = 1/(C_{\varepsilon2}-1)$, the latter of which leads to:
\begin{align}
 C_{\varepsilon2} = \frac{n+1}{n}.
\end{align}
The standard value for $n$ is such that $C_{\varepsilon2} = 1.92$. However, this is by no means a hard requirement and other models do use different values for $C_{\varepsilon2}$. For instance, the RNG $k$--$\varepsilon$ model uses a modified value $\tilde{C}_{\varepsilon 2}= 1.68$, and the $k$--$\tau$ model (essentially a $k$--$\varepsilon$ model rewritten in terms of $\tau=k/\varepsilon$) uses $C_{\varepsilon 2}=1.83$~\citep{speziale1992critical}.  Nevertheless, experimental results suggest that most data agrees with $n=1.3$, which corresponds to $C_{\varepsilon 2} = 1.77$~\cite{mohamed1990decay}.

The coefficient $C_\mu$ is calibrated by considering the approximate balance between production and dissipation which occurs in free shear flows or in the inertial part of turbulent boundary layers. This balance can be expressed as
\begin{align} \label{eq:balance}
 \mathcal{P}_k = \nu_t\left(\frac{\partial U_1}{\partial x_2}\right)^2 = C_\mu\frac{k^2}{\varepsilon}\left(\frac{\partial U_1}{\partial x_2}\right)^2 = \varepsilon.
\end{align}
Equation (\ref{eq:balance}) can be manipulated together with the turbulent-viscosity hypothesis $\tau_{12} = \nu_t{\partial U_1}/{\partial x_2}$  to yield $\tau_{12}=\varepsilon (\partial U_1/\partial x_2)^{-1}$, which in turn yields
\begin{align}
  C_\mu = \left(\frac{\tau_{12}}{k}\right)^2.
\end{align}
\noindent
DNS data~\citep{kim1987turbulence} were used to show that $\tau_{12} \approx -0.30k$ (except close to the wall),  and thus $C_\mu = 0.09$ is the recommended value~\citep{pope2000turbulent}. Again, however, different models use different values for $C_{\mu}$. For example, $C_\mu \approx 0.085$ in the case of the RNG $k$--$\varepsilon$ model.

Another fundamental flow to be considered is the fully developed plane channel flow, which implies that $Dk/Dt = D\varepsilon/Dt =0$. The resulting simplified governing equations leads to the following constraint among several parameters~\citep{pope2000turbulent}:
\begin{align} \label{eq:constraint1}
  \kappa^2 = \sigma_\varepsilon C_\mu^{1/2}\left(C_{\varepsilon 2} - C_{\varepsilon 1}\right),
\end{align}
where $\kappa$ is the von Karman constant. It should be noted that the nominal coefficients in the $k$--$\varepsilon$ model satisfy this constraint only approximately, leading to $\kappa \approx 0.43$, instead of the standard value of $\kappa = 0.41$.  However, even the ``standard values'' has been questioned recently, with $\kappa$ determined to fall in the range [0.33, 0.45] based on experimental data in the literature~\cite{zanoun2003evaluating}.

The following constraint between $C_{\varepsilon1}$ and $C_{\varepsilon2}$ can be found by manipulating the governing equations of uniform (i.e., $\partial{U}_1/\partial{x_2} = \mathrm{constant}$) shear flows \citep{pope2000turbulent}
\begin{align} \label{eq:constraint2}
 \left(\frac{\mathcal{P}_k}{\varepsilon}\right) = \frac{C_{\varepsilon 2} - 1}{C_{\varepsilon 1} - 1}.
\end{align}
\noindent
  \citet{tavoularis1989further} measured $\mathcal{P}_k/\varepsilon$ for several uniform shear flows and reported values between $1.33$ and $1.75$, with a mean around $1.47$. Note however, that Equation~(\ref{eq:constraint2}) becomes $2.09$ with the standard values for $C_{\varepsilon1}$ and $C_{\varepsilon2}$, which is significantly different from the mentioned experimental values.  Note that, regardless of the uncertainties, the coefficients have to satisfy the constraint $C_{\epsilon 2} > C_{\epsilon 1}$ as has been shown through numerical experiments by~\citet{ray2018learning}.  The physical reason behind this delineation is that the ratio $C_{\epsilon 2}/C_{\epsilon 1}$ corresponds to the spreading rate of a free jet. A ratio of $C_{\epsilon 2} / C_{\epsilon 1} < 1$, or equivalently $C_{\epsilon 2} < C_{\epsilon 1}$, would lead to contracting jet, which is nonphysical~\citep{durbin2017personal}.

  The parameter $\sigma_k$ can be considered a turbulent Prandtl number and represents the ratio of the momentum eddy diffusivity and the TKE diffusivity. These quantities are usually close to unity, which is why the standard value for $\sigma_k$ is assumed to be $1.0$.  However, no experimental data can be found to justify this assumption~\citep{platteeuw2008uncertainty}, leading to a range of recommended values among the different variations of the $k$--$\varepsilon$ model. For instance, the RNG $k$--$\varepsilon$ model uses $\sigma_k = 0.72$~\citep{wilcox2006turbulence}.

The parameter $\sigma_\varepsilon$ controls the diffusion rate of $\varepsilon$, and its value can be determined by using the constraint imposed by Equation~(\ref{eq:constraint1}), i.e.
\begin{align}
 \sigma_\varepsilon =  \frac{\kappa^2 }{C_\mu^{1/2}\left(C_{\varepsilon 2} - C_{\varepsilon 1}\right)}.
\end{align}
Similar uncertainties affect the coefficients of other turbulence models.  \citet{margheri2014epistemic} discuss in further detail the uncertainties in the coefficients of the $k$--$\varepsilon$ model and Menter's SST $k$--$\omega$ model~\cite{menter1994two-equation} and characterized their probability distributions by using generalized polynomial chaos approximations of extensive literature databases. Recently, \citet{schaefer2017uncertainty} also investigated the uncertainties in the coefficients of the SA model~\cite{spalart1992one-equation}, Wilcox' $k$--$\omega$ model, and the SST $k$--$\omega$ model, pointing out the large epistemic intervals on their values.

\subsection{Parametric uncertainty in RANS models: forward approaches}
\label{sec:parametric-forward}

\begin{figure}[!htbp]
   \subfloat[$C_f$ and uncertainty margins] {\includegraphics[scale = 0.37]{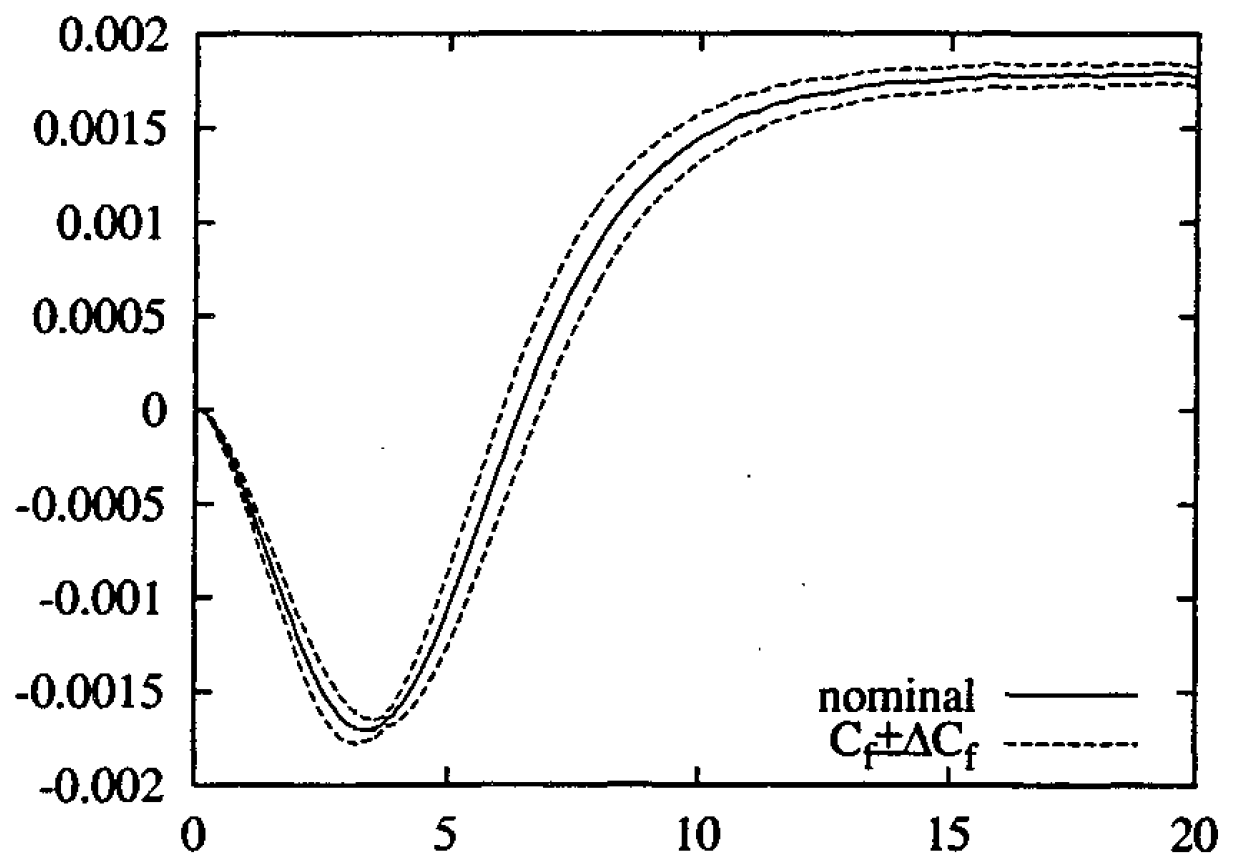}}
   \subfloat[Scaled sensitivities] {\includegraphics[scale = 0.35]{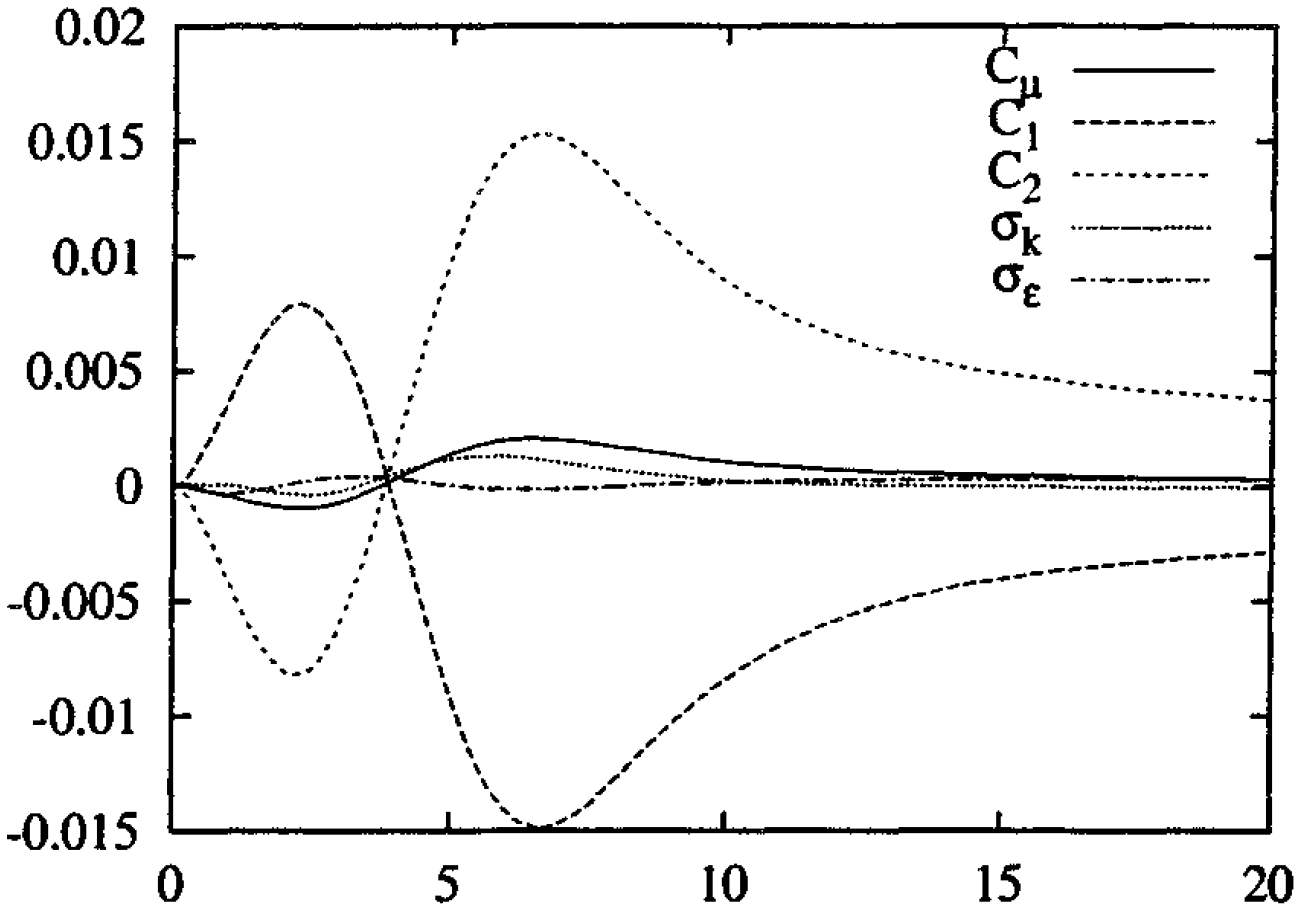}}
   \caption{Flow past a backward facing step at $Re_h=50000$. Sensitivity of the $k$--$\varepsilon$ model to the closure coefficients. Plots of the skin friction and its sensitivities versus the longitudinal position behind the step. Figures reproduced with permission from \citet{turgeon2001application}.}
  \label{fig:turgeon2002}
\end{figure}

Given the scattering in closure coefficients of RANS models as reviewed above, several uncertainty quantification (UQ) analyses have focused on quantifying the effect of such uncertainties on the output quantities of interest (QoI).  The first forward sensitivity analysis of RANS models can be ascribed, to the authors' knowledge, to \citet{turgeon2001application}. They investigated the effect of uncertainty in the $C_{\mu}$, $C_{\varepsilon 1}$, $C_{\varepsilon 2}$, $\sigma_{k}$ and $\sigma_{\varepsilon}$ of the standard $k$--$\varepsilon$ turbulence model (combined with wall functions) on the solution output. The uncertainty analysis was based on a generalized sensitivity equation method~\citep{turgeon2004general}, i.e. using sensitivity derivatives to propagate uncertainties in the turbulence model coefficients to the solution.  In these papers, the uncertainty intervals of the turbulence coefficients are taken arbitrarily, since finding information about the range of uncertainty in the coefficients is not straightforward. The results presented for the flow past a flat plate and past a backward facing step, a severe configuration for RANS models, show that the uncertainty in the model coefficients is not sufficient to account for the observed discrepancies between the predictions and the measurements. An interesting by-product was the identification of the most influential parameters based on the scaled sensitivities.  For the flow over a backward-facing step, parameters $C_{\varepsilon 1}$ and $C_{\varepsilon 2}$ are found to exert the strongest influence on the wall friction coefficient $C_f$ and thus on the reattachment point location. Figure \ref{fig:turgeon2002} shows the nominal prediction and the uncertainty range for the distribution of $C_f$ downstream of the step (panel a) and of its scaled sensitivities (panel b), defined as 
\[
C^{'}_f=\displaystyle\frac{\partial C_f}{\partial \theta_j}\theta_{nom,j},
\]
where $\theta_j$ is the $j$-th model parameter and $\theta_{nom_j}$ is its nominal value.  The method was finally applied to an airfoil flow, showing the increasing sensitivity of the solution to the RANS coefficient for larger angles of attack.

\begin{figure}[!htbp]
   \centering
   \includegraphics[scale = 0.45]{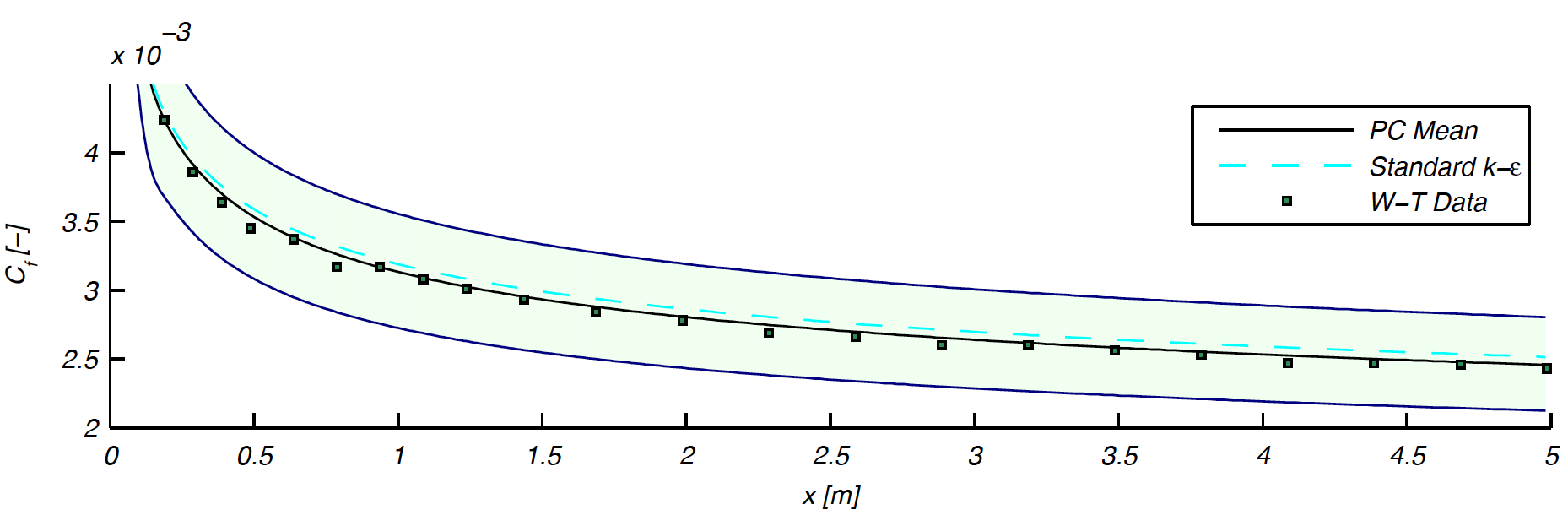}
    \caption{Distribution of $C_f$ for the flow along a semi-infinite flat plate with zero pressure gradient and 99\% uncertainty interval.
    Sensitivity of the $k$--$\varepsilon$ model to the von Karman constant $\kappa$. Figure reproduced with permission from \citet{platteeuw2008uncertainty}.}
  \label{fig:platteeuw2008}
\end{figure}

Sensitivity-based analyses provide only an uncertainty band around the nominal solution.  To obtain more information about the uncertainty of the solution, and specifically its full probability distribution given some input joint probability of the model parameters,  UQ techniques (e.g., the MC method presented in \ref{app:mc}) can be used to propagate an assigned joint distribution on the closure coefficients across the model.  For instance, \citet{platteeuw2008uncertainty} used experimental databases and DNS results, along with physical constraints on the coefficients to construct realistic a priori approximations of the input distributions for the different coefficients of the standard $k$--$\varepsilon$ with wall functions~\citep{launder1974application}. 
Their final set of uncertain coefficients includes the model parameters $C_{\mu}$, $C_{\varepsilon 2}$, $\sigma_{k}$, the wall function parameters $\kappa$ (i.e. the von Karman constant) and the log-law constant, as well as the turbulence intensity imposed at the free-stream boundary. A probabilistic collocation method was used to efficiently propagate the input joint distribution through a zero-gradient flat plate flow configuration.  Mean flow variations as a consequence of the turbulence model uncertainty were found to be large enough (at least compared to numerical errors) to encompass the experimental data available for the friction coefficient distribution along a flat plate.  They also carried out a sensitivity analysis of the output QoI, showing that the solution was more sensitive to the wall function parameters than to other model parameters.  Figure~\ref{fig:platteeuw2008} shows the uncertainty range obtained by assigning a normal probability density to the von Karman constant, $\kappa \sim \mathsf{N}(0.417,0.0127)$, while keeping other parameters constant. The most probable solution is in slightly better agreement with the experimental data. On the other hand, the predicted uncertainty interval encompasses the data.

Forward UQ for the $k$--$\varepsilon$ turbulence model with wall functions was also carried out by using the Latin hypercube sampling method~\cite{dunn2011uncertainty}.  This was used to propagate distributions of the input coefficients estimated from the data from \citet{pope2000turbulent} for the flow past a backward-facing step,  and the mean values were reported for the flow output parameters of interest along with their associated uncertainties.  The results showed that model coefficient variability had significant effects on the streamwise velocity component in the recirculation region near the reattachment point and turbulence intensity along the free shear layer.  The reattachment point location, pressure, and wall shear were also significantly affected.

\begin{figure}[!htbp]
   \centering
   \includegraphics[scale = 0.5]{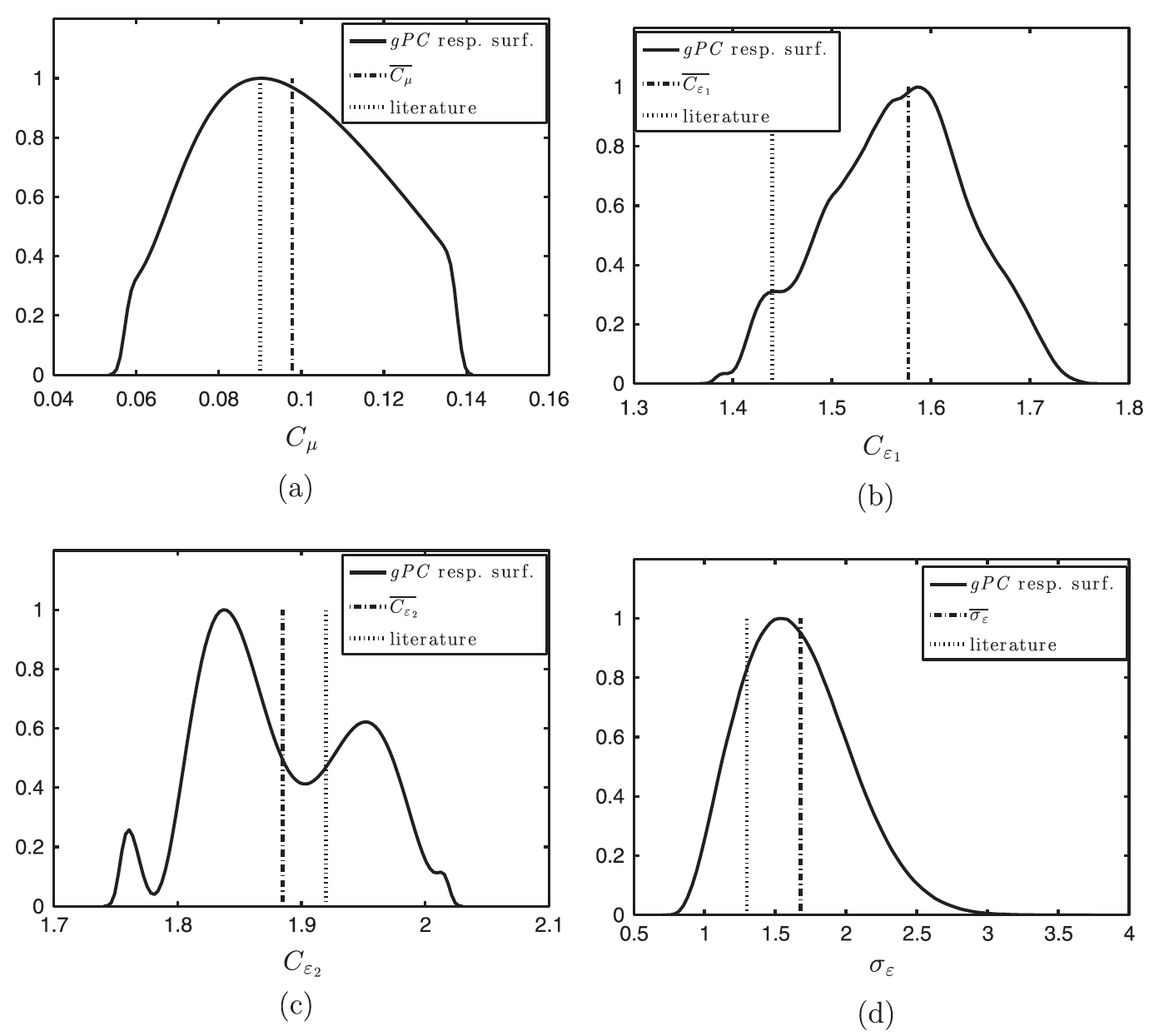}
    \caption{Normalized probability density function (pdf/max(pdf)) of the Launder--Sharma $k$--$\varepsilon$ model coefficients recovered through gPC. Figures reproduced with permission from \citet{margheri2014epistemic}.}
  \label{fig:margheri2014}
\end{figure}

In the above-mentioned works, the uncertainty distributions of the input parameters were all obtained in a largely subjective manner. The specification of such prior distribution has an impact on the output probability distributions.  To reduce such uncertainties it is possible to use analytical relationships allowing to express the closure coefficients in terms of basic properties of canonical flows (e.g., the power-law exponent of the free decay of turbulent kinetic energy in isotropic turbulence). Following this idea, \citet{margheri2014epistemic} carried out an extensive literature survey and collected a large amount of experimental and numerical data characterizing the input coefficient distributions for the Launder--Sharma low-Reynolds number $k$--$\varepsilon$ and Wilcox $k$--$\omega$ models.  The collected data exhibited a significant scattering, which confirmed the hypothesis that the uncertainties in the measured or computed basic flow properties leads to uncertainties in the RANS model coefficients.  Figure~\ref{fig:margheri2014} reports the resulting input probability density function (pdf) for the parameters of the $k$--$\varepsilon$ model, which are reconstructed by using the generalized polynomial chaos (gPC) expansion~\citep{xiu2002wiener-askey}.  The input distributions were propagated through the RANS equations applied to a turbulent channel flow for two different friction Reynolds numbers, $Re_{\tau}=950$ and $Re_{\tau}=2000$, showing that both models give inaccurate predictions of the intensity and peak location of the turbulent kinetic energy.  The observed inaccuracies were ascribed to structural uncertainties of turbulence models, which are not accounted for by the parametric data-free approaches.

\subsection{Parametric uncertainty in RANS models: backward approaches}
\label{sec:parametric-backward}

\subsubsection{Statistical inference of model parameters}
\label{sec:para-backward-plain}

Forward parametric approaches strongly rely on the availability of reliable data for constructing the coefficient probability intervals or joint distributions.  Unfortunately this information is inevitably incomplete and subject to errors.  Additionally, it remains restricted to rather simple flow configurations, and it is difficult to extend such data for robust predictions of different flows.   Finally, data are only available for observable quantities (e.g., pressures and velocities)  and not for the closure coefficients themselves. However, an inverse statistical problem can be solved to infer on the input coefficients and possibly their uncertainties.  Once obtained, this information can be propagated back through the model to estimate uncertainty intervals on the output QoIs.

The inverse statistical problem can be solved by using a deterministic or a probabilistic approach. In the deterministic approach, a set of optimal closure coefficients is obtained by minimizing the model error with respect to some reference data.  For instance, \citet{margheri2014epistemic} utilized the gPC response surfaces generated for their forward UQ analyses to find optimal combinations of model coefficients that lead to minimum global error on the mean and friction velocities with respect to DNS data for the turbulent channel flow case. Their findings suggest that the values of the model coefficients recommended in literature, which are generally set as default in commercial and open-source CFD codes, do not fall within the best-fit range.  Note however that such deterministic estimates do not provide information on the variability of the optimal coefficients or their validity for a different flow case.

In order to quantify and reduce the uncertainties on model coefficients while simultaneously providing an estimate of model-form uncertainties, it is possible to use Bayesian inference techniques as in Section \ref{sec:prob-theory-da}. In such an approach, \textit{a priori} knowledge or assumptions about the coefficients is updated by using available data. When data are highly uncertain or sparse, the updated information will exhibit little difference from the prior distribution. As more data arrive, it is possible to further update the model, thus refining the initial estimate. In the Bayesian calibration process, a key ingredient is the likelihood function in Equation~(\ref{eq:bayes}), which may carry information about observational noise on the data and model-form uncertainty. The latter being the gap between the average model predictions and the ``truth", as will be discussed later in Section~\ref{sec:para-stochastic}.

\begin{figure}[!htbp]
\centering
   \subfloat[Marginal posteriors.] {\includegraphics[scale = 0.4]{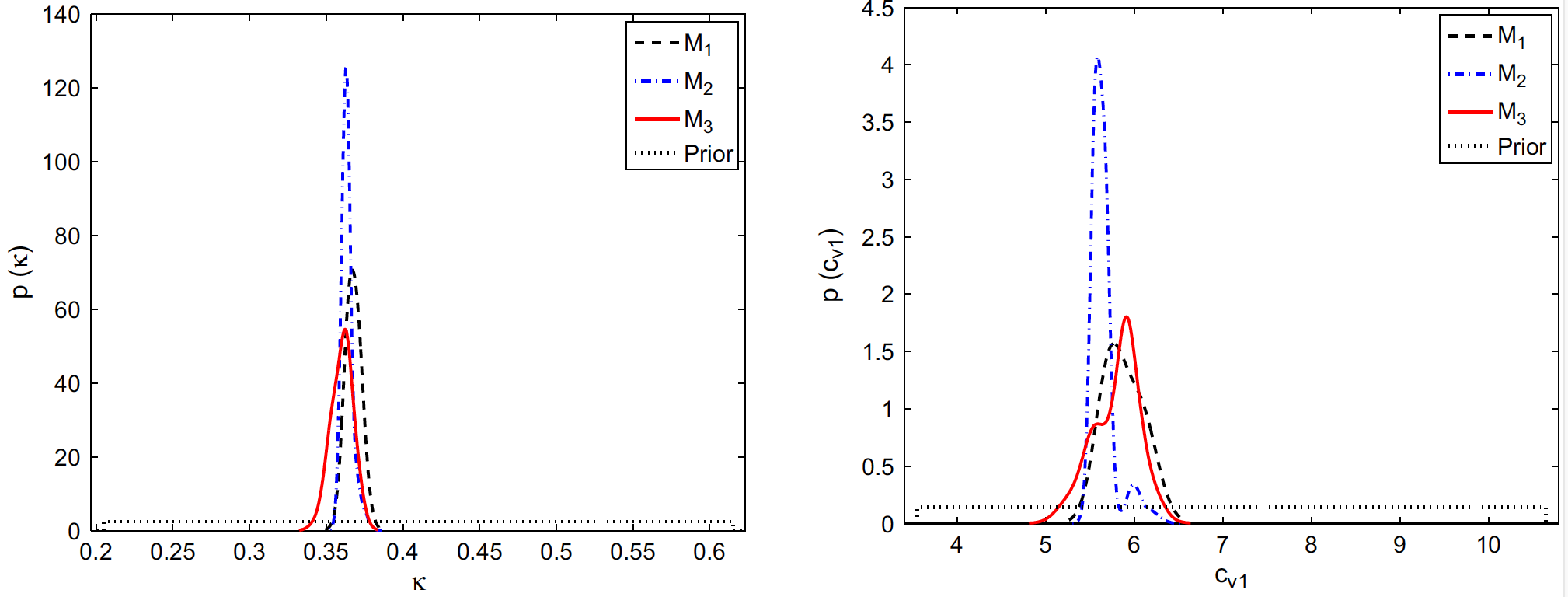}}\\
   \subfloat[Posterior joint scatter plot] {\includegraphics[scale = 0.4]{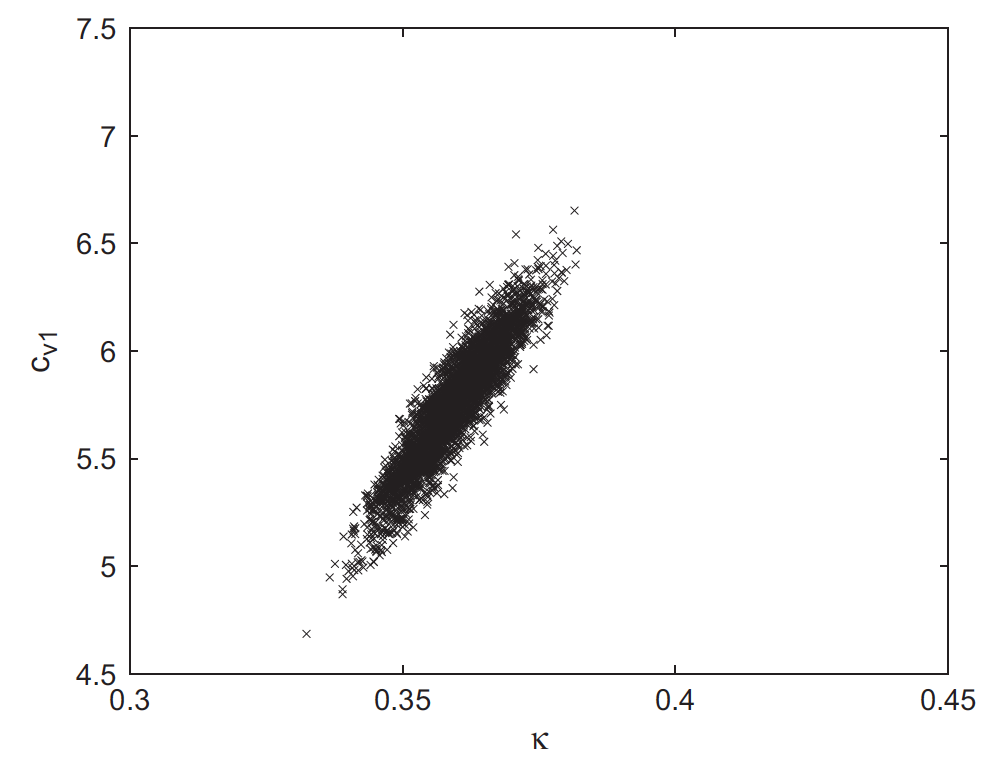}}
   \caption{Calibration of the Spalart--Allmaras model from the flat plate flow data, showing (1) the posterior distributions and (2) scatter plots of the inferred parameters $\kappa$ and $C_{\nu 1}$ by using different statistical models for the inadequacy term. Figures reproduced with permission from \citet{cheung2011bayesian}.}
  \label{fig:cheung2011}
\end{figure}

\citet{cheung2011bayesian} performed the first application of Bayesian uncertainty quantification techniques for calibrating turbulence models and making probabilistic predictions for new flows.  They used MCMC sampling to carry out Bayesian calibration of the Spalart-Allmaras model from velocity and skin friction data for three boundary layers with zero, adverse, and favorable pressure gradients. This effort enabled the estimation of the whole posterior joint probability distribution of the coefficients (instead of deterministic values) as well as a comparison of competing models for the likelihood function (noted $M_1$, $M_2$, and $M_3$) relating the observed data to the model output.  As an example,  Figure~\ref{fig:cheung2011} shows the marginal posterior distributions obtained for the von Karman constant $\kappa$ and the coefficient $c_{\nu,1}$, along with their joint scatter plot when using the stochastic model M3. Bayesian calibration is able to discover a posterior correlation between these two parameters, showing the importance of calibrating all parameters simultaneously.  The MCMC-based calibration process involved a large number of boundary layer calculations (32,768 samples), each based on a full Navier--Stokes incompressible flow solver.  Ray and co-workers~\cite{lefantzi2015estimation,ray2016bayesian,ray2018learning,ray2018robust} used a similar approach to infer the model coefficients for a more complex configuration, namely, a jet-in-cross-flow. For example, experimental data were used to calibrate the parameters in a nonlinear eddy viscosity model~\cite{ray2018learning}, where surrogate models were used to reduce the computational burden of the MCMC sampling.

\citet{kato2013approach} used ensemble Kalman filtering  \citep{evensen2003ensemble,evensen1994sequential} to determine the values of the parameters of the Spalart--Allmaras turbulence model for the zero-pressure gradient flat plate boundary layer at $M = 0.2$ and $Re = 5 \times 10^6$. The data were velocity profiles and wall pressures generated by the same model using a known set of coefficients (equal to the nominal ones).  An advantage of using synthetic data is to remove structural uncertainty, since the trained model is the same used to generate the data.  The results show the ability of the EnKF method to identify the correct model parameters for a relatively low computational cost (ensembles of 100 function evaluations, i.e. CFD calculations). The approach has been extended to more complex flows around airfoils~\cite{kato2015data}, establishing a general framework for combining experimental fluid dynamics and CFD for predictions.

An even more efficient way of finding the optimal coefficients is to maximize the likelihood function by using gradient-based methods. This corresponds to finding the set of closure coefficients corresponding to the maximum probability of observing the data.  The main drawback of this approach is that only deterministic sets of coefficients are obtained as an outcome of the calibration.  \citet{papadimitriou2015bayesian} obtained variance estimates of the optimal coefficients by using the Hessian of the likelihood function with respect to the parameters $\theta$.  They found that the posterior variance due to the overall observational uncertainty (e.g. to the discrepancy between the model output and the data) plays a dominant role.
This indicates that coefficient calibration alone is not sufficient to match the data, and that the bias introduced by the model structure is mostly responsible for the discrepancy.  Unfortunately, Hessian calculations require computing the second sensitivity derivatives of the model with respect to the parameters, which is a highly intrusive and delicate task and is not compatible with black-box Navier--Stokes solvers.

Bayesian strategies similar to that of \citet{cheung2011bayesian} can also provide estimates of the uncertainty associated with the model form, grounded in uncertainties in the space of model closure coefficients.
This can be achieved by calibrating the model separately against several sets of data.
The spread in the posterior estimates of closure coefficients across calibration scenarios provides a measure of the need for readjusting the model coefficients to compensate for the inadequacy.
An example of  such a sensitivity study is given by \citet{edeling2014bayesian}, where the Launder--Sharma model was calibrated separately against 13 sets of flat-plate boundary layer profiles from \citet{kline1969computation}.
The results showed a significant variation in the most-likely closure-coefficients values for the different pressure gradients, despite the relatively restricted class of flows (flat plate boundary layers) considered for the calibrations.

\begin{figure}[!htbp]
  \centering
    \subfloat[Marginal distributions of $\kappa$ for various calibration scenarios] {\includegraphics[scale = 1.2]{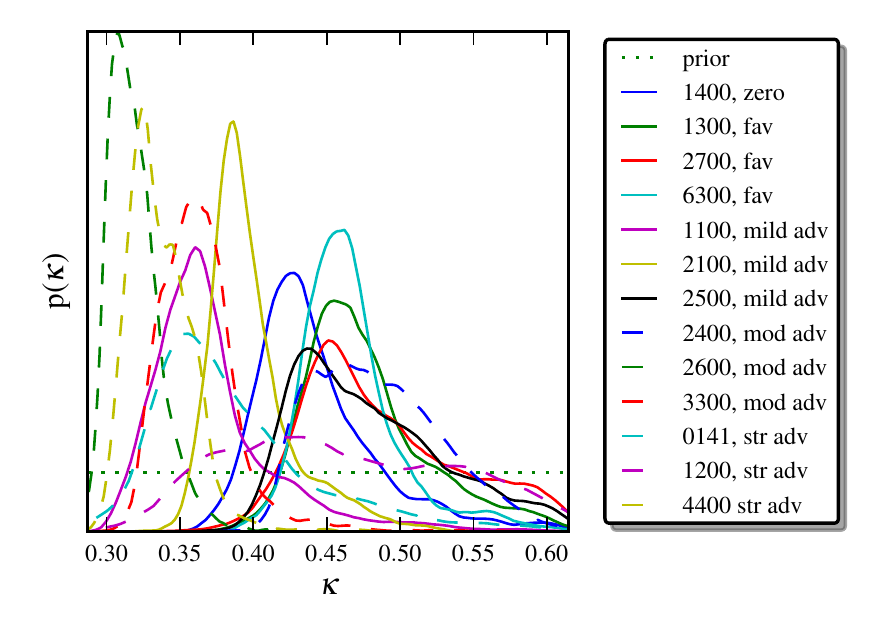}} \\
   \subfloat[p-box of $u^+$ at $y^+=46.2$] {\includegraphics[scale = 1]{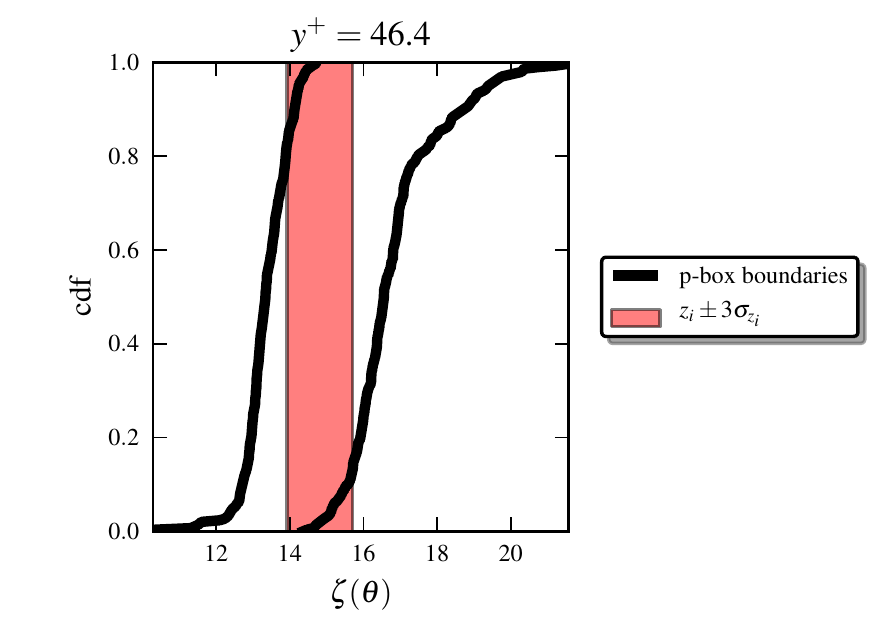}}
   \subfloat[90\% confidence intervals of the velocity profile] {\includegraphics[scale = 0.85]{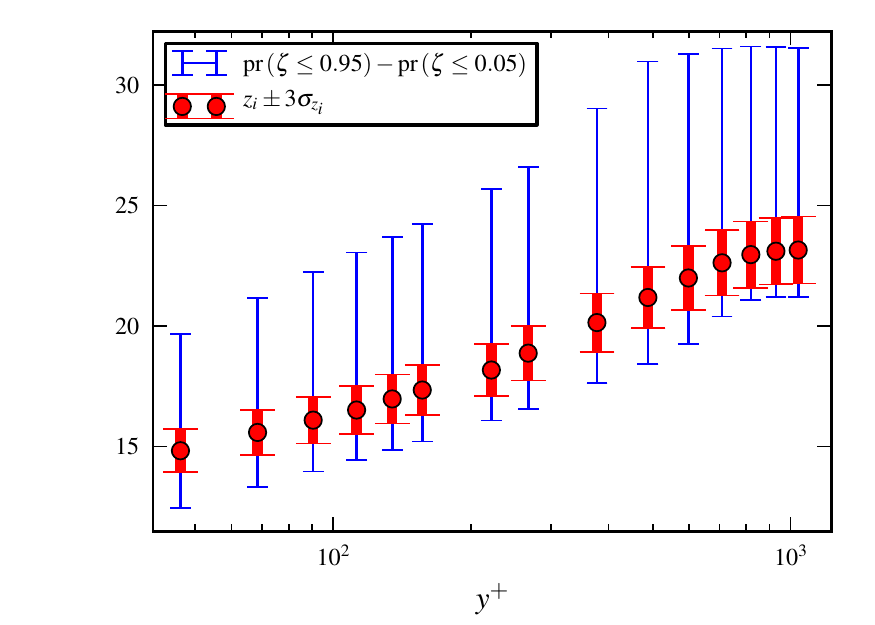}}\\
   \caption{Sample posterior distributions and p-box predictions of a new flow based on 13 separate calibrations of the $k$--$\varepsilon$ model. Figures reproduced with permission from \citet{edeling2014bayesian}.}
  \label{fig:edeling2014a}
\end{figure}

The main lessons learned from the preceding exercise are: (i) there are no universal values for the closure parameters of the turbulence models; (ii) the parameters need to adjust continuously when changing the dataset to compensate the intrinsic inadequacy (simplifying modeling assumptions) of the chosen model (see, e.g., the variation of the marginal posterior pdf for $\kappa$, reported in Figure~\ref{fig:edeling2014a}a); and (iii) as a result, closure coefficients obtained by calibrating the model against a given boundary layer are generally not valid for the prediction of a different one.

The variability of closure coefficients for the observed flow can however be used as a measure of model inaccuracy when predicting a new flow.
In \citet{edeling2014bayesian}, this is done by summarizing the posterior variability of the parameters within and in between calibration datasets (called hereafter \emph{scenarios}) by means of probability boxes (p-boxes), commonly used in Bayesian statistics to summarize the joint effect of parametric and epistemic model-form uncertainties~\citep{oberkampf2010verification}. P-boxes are constructed as the envelope of the empirical  cumulative distribution functions of the  output predicted using different posteriors of the parameters. An example of p-box for the nondimensional velocity $u^+$ predicted at a non-dimensional wall distance $y^+=46.2$ for a pipe flow boundary layer  is given in Figure~\ref{fig:edeling2014a}b. Analogous results are obtained at various locations across the boundary layer, thus leading to an estimate of the uncertainty bounds on the predicted velocity profile (Fig.~\ref{fig:edeling2014a}c).
The p-box prediction is found to encompass the experimental uncertainty intervals, leading however to an overly conservative estimate of the uncertainty bounds.

\subsubsection{Accounting for structural uncertainties in RANS models}
\label{sec:para-stochastic}

A delicate step in Bayesian calibration is the construction of a statistical model relating the true (unseen) process to the data via the model, which is directly related to the definition of the likelihood function.
This should consider at least the fact that the observed quantities differ from the true ones by the experimental (observational) noise, which  may be expressed through the relation:
\begin{align}
z=\zeta+\epsilon
\label{experimental:noise}
\end{align}
with $\zeta$ the true value for $z$ and $\epsilon$ a random vector representative of the experimental noise.  The experimental data noise $\epsilon$ is often assumed to be independently distributed without spatial correlation, and it is modeled as a Gaussian process with diagonal covariance matrix, i.e., $\epsilon \sim \mathsf{N}\left(\mathbf{0},\Lambda \right)$~\cite{cheung2011bayesian,edeling2014bayesian}.

Theoretically, the true value for $\zeta$ could be obtained as an output of the model $y$, once a suitable set of parameters $\theta$ has been identified, i.e. $\zeta=y(\theta)$.  In practice however, no model is perfect.  Even if there is no parameter uncertainty, so that we know the true values of all the inputs required to make a particular prediction of the process being modeled, the predicted value will not equal the true value of the process~\citep{brynjarsdottir2013learning}. The discrepancy is due to model inadequacy.  It is even possible that the physically true value of a calibration parameter gives a worse fit and less accurate future prediction than other values, simply because of the simplifying assumptions upon which the model has been built.  Conversely, it is dangerous to interpret calibration results as estimates of the true physical values of those parameters.

A general framework to include the model inadequacy term in the stochastic model was first proposed in \citet{kennedy2001bayesian}.  Model discrepancy can be taken into account by introducing an additional error term to the statistical model as in Equation~\eqref{experimental:noise}, which could be of additive nature, i.e.,
\begin{align}
z = \zeta + \epsilon = y(\theta) + \eta + \epsilon
\label{additive:error}
\end{align}
or of  multiplicative nature:
\begin{align}
z = \zeta + \epsilon = \eta \circ y(\theta)+\epsilon
\label{multiplicative:error}
\end{align}
The symbol $\circ$ denotes the Hadamard (element-wise) multiplication. Note that all quantities above, $z$, $y$, $\eta$, and $\epsilon$, are spatial fields and should be written as $z(\bm{x})$, $y(\bm{x})$, $\eta(\bm{x})$, and $\epsilon(\bm{x})$, respectively. The spatial dependence is omitted for brevity.  The choice of model-inadequacy formulation largely depends on the nature and prior knowledge about the observed quantity $z$. In Equations~(\ref{additive:error}) and~(\ref{multiplicative:error}), $\eta$ is a random field representative of the model inadequacy, i. e., of the fact that true value is not equal to the code output but with some systematic deviations.  For instance, \citet{cheung2011bayesian} chose a multiplicative error model for relating the measured and computed velocity profiles, so that the no slip boundary condition at the solid wall is satisfied by any realization of the stochastic model. 

When an additive model inadequacy term is used, it becomes difficult to separate its effect from that of the observational error. As a consequence, both terms are often merged together. In all cases, the random variable $\eta$ may involve additional parameters proper to the statistical model introduced for describing the error behavior, referred to as \emph{hyperparameters}.  Sometimes these are known before hand or are estimated independently based on likelihood maximization criteria~\cite{nitschke2018model} but most often they need to be calibrated from the data along with the physical model parameters $\bm{\theta}$.  Another important point is that $\eta$ is expected to correlate modeling errors for a QoI evaluated at different locations in the flow field or for even various QoI for various datasets.  For instance, \citet{cheung2011bayesian} introduced a multiplicative term to calibrate the Spalart--Allmaras model from velocity profiles and skin friction distributions for three boundary layer data sets.  In their work, all the competing multiplicative statistical models describe the inadequacy term as Gaussian process, i.e., $\bm{\eta}\sim \mathsf{N}\left(\bm{1},\Sigma_{\bm{\eta}}\right)$. Consequently, the observations can also be modeled as a Gaussian process, and thus the likelihood function can be written as follows:
\begin{align}
\label{eq:likelihood-fn}
\begin{split}
p(\bm{z}|\bm{\theta})=\frac{1}{\sqrt{(2\pi)^N|\Sigma_{z}|}}\textrm{exp}\left[-\frac{1}{2}\bm{d}^T\Sigma_{z}^{-1}\bm{d}\right] \\
\text{with} \qquad \bm{d}=\bm{z}-\bm{y}\left(\bm{\theta}\right) \quad
\text{and} \quad
\Sigma_{z}=\Lambda+\Sigma_{\bm{\zeta}} ,
\end{split}
\end{align}
where the covariance matrix of the true process is $\Sigma_{\bm{\zeta}}=\textrm{diag}(\bm{y})\,\Sigma_{\eta}\,\textrm{diag}(\bm{y})$ based on the definition $\zeta = \eta \circ y(\theta)$ in Equation~\eqref{multiplicative:error}. It can be seen that the covariance matrix $\Sigma_{\eta}$ of the modeled multiplicative term has impact upon the true process $\zeta$ and thus influences the formulation of the likelihood function as shown in Equation~\eqref{eq:likelihood-fn}.  Different statistical models can be obtained for different choices of the covariance matrix, and more complex choices for the inadequacy term have also been investigated~\citep{oliver2011bayesian}. Their results showed the choice of spatial correlation structure for the modeling inadequacy played an important role in the Bayesian model selection.

Although the use of model inadequacy terms such as those of Equations~(\ref{additive:error}) and (\ref{multiplicative:error}) is helpful in alleviating parameter overfitting problems and in estimating how well the calibrated model is able to fit the data,  the approach suffers from several limitations:
\begin{itemize}
\item The correction terms are specific to the observed QoI and cannot be re-used for the prediction of a different (unobserved) QoI.
\item The correction  often depends on the spatial distribution of observed data for the calibration scenario, and can be hardly applied to, e.g., a different geometry
\item Even if the same QoI and geometrical configuration are considered, the validity of the  inadequacy terms calibrated for a given dataset for the prediction of a different scenario (e.g. operating condition) must be considered carefully.
\end{itemize}
The non universality of the inadequacy term is well illustrated by the results of \citet{edeling2014bayesian}, who used a statistical model involving a multiplicative model-inadequacy term similar to the correlated model $M_3$ of \citet{cheung2011bayesian}. It was observed that the expected value of the model inadequacy term, as calibrated from data, varies significantly from case to case. Additionally, for some of the calibration datasets the posterior values taken by the correction term are much higher than for the other cases, indicating that parameter adjustment was not sufficient for the model output to capture the data.

\subsubsection{Accounting for multiple models: Bayesian model selection and averaging}
\label{sec:bsma}

An interesting outcome of Bayesian calibration is the possibility of deriving statistical criteria for model selection, i.e., for choosing the best model in some statistical sense among a class of competing models.  This consists in providing estimates of the posterior probability of each model in the considered set of models $\mathcal{M} = \{M_1, M_2, \cdots, M_I\}$ given the observed data. The ``model'' here should be interpreted in a broader sense, including not only physical models (e.g., $k$--$\varepsilon$, $k$--$\omega$, and Reynolds stress models) with associated coefficients but also statistical models (e.g., covariance kernel used to construct likelihood functions~\citep{cheung2011bayesian,oliver2011bayesian} as in Equations~(\ref{additive:error})--(\ref{eq:likelihood-fn})).  Model probabilities are obtained as an outcome of parameter calibration introduced above. First, each model in the set $\mathcal{M}$ is assigned a probability $\mathbb{P}(M_j)$, $j=1, \cdots, I$, based on prior knowledge (e.g. from expert elicitation) or the lack thereof, in which case a noninformative, uniform distribution is chosen.  Additionally, the prior distributions for the closure coefficients $\theta$ or statistical hyperparameters associated with each model are also specified.  If data $\mathcal{D}$ are available, the prior probability mass function (pmf) can be updated according to Bayes' theorem, leading to the {\it posterior pmf} of model $M_j$:
\begin{align}
\label{eq:model-probability}
\mathbb{P}(M_j|\mathcal{D})=\frac{p(\mathcal{D}|M_j) \, \mathbb{P}(M_j)}{\sum_{i=1}^I p(\mathcal{D}|M_i) \, \mathbb{P}(M_i)} \qquad j = 1, \cdots, I
\end{align}
where $p(\mathcal{D}|M_j)$ is the evidence for model $M_j$ that normalizes the posterior pdf of the model parameters $\theta$, as in Equation~(\ref{eq:bayes}).  The evidence can be computed at the end of the calibration by numerically integrating the numerator of Equation~(\ref{eq:bayes}), using the posterior samples of $\theta$. This can be a challenging process requiring special techniques \citep[e.g.,][]{prudencio2012parallel}). The estimated pmf of the models can subsequently be used for predictions by choosing the model with maximum posterior probability in the case of \emph{model selection}, or alternatively by weighting the various posterior predictive distributions for the QoI with the posterior pmf in the case of \emph{model averaging}.

\begin{figure}[!htbp]
  \centering
    \includegraphics[width=0.6\textwidth]{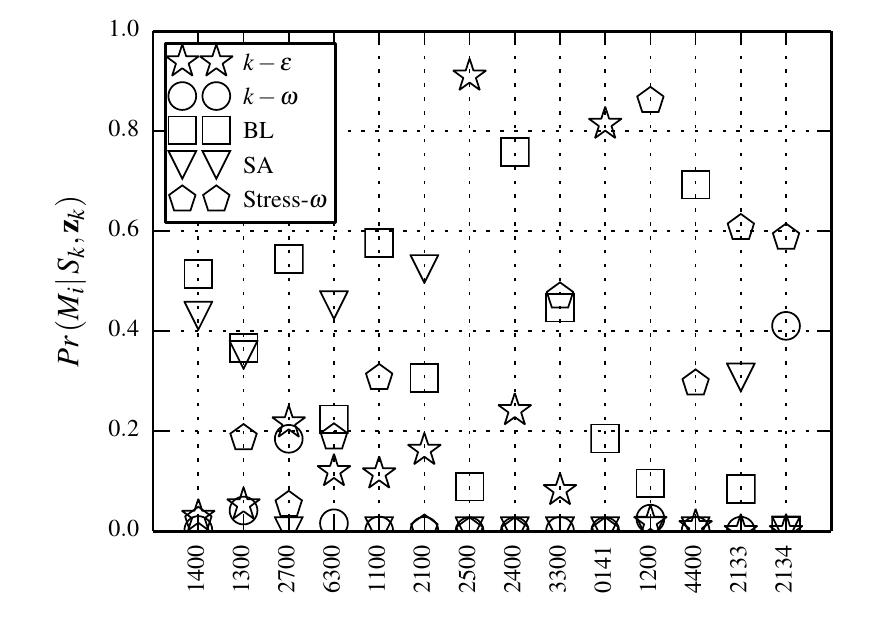}
    \caption{Posterior probabilities $\mathbb{P}(M_i | \mathcal{D}_k)$ of five turbulence models for 13 calibration datasets (boundary layers of various external pressure gradients). The set of models includes a simple algebraic model (Baldwin and Lomax~\citep{baldwin1978thin}), one-equation and two-equation eddy viscosity models (SA model~\citep{spalart1992one-equation}, $k$--$\varepsilon$ model, and $k$--$\omega$ model~\citep{wilcox2006turbulence}), and a Reynolds stress model (stress--$\omega$ model \citep{wilcox2006turbulence}). Numbers on the horizontal axis denote identification codes for datasets (flow configurations). Figure reproduced with permission from \citet{edeling2014predictive}.}
  \label{fig:posterior}
\end{figure}

The approach above has been used for Bayesian model selection and calibration in RANS simulations. It was found difficult to identify a single best model for a range of flows. Consequently, predicting new (unobserved) flow scenarios based on a single closure model calibrated on a limited dataset may lead to biased results, and thus Bayesian model selection is insufficient.  \citet{oliver2011bayesian} calibrated the combination of four eddy viscosity models and three statistical models by using DNS data of plane channel flows and compared the posterior probabilities and predictive capabilities. The results showed that the considered data slightly favored Chien's $k$--$\varepsilon$ model~\citep{chien1982predictions} with an inhomogeneous stochastic model for the inadequacy, but no clear winner emerged with a dominantly high posterior probability.  \citet{edeling2014predictive} systematically demonstrated the difficulty of identifying a single best model without ambiguity. They used Bayesian inference to compute the posterior probabilities of five turbulence models ranging from simple algebraic eddy viscosity models to sophisticated Reynolds stress models by using DNS data of 13 boundary layer flows of various configurations. The posterior pmf for each dataset are presented in Fig.~\ref{fig:posterior}, which suggests that none of the models has a consistently higher probability than other models for all datasets, and the probabilities of all models are highly flow-dependent. As a consequence, it was not possible to select a single best model valid for all flow configurations.  Moreover, somewhat surprisingly, the Reynolds stress model was not the most plausible one for all flows despite its theoretical superiority; on the other hand, after calibration the algebraic model performed rather well over a wide range of flow configurations.

The difficulty of making predictions with a single calibrated model clearly calls for a framework based on multi-model ensembles. Multi-model approaches have been used in aerodynamics~\citep{poroseva2006improving} and many other applications~\citep{diomede2008discharge,duan2007multi-model,tebaldi2007use}.  Bayesian modeling averaging is among the most widely used multi-model approaches, where the posterior of the predicted quantity $\Psi$ is~\cite{draper1995assessment,hoeting1999bayesian}:
\begin{align}
  p(\Psi \mid {\cal D}, {\cal M}) = \sum_{i=1}^I  p(\Psi\mid M_i) \;\, \mathbb{P}(M_i \mid \mathcal{D}) ,
 \label{eq:bma-post}
\end{align}
given calibration data $\mathcal{D}$ and a set of models $\mathcal{M}$.  In this framework the posterior of $\Psi$ is an average of $I$ posterior predictive distributions corresponding to $I$ competing models weighted by their respective model posterior as computed from Equation~(\ref{eq:model-probability}).

A significant recent development is the Bayesian model--scenario averaging (BMSA), which is an extension of the classical Bayesian model averaging as shown in Equations~(\ref{eq:model-probability}) and (\ref{eq:bma-post}) above.  BMSA accounts for uncertainties on the choice of the calibration flow configuration (referred to as \emph{scenario}).  It predicts the QoI for a new scenario $\tilde{S}$ (not used for model calibration) as a weighted average of the predictions provided by a set of models ${\cal M} = \{M_i\}_{i=1}^I$, each model being previously calibrated against a set of scenarios $\mathcal{S} = \{S_k\}_{k=1}^K$ with corresponding datasets ${\cal D} = \{\mathcal{D}_k\}_{k=1}^K$.  Specifically, BMSA yields the posterior distribution of $\Psi$ as follows:
\begin{align}
  p(\Psi \mid \tilde S; {\cal D}, {\cal M}, {\cal S}) = \sum_{k=1}^K \sum_{i=1}^I p(\Psi\mid \tilde{S};   M_i,  S_k, \mathcal{D}_k) \; \underbrace{\mathbb{P}(M_i\mid \mathcal{D}_k, S_k)}_{\text{model posterior}}
  \; \underbrace{ \mathbb{P}(S_k)}_{\text{ scenario prior}}
 \label{eq:pbmsa}
\end{align}
which is an average of the $I\times K$ posterior predictive distributions $p(\Psi\mid \tilde S; \mathcal{D}_k, M_i, S_k)$,  each corresponding to the forward propagation of the parameter posterior obtained by calibration of model $I$ against scenario $K$ through the new prediction scenario $\tilde{S}$. The average is weighted by the corresponding posterior model probability $\mathbb{P}(M_i \mid \mathcal{D}_k, S_k)$ and prior scenario probability $\mathbb{P}(S_k)$. It is important to stress here that, for nonlinear systems, averaging the posterior predictive distributions of the QoI obtained by propagating the posterior pdf of the parameters for various scenarios through each model, as in Equation~(\ref{eq:pbmsa}), is radically different than creating a mixture of the $K$ pdfs for the closure coefficients and propagating it through the model. Specifically, \citet{ray2018robust} showed that latter provided unsatisfactory predictions, albeit being less expensive computationally.

In the BMSA prediction, the posterior probability of model $M_i$ is the outcome of the multiple calibration process after application of Equation~(\ref{eq:model-probability}). On the other hand, the scenario probability $\mathbb{P}(S_k)$ needs to be specified \textit{a priori} and represents the user's belief about the similarity between calibration scenario $S_k$ to prediction scenario $\tilde{S}$ when the prediction of $\Psi$ is concerned.  When a physically justified prior is not available, a non-informative, uniform pmf can be used, implying equal probabilities for all scenarios. However, this may overestimate the posterior variance for $\Psi$, which leads to an overly pessimistic estimate of the prediction uncertainty~\citep{edeling2014predictive}. To address this issue, \citet{edeling2014bayesian} proposed an empirical scheme for choosing the scenario prior, with $\mathbb{P}(S_k)$ being inversely proportional to the scattering of all models trained on scenario $S_k$ when predicting the QoI for $\tilde{S}$.  The rationale is that if a calibration scenario $S_k$ is similar to the prediction scenario $\tilde S$, the models would give similar predictions of the QoI.

A major drawback of BMSA is its high computational cost, since it requires $I\times K$ stochastic calculations, each requiring forward propagation of a posterior parameter pdf (corresponding to a model/scenario combination) through the CFD model. The computational cost can be drastically reduced to~$I\times K$ deterministic CFD simulations by propagating though $\tilde{S}$ only the set of parameters  with maximum posterior probability for each model and calibration scenario~\citep{edeling2018bayesian}, instead of the full pdf. With this simplification, the BMSA approach was applied to complex flow configurations such as the transonic three-dimensional flow around the ONERA M6 wing.

As noted by \citet{draper1995assessment}, multi-model approaches still introduce biases in the prediction because of the subjective selection of a finite set of models. However, they play a useful role in reducing the bias compared to predictions based on a single model. An averaged model is a way of obtaining a conservative prediction for an unseen configuration.  Indeed, the result will not be as good as the (\textit{a priori} unknown) best model but will not be as bad as the worst one. Additionally, BMSA provides an estimate of the solution variance based on the solution variability among the competing models.

\section{Non-parametric approaches}
\label{sec:nonparametric}

\subsection{Motivation and overview}
\label{sec:nonpara-intro}
\begin{figure}[!htbp]
  \centering
  \includegraphics[width=0.5\textwidth]{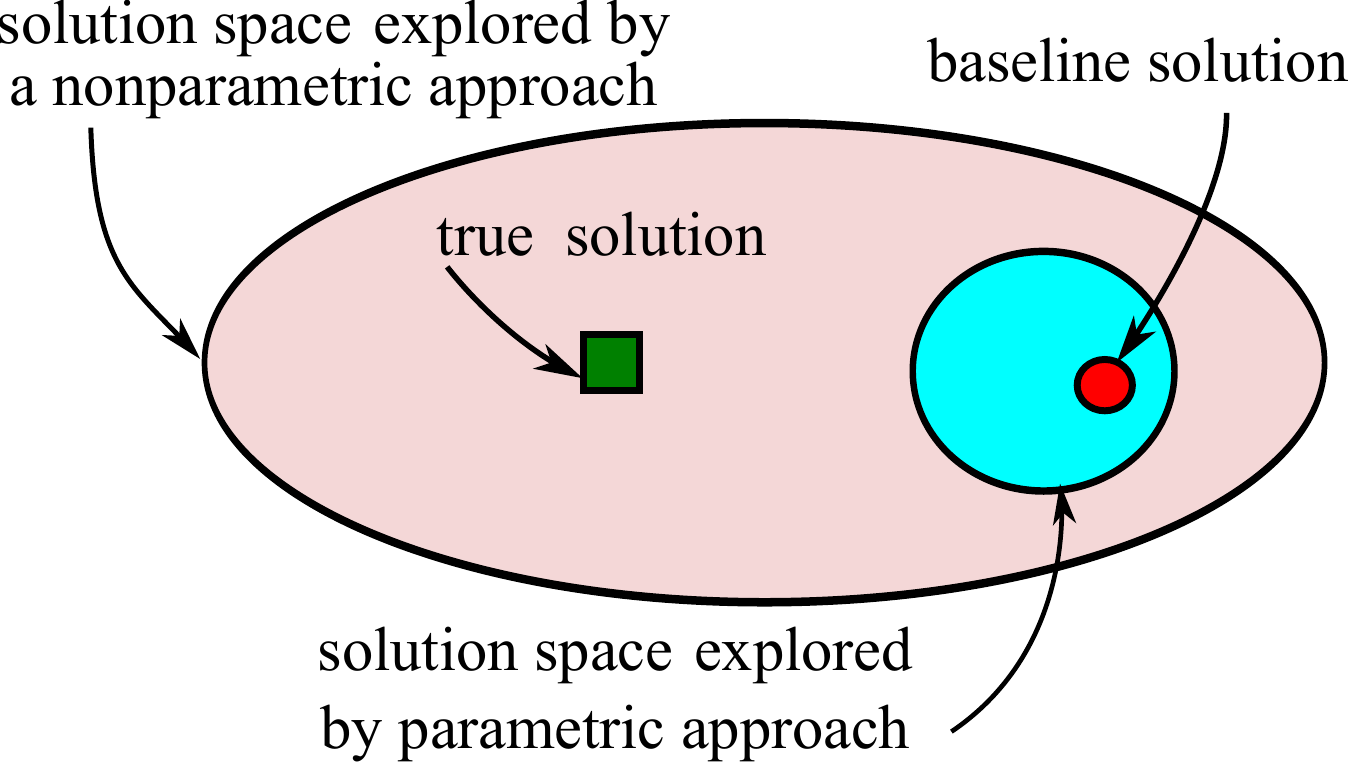}
  \caption{A conceptual illustration of the merit of the non-parametric approach in RANS model uncertainty quantification, i.e., the ability to explore the solution space more thoroughly.  Figure inspired by \citet{soize2005comprehensive}.}
   \label{fig:space}
\end{figure}

The parametric and multi-model approaches introduced in Section~\ref{sec:parametric} explore the uncertainties in the model coefficients and in the model choices. However, it is possible that the true solution lies outside the region in the solution space reachable by the parametric approaches. For example, it is well-known that linear eddy viscosity models are intrinsically not capable of predicting the secondary flows in a square duct. Such a feature is driven by the anisotropy of the Reynolds stresses, but the Boussinesq assumption that is inherent to linear eddy viscosity models excludes this part of the solution space. This intrinsic deficiency cannot be remedied by the calibration of coefficients. An ensemble or averaging of linear eddy viscosity models would not be able to predict such a feature either, because all the models would strongly agree on the wrong solution. A larger portion of the solution space could be covered by introducing a wider variety of models (namely, non-Boussinesq) in the multi-model ensemble. However, the choice of the set of models remains subjective and the selection of a finite set of models prevents the approach from exploring the entire solution space, limiting it to only the portion spanned by the chosen model ensemble. In order to go beyond these limitations, an intriguing possibility is to introduce uncertainties directly into the turbulent transport equations or the modeled terms such as the Reynolds stress or eddy viscosity. Such non-parametric approaches allow for more general estimates of the model inadequacy than the parametric approaches.  As illustrated conceptually in Figure~\ref{fig:space},  the solution space explored by parametric approaches is a subspace of that explored by nonparametric approaches, and the true solution may lie outside the former space.

\begin{figure}[!htbp]
  \centering
  \includegraphics[width=0.55\textwidth]{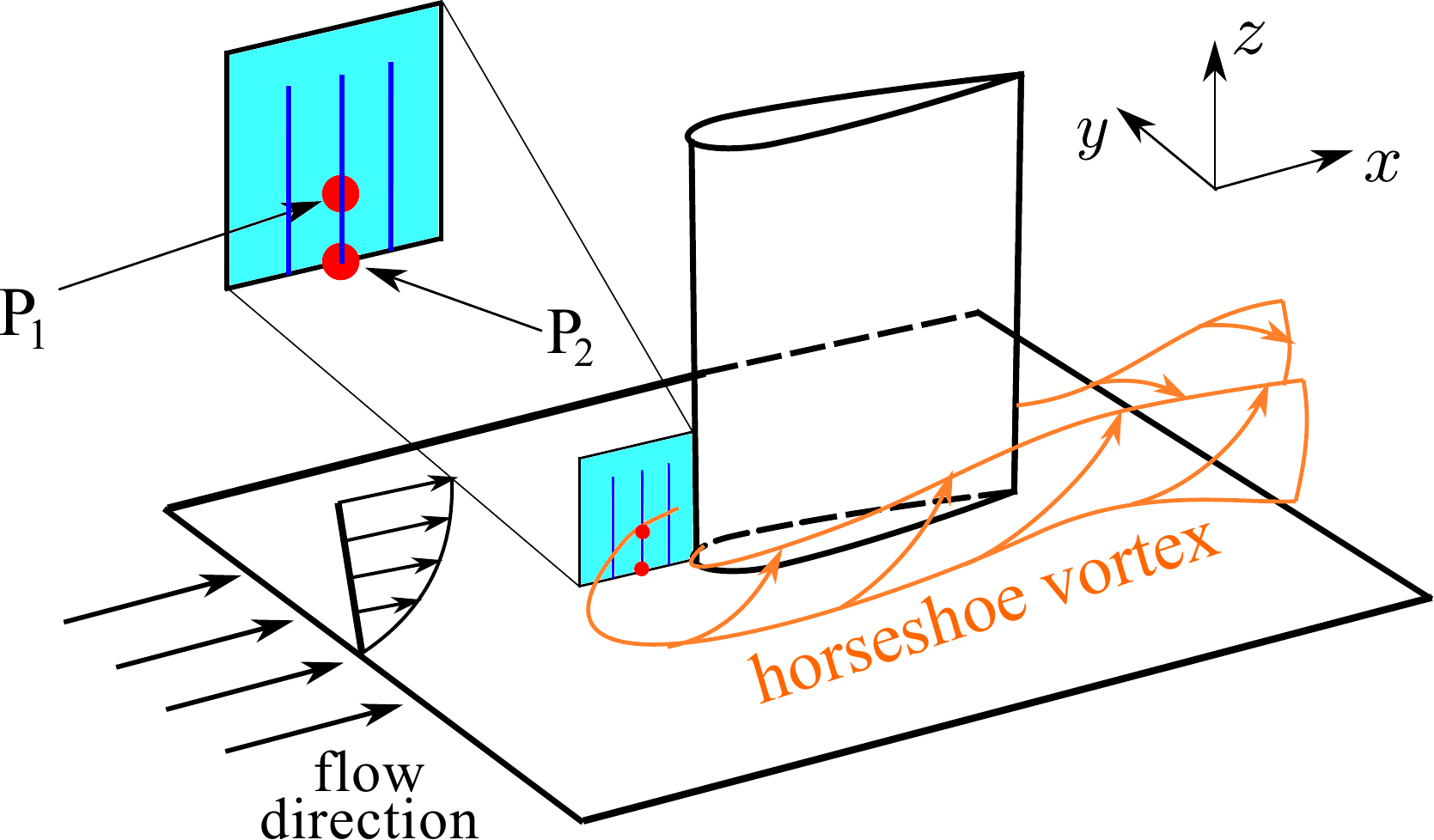}
   \caption{Configuration of the wing--body junction flow, illustrating the points where orientations of the Reynolds stress tensors (Figure~\ref{fig:orientation}) are presented and three lines where the velocity profiles (Figure~\ref{fig:wing-body}) are presented.
   }
   \label{fig:wing-body-schematic}
\end{figure}

\begin{figure}[htbp]
  \centering
  \includegraphics[width=0.7\textwidth]{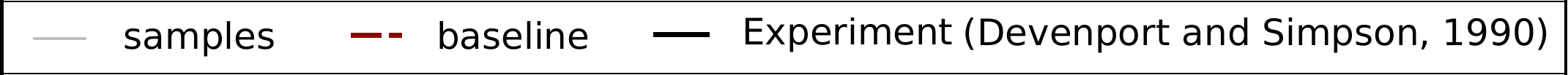}\\
  \vspace{1em}
  \subfloat[ $\mathbf{v}_1$ at point  ${P}_1$]{\includegraphics[width=0.245\textwidth]{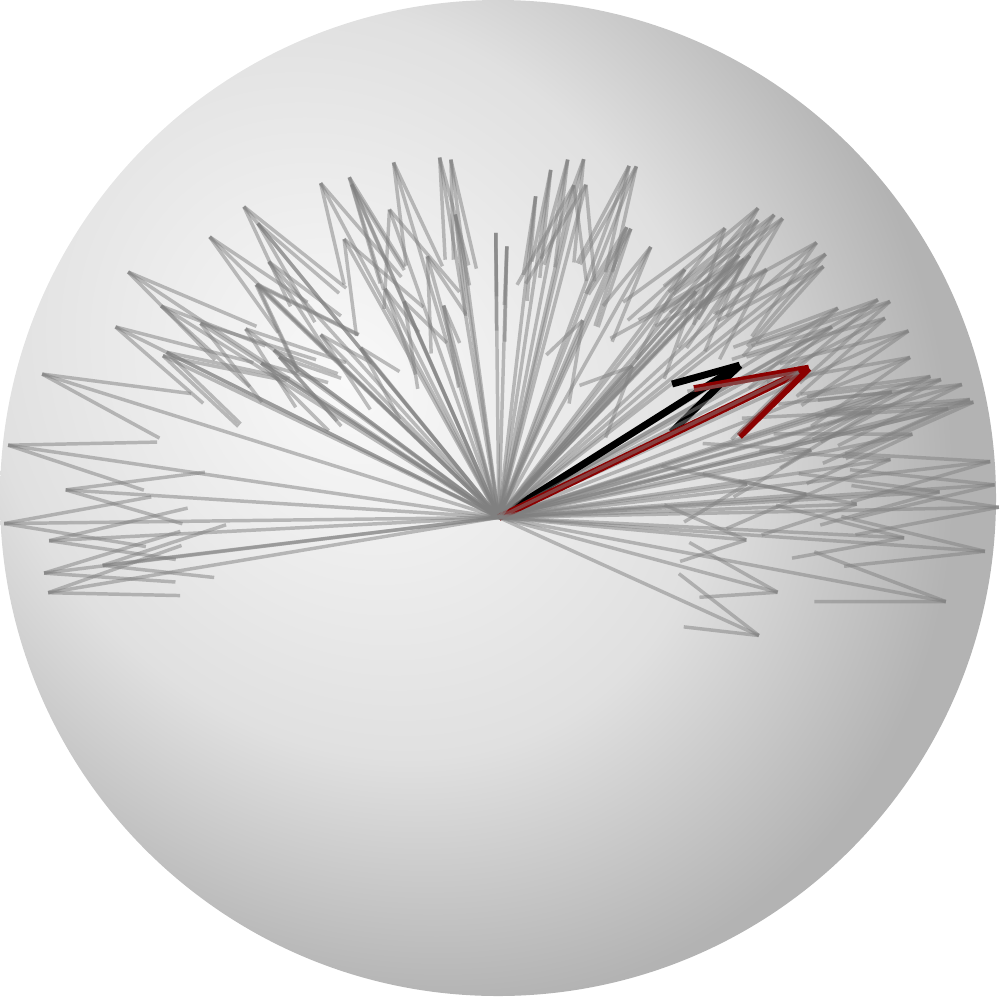}}
  \subfloat[ $\mathbf{v}_2$ at point ${P}_1$]{\includegraphics[width=0.245\textwidth]{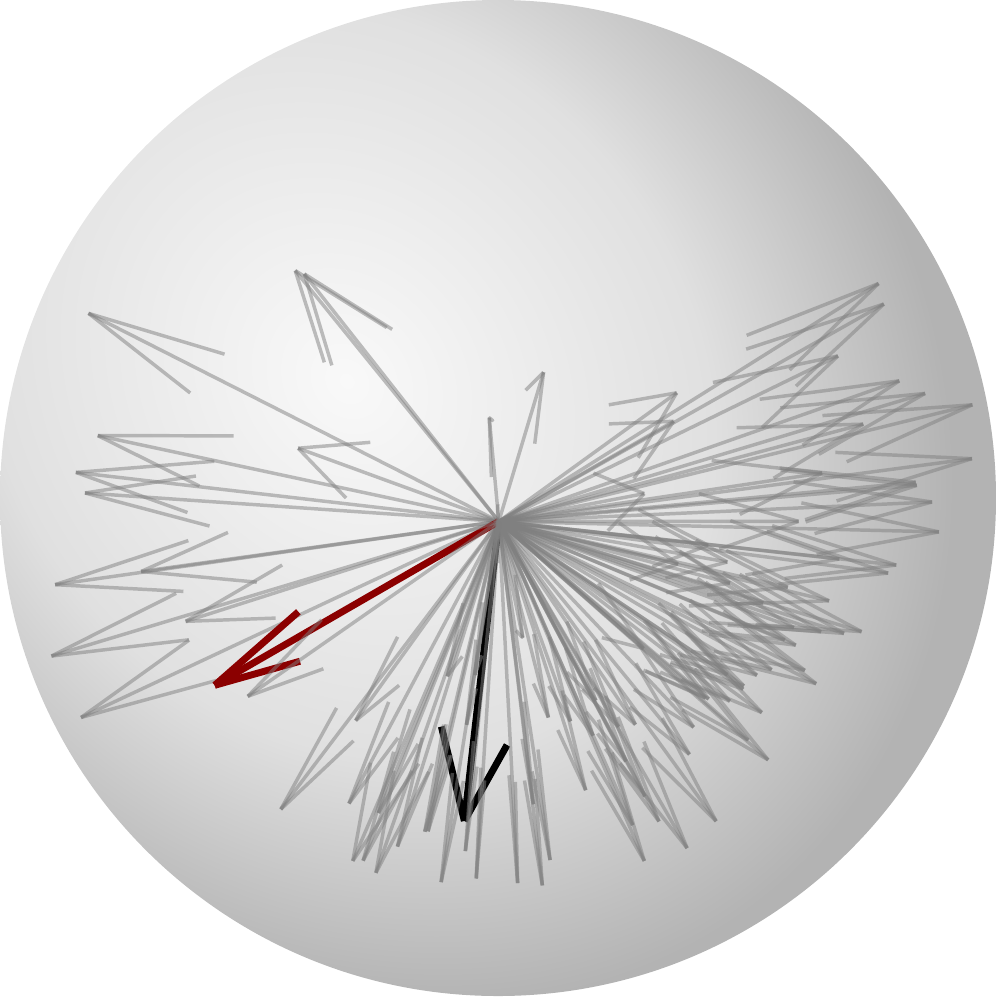}}\\
  \subfloat[ $\mathbf{v}_1$ at point~${P}_2$]{\includegraphics[width=0.245\textwidth]{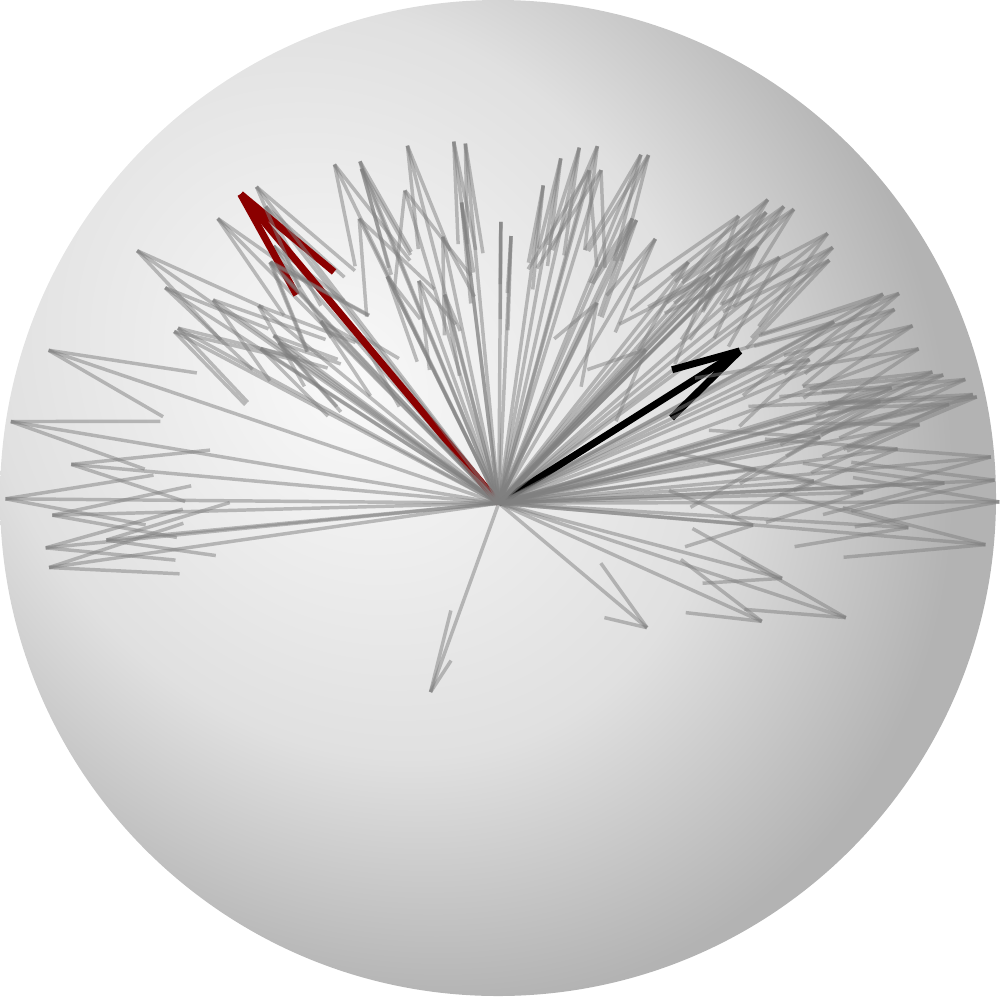}}
  \subfloat[ $\mathbf{v}_2$ at point ${P}_2$]{\includegraphics[width=0.245\textwidth]{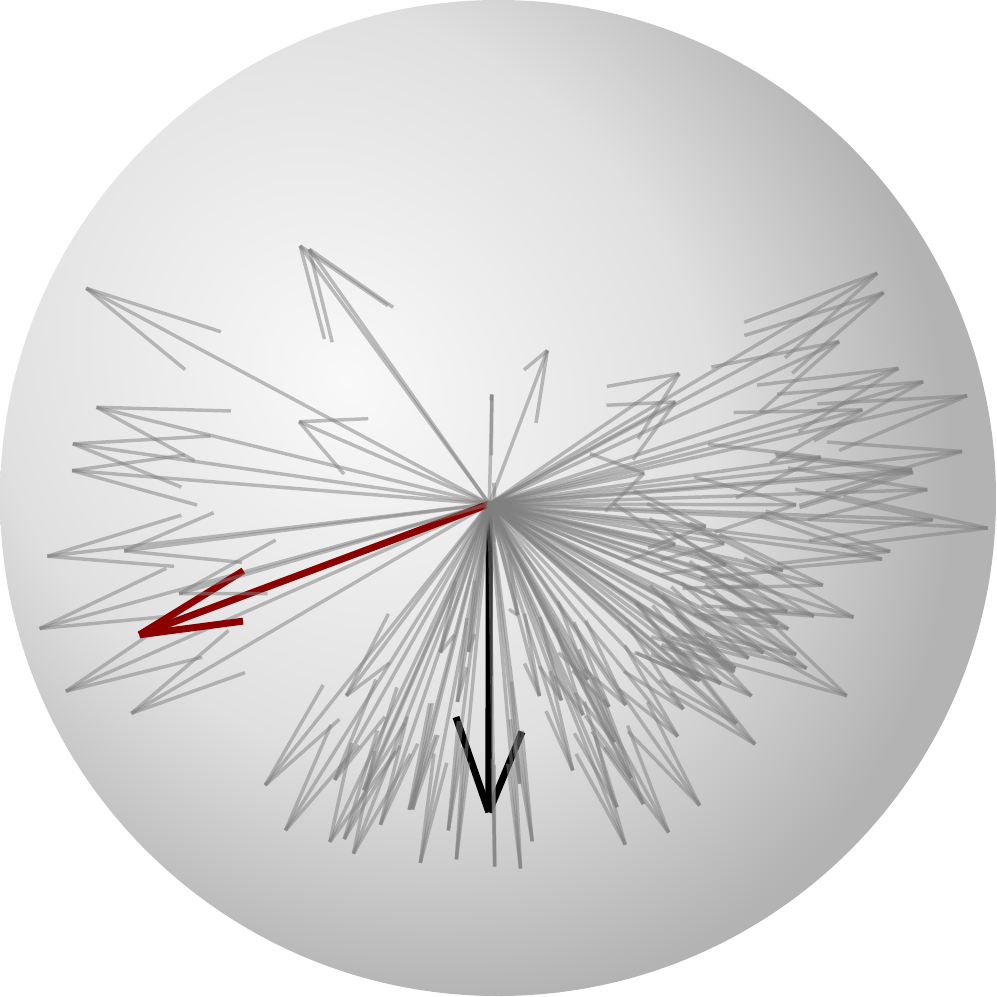}}
  \caption{Comparison of orientations (as indicated by two eigenvectors $\mathbf{v}_1$ and $\mathbf{v}_2$) of Reynolds stresses tensor $\bm{\tau}$ from RANS modeling (with  SST $k$--$\omega$ model) and experimental measurement of \citet{devenport1990time-dependent} at freestream locations $P_1$ (a and b) and near-wall location $P_2$ (c and d). Grey arrows indicate the perturbations on the eigenvectors for exploring uncertainties in RANS-predicted Reynolds stresses, which is a \textbf{non-parametric approach}.  The third eigenvector $\mathbf{v}_3$ of $\bm{\tau}$ and $\mathbf{S}$ can be uniquely determined from $\mathbf{v}_3 = \mathbf{v}_1 \times \mathbf{v}_2$ and are thus omitted. Figures reproduced from~\citet{wu2016quantifying} (unpublished manuscript).}
  \label{fig:orientation}
\end{figure}

\begin{figure}[!htbp]
  \centering
   \subfloat[perturbing turbulent kinetic energy only] {\includegraphics[width=0.5\textwidth]{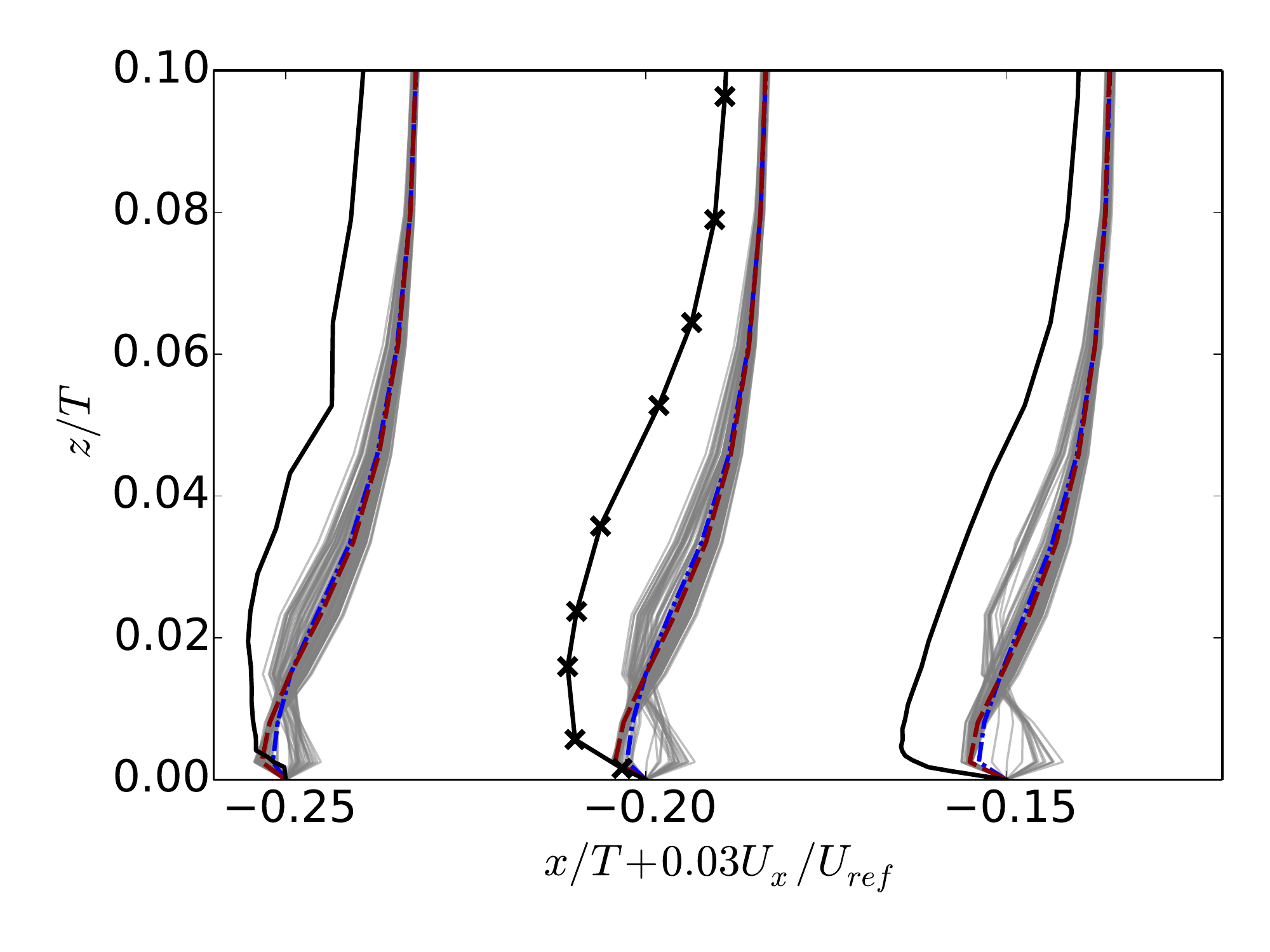}}
   \subfloat[perturbing full Reynolds stress] {\includegraphics[width=0.5\textwidth]{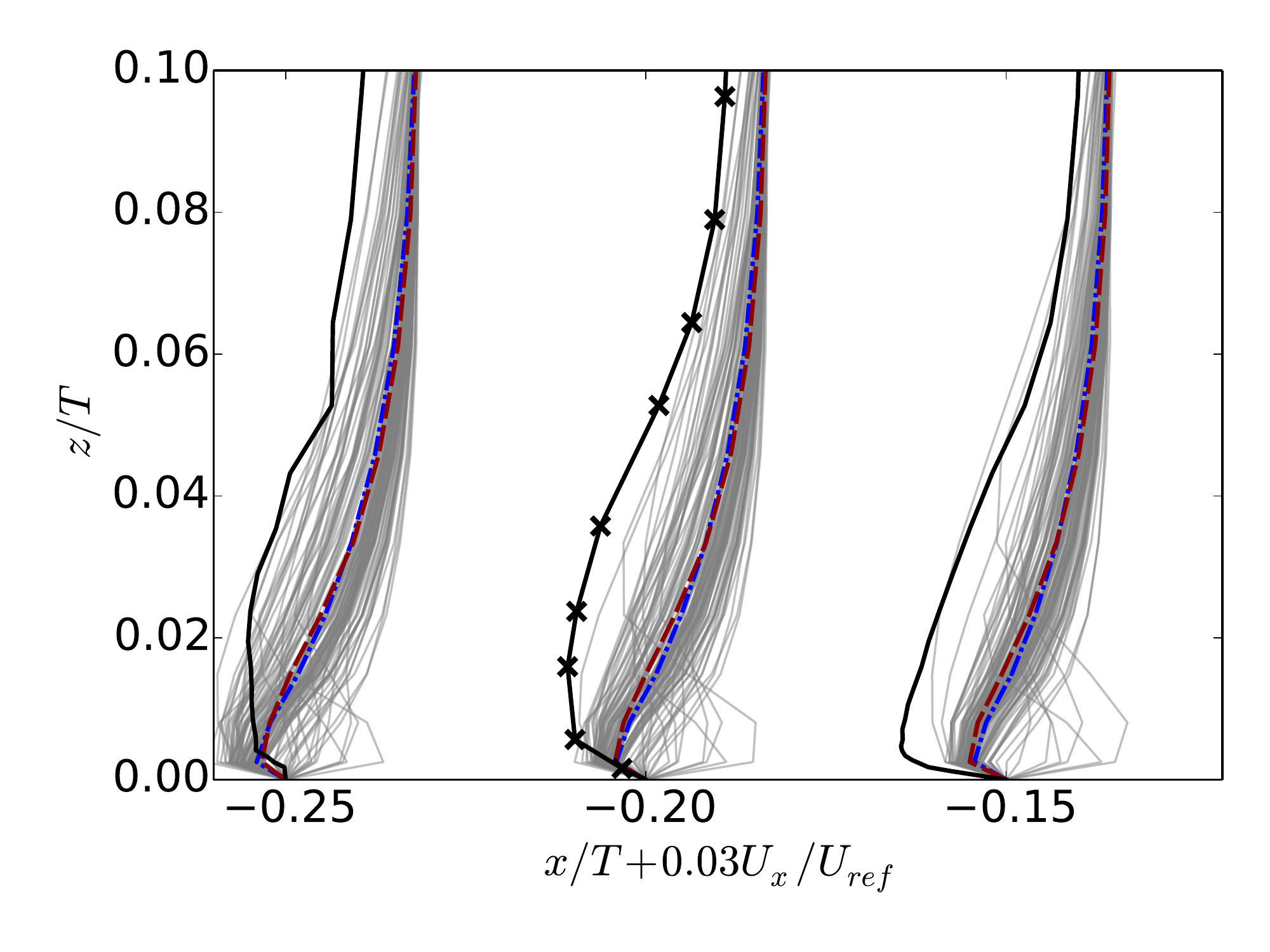}}
   \caption{ Comparison of parametric and nonparametric approaches for model-form uncertainty by using RANS simulations (with  SST $k$--$\omega$ model) of a wing--body junction flow as an example.  This figure compares the mean velocities at three locations (shown in Figure~\ref{fig:wing-body-schematic}) in front of the leading edge of a wing--body juncture obtained by (a) perturbing turbulent kinetic energy only and (b) perturbing the full Reynolds stress, corresponding to parametric and nonparametric approaches, respectively.  Figures reproduced from~\citet{wu2016quantifying} (unpublished manuscript).  }
\label{fig:wing-body}
\end{figure}

We use the wing--body juncture flow as an example to motivate the use of nonparametric approaches in exploring solution spaces for RANS model uncertainty quantification. This configuration consists of an airfoil attached to a flat plate, which is representative of the flows at the wing--fuselage connection of fixed-wing aircraft and blade--hub assembly in turbomachinery.  This flow features an abrupt stagnation of the mean flow at the leading edge and a horseshoe vortex around the juncture of the wing and the body as shown in Figure~\ref{fig:wing-body-schematic}.  Due to the high non-equilibrium turbulence, the Reynolds stress $\bm{\tau}$ and strain rate $\mathbf{S}$ at the leading edge region are not aligned with each other, and thus the Boussinesq assumption fails.  Figure~\ref{fig:orientation} shows clearly the misalignment between orientations of RANS-modeled Reynolds stress (with  SST $k$--$\omega$ model)~\cite{wu2016quantifying} and the experimentally measured Reynolds stress~\citep{devenport1990time-dependent} at two locations, particularly at the near-wall point $P_2$ (see Figure~\ref{fig:wing-body-schematic}).  Consequently, when exploring uncertainties in the RANS simulations for this flow, the velocity samples obtained with parametric approaches (gray lines in Figure~\ref{fig:wing-body}a) based on the Boussinesq assumption, are not able to encompass the truth ($\times$ symbols). This is because such a parametric approach is not able to account for the different eigen-directions of the RANS-modeled and the true Reynolds stresses.  In contrast, a nonparametric approach that perturbs the RANS-modeled Reynolds stresses, including their \emph{eigen-directions} (see gray arrows in Figure~\ref{fig:orientation}), can effectively span a range covering the true solution~\cite{wu2016quantifying} (Fig.~\ref{fig:wing-body}b).

A number of nonparametric approaches have been proposed to quantify model uncertainties in RANS simulations, which can be broadly classified into two categories:
\begin{enumerate}[(1)]
\item those introducing uncertainties into the \emph{model forms}, e.g., turbulent transport equations (for fields $k$, $\omega$, or $\bm{\tau}$)~\citep{singh2016using,parish2016paradigm}, and
\item those introducing uncertainties into the \emph{model outputs}, e.g., the turbulent viscosity field~\citep{dow2011quantification} or the Reynolds stress field~\citep{emory2011modeling,emory2013modeling,xiao2016quantifying,ling2016reynolds}.
\end{enumerate}
At the algorithmic level, the different parametric and nonparametric approaches outlined above target different stages of the algorithms in turbulence modeling, i.e., the parametric level, the PDE level, and the intermediate field level. This is illustrated in Figure~\ref{fig:hierarchy} by using linear eddy viscosity models and Reynolds stress transport models as examples. The intermediate fields and PDEs where uncertainties are introduced are highlighted in shaded (orange) boxes in Figure~\ref{fig:hierarchy}. At a fundamental level, however, they differ from each other in their respective assumptions on where the RANS model uncertainties originate from: the coefficients, the model form of the transport equations, the eddy viscosity field, or the Reynolds stress itself.  As reviewed above, even for a specific flow it is difficult to identify the exact source of the model uncertainty (see Section~\ref{sec:origin-rans-uq}) due to the coupling among various levels of uncertainties. As such, any such statements on the relative importance of different sources of uncertainties are likely to be not only flow-specific but also weak and inconclusive~\citep{ray2018robust}. Consequently, the relative advantages of various approaches are far from clear as of now.

Concerning the comparison between parametric and nonparametric approaches, the parametric approaches allow straightforward extrapolation of the calibrated coefficients to additional flow configurations that are not in the calibration dataset. However, naive extrapolation  may lead to an over-fitted model with reduced predictive capability, particularly when the generalization of the coefficients cannot be justified. On the other hand, extrapolating a calibrated field from nonparametric approaches (generally dependent on space and time coordinates) is a much more delicate task.  As to the comparison between model-form-based and model-output-based UQ approaches, research so far suggests that model-form based approaches are more robust as they involve only mild perturbations of equations in the original models~\cite{singh2016using}.  On the other hand, model-output perturbation approaches make it easier to utilize benchmark (DNS, LES, or experimental) data for the  Reynolds stress or turbulent viscosity, because the quantities being perturbed or inferred have better physical anchoring.  Both categories of approaches will be reviewed and compared below.

\subsection{Introducing uncertainties in turbulent transport equations}
\label{sec:nonpara-beta}

\begin{figure}[!htbp]
  \centering
  \includegraphics[width=0.45\textwidth,height=0.39\textwidth]{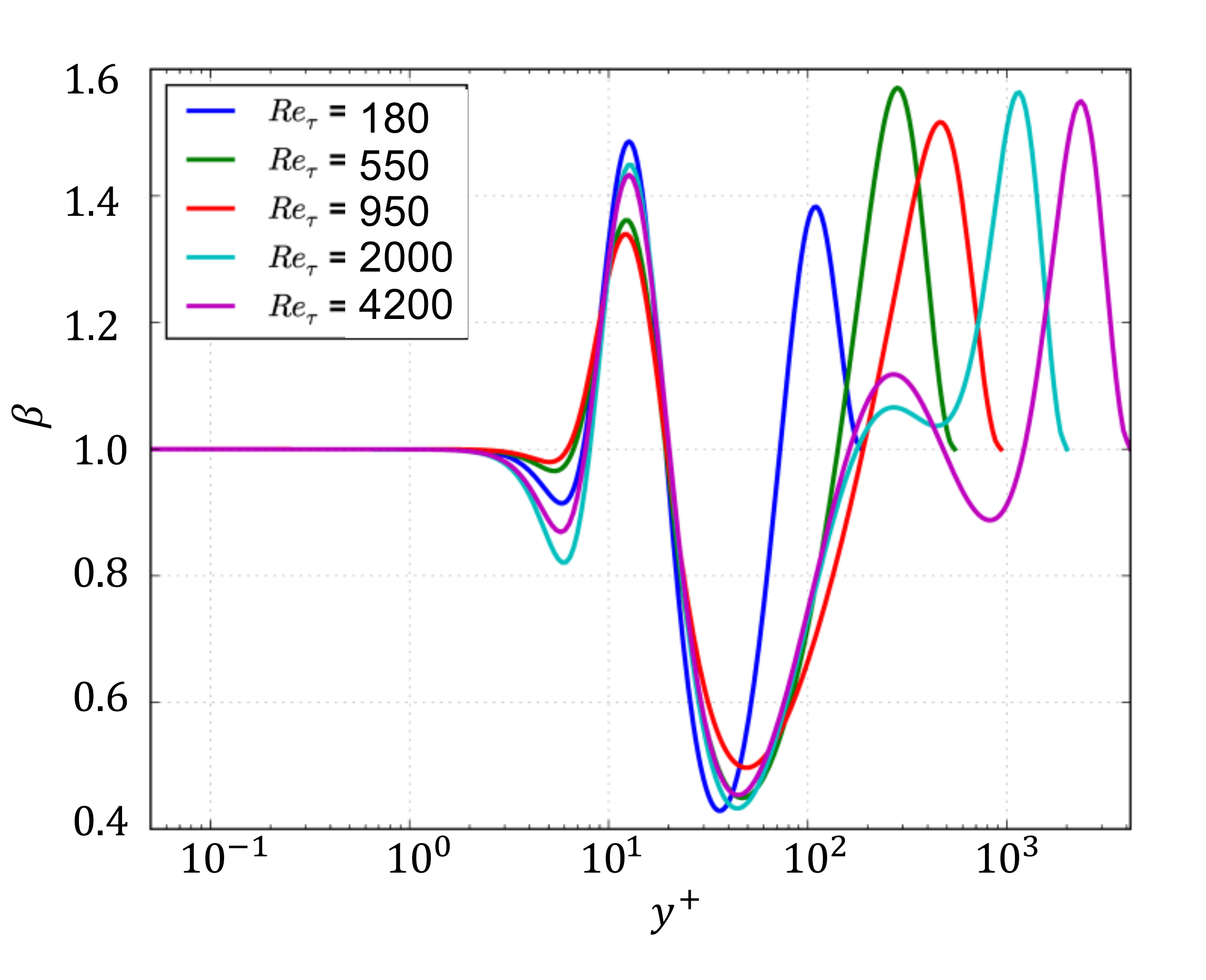}
    \caption{
    Profiles of the inferred correction function $\beta$ in plane channel flows at Reynolds numbers $Re_\tau$ ranging from 180 to 4200.  Figure reproduced with permission from \citet{singh2016using}.  
    }
  \label{fig:kdur}
\end{figure}

The parametric and multi-model approaches are restricted to the chosen baseline models. An immediate extension of these approaches is to perturb the model forms in a non-parametric way, i.e., by modifying the source terms in the turbulent transport equations (e.g., for $k$, $\omega$, and $\bm{\tau}$).  This choice is based on the assumption that errors in the turbulent transport equations rather than the structural uncertainties, e.g., those associated with the Boussinesq assumption, are the dominant source of the prediction errors in RANS simulations.  The uncertainties introduced in this approach depend on the specific form of the baseline turbulence model.  Taking the $k$--$\omega$ equation for example, a multiplicative discrepancy field $\beta(\bm{x})$ is introduced to the source terms of the $\omega$ transport equation by \citet{singh2016using}:
\begin{equation}
    \label{eq:omega}
    \frac{D \omega }{D t} = \beta(\bm{x}) \mathcal{P}_\omega(k, \omega, U_i) - D_\omega(k, \omega, U_i) + T_\omega(k, \omega, U_i)
\end{equation}
where $\omega$ is the turbulent frequency; $ \mathcal{P}_\omega$, $D_\omega$, and $T_\omega$ indicate production, dissipation, and transport, respectively, of $\omega$.  This formulation is equivalent to introducing an additive discrepancy $\delta_\omega = (\beta(\bm{x}) -1 ) \mathcal{P}_\omega $ but has better conditioning than the latter~\cite{singh2016using}. The discrepancy field $\beta(\bm{x})$ can be inferred by using DNS or experimental data of velocities or other quantities of interest, e.g.,  drag, lift,  pressure coefficient, and surface friction.  Assuming the velocity is the data to be used,  the inference can be cast as the following optimization problem:
\begin{equation}
\label{eq:beta-opt}
  \beta^{\text{opt}} = \arg \min_{\beta} J \; ,  \qquad \text{with} \quad J = \| \mathbf{U}(\beta) - \mathbf{U}^{\text{dns}}  \|_{L^2}
\end{equation}
where $\| \cdot \|_{L^2}$ indicates $L^2$ norm. In cases where other derived quantities $\bm{g}$ (e.g., drag and lift) are used in the optimization, an observation operator $\mathsf{H}$ is needed to map the solution to these quantities, i.e., $\bm{g} = \mathsf{H}[\mathbf{U}]$, and the cost function would be $J = \| \bm{g}(\beta) - \bm{g}^{\text{dns}} \|_{L^2}$.  The inferred discrepancy $\beta^{\text{opt}}(\bm{x})$ is a correction that allows the baseline $k$--$\omega$ model to agree with the data.  The discrepancy field $\beta$ resides in a space of very high dimensions with a dimension equal to the number of cells in the CFD mesh, and thus the optimal solution is not unique. In the terminology of inverse modeling, this problem is ill-posed and needs to be regularized.  The deviation of $\beta$ from $1$ is used as a penalty to regularize the problem, which leads to the following cost function~\cite{singh2016using}:
\begin{equation}
  \qquad \text{with} \quad J = \| \mathbf{U}(\beta) - \mathbf{U}^{\text{dns}}  \|_{L^2} + \gamma \| \beta(\bm{x}) - 1 \|_{L^2}
\end{equation}
where $\gamma$ is a regularization parameter.
The second term, $\beta(\bm{x}) - 1$, prevents the corrected model from deviating too much from the baseline model. With such a regularization, the corrected model is constrained to explore only the \emph{vicinity} of the baseline solution, which greatly reduces the dimension of the search in the high-dimensional space of possible discrepancy fields $\beta$.  The inferred discrepancy field can be subsequently used to guide the improvement of the baseline model and to develop data-driven correction schemes. \citet{singh2016using} used velocities from DNS databases to infer the discrepancy field in plane channel flows at frictional Reynolds numbers ranging from $Re_\tau = 395$ to $4200$. The results are shown in Figure~\ref{fig:kdur}. It can be seen that the profiles of discrepancies $\beta$ for different Reynolds numbers are qualitatively similar. This \textit{a priori} study suggests that the knowledge gained in one flow can be extended to other flows of similar configurations where data are not available.  The end product is a data-driven correction function $\beta(\bm{q})$ for the baseline model obtained by posing the discrepancy term $\beta$ as a function of non-dimensionalized mean flow variables (e.g., $\mathbf{S}$ and $\bm{\Omega}$, both properly normalized with local quantities~\cite{ling2015evaluation}, as well as the ratio $\mathcal{P}_\omega/D_\omega$ between production and dissipation~\cite{singh2017machine-learning-augmented}). Choosing flow variables $\bm{q}$ rather than spatial coordinates $\bm{x}$ as the input of the regression enables generalization of the learned function in different flows, possibly at different spatial scales. \citet{singh2017machine-learning-augmented} showed predicted pressure coefficient of the S809 airfoil at $Re=2 \times 10^6$ by using the SA model augmented by the correction function, which was trained with the inferred discrepancy field by using the data from the S814 airfoil at Reynolds numbers $Re = 1 \times 10^6$ and $Re=2 \times 10^6$.

Although the correction scheme is applied on a few specific models ($k$--$\omega$ or SA model), generalization to additional models (e.g., $k$--$\varepsilon$ model or Reynolds stress transport model) is straightforward. On the other hand, since the corrected model is obtained by perturbing the transport equations in the baseline model, it is still constrained by the limitation of the latter.  For example, if a linear eddy viscosity model is chosen as baseline, the corrected model would still be limited by the Boussinesq assumption.

\subsection{Introducing uncertainties in turbulent viscosity}
\label{sec:nonpara-nut}

Most of the widely used turbulence models (e.g., $k$--$\varepsilon$, $k$--$\omega$, and SA models~\citep{spalart1992one-equation}) are linear eddy viscosity models, which model the Reynolds stress $\boldsymbol{\tau}$ in the form $\boldsymbol{\tau} - \frac{2k}{3} \mathbf{I} = 2 \nu_t \mathbf{S}$ as in Equation~\eqref{eq:evm-tau}, where $\nu_t$ is the turbulent eddy viscosity. It can thus be assumed  that the model uncertainty in RANS simulations can be attributed to the discrepancies in the predicted eddy viscosity field and subsequently introduce uncertainties thereon. For flows with ground truth of mean velocities (e.g., $\mathbf{U}^{\text{dns}}$ from DNS data), one can define an optimal eddy viscosity field $\nu_t^{\text{opt}}$ that minimizes the discrepancy between the computed velocity $u$ and the ground truth velocity. Finding the optimal viscosity $\nu_t^{\text{opt}}$ amounts to solving the following optimization problem~\cite{dow2011quantification}:
\begin{equation}
\label{eq:nu-opt}
  \nu_t^{\text{opt}} = \arg \min_{\nu_t} J \; ,  \qquad \text{with} \quad J = \| \mathbf{U}(\nu_t) - \mathbf{U}^{\text{dns}}  \|_{L^2}
\end{equation}
where $\mathbf{U}(\nu_t)$ indicates the dependence of the velocity field on the eddy viscosity field through the RANS equations. The optimization is further constrained by the positivity and smoothness of $\nu_t$, which can be built into the cost function or enforced in the optimization procedure.  For example, Dow and Wang~\cite{dow2011quantification} used the following cost function:
\begin{equation}
 \label{eq:nu-opt-mod}
 J = \| \mathbf{U}(\nu_t) - \mathbf{U}^{\text{dns}}  \|_{L^2} + \gamma \| \nabla \nu_t \|_{L^2}
\end{equation}
where a regularization term with $\nabla \nu_t$ is incorporated to promote smoothness of the viscosity field with parameter $\gamma$ controlling the desired smoothness.  The optimization problem can be solved with gradient descent methods, where the gradient ${\partial J}/{\partial \nu_t}$ of the cost function with respect to the control variable $\nu_t(\bm{x})$ can be obtained very efficiently by using adjoint methods. Alternatively, the optimization problem in Equation~\eqref{eq:nu-opt} can also be solved by the iterative ensemble Kalman method~\cite{iglesias2013ensemble-kalman}, which can be considered a derivative-free optimization that uses the state covariance, estimated from Monte Carlo samples, instead of the Jacobian. The iterative Ensemble Kalman method has been used to infer the Reynolds stresses discrepancies by using sparse observation data of velocities~\cite{xiao2016quantifying} (see Section~\ref{sec:nonpara-tau}).

The viscosity obtained by using optimization methods can be potentially used in two ways for flows whose configurations are similar to that from which data is available:
\begin{enumerate}[(1)]
\item to \emph{improve} predictions and \emph{reduce} uncertainties, or
\item to \emph{quantify} uncertainties by building statistical models for the discrepancies in the RANS-modeled eddy viscosity.
\end{enumerate}
 The first approach would involve building a functional mapping from the mean flow field to the eddy viscosity or its discrepancies. However, as of the writing of this review, the authors are not aware of any published research pursuing this approach. A machine-learning based approach to predict discrepancies of RANS-modeled Reynolds stresses has been investigated~\citep{wang2017physics-informed, wu2018data-driven}(see Section~\ref{sec:nonpara-tau}), and one can envision a similar approach to be used on the eddy viscosity. On the other hand, the second approach has been pursued by~\citet{dow2011quantification}, which is detailed below.

\begin{figure}[!htbp]
  \centering
    \subfloat[Samples of eddy viscosity]{\includegraphics[width=0.47\textwidth,height=0.36\textwidth]
    {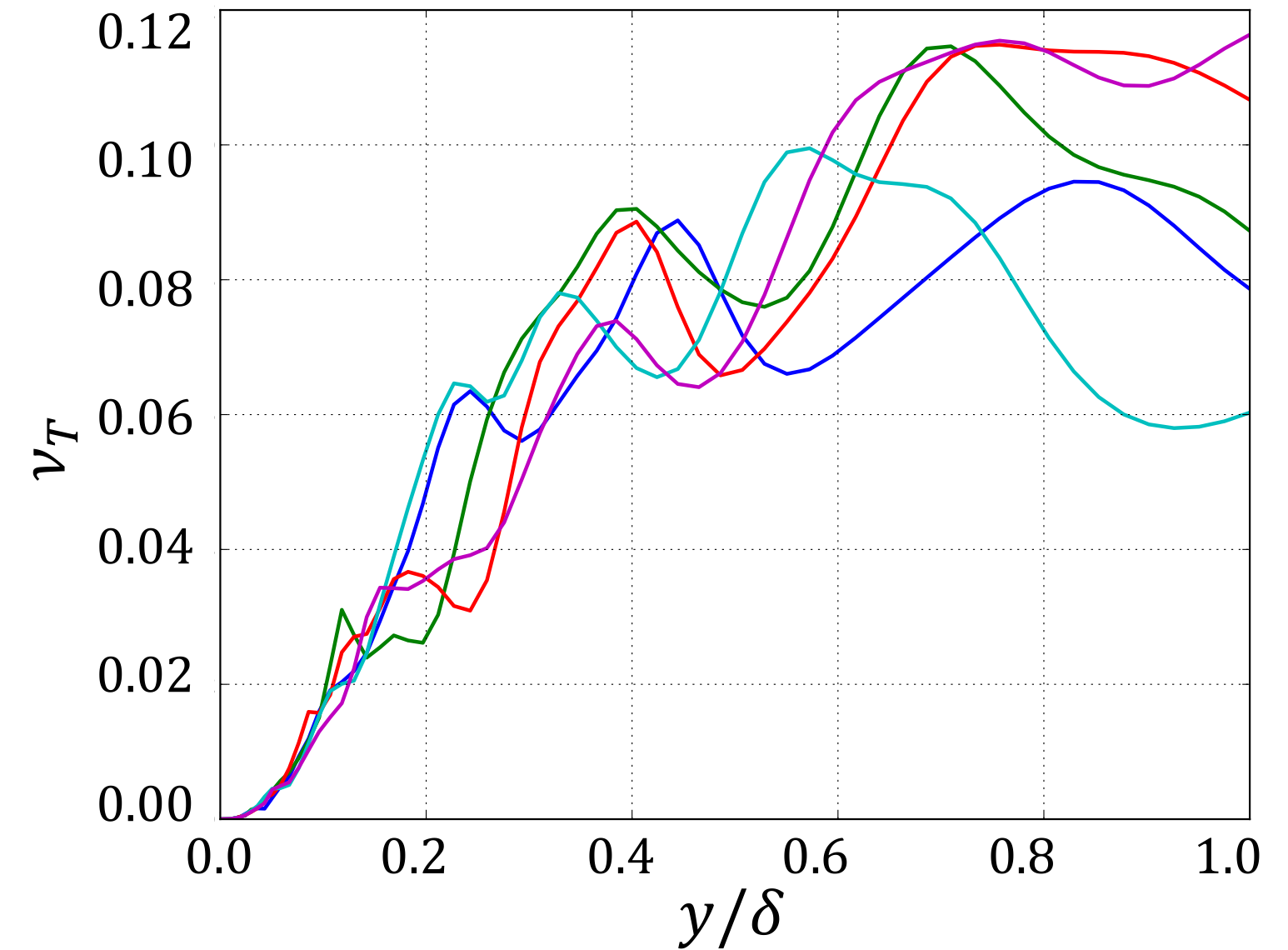}}
    \subfloat[Velocities obtained from $\nu_t$]
    {\includegraphics[width=0.47\textwidth,height=0.36\textwidth]{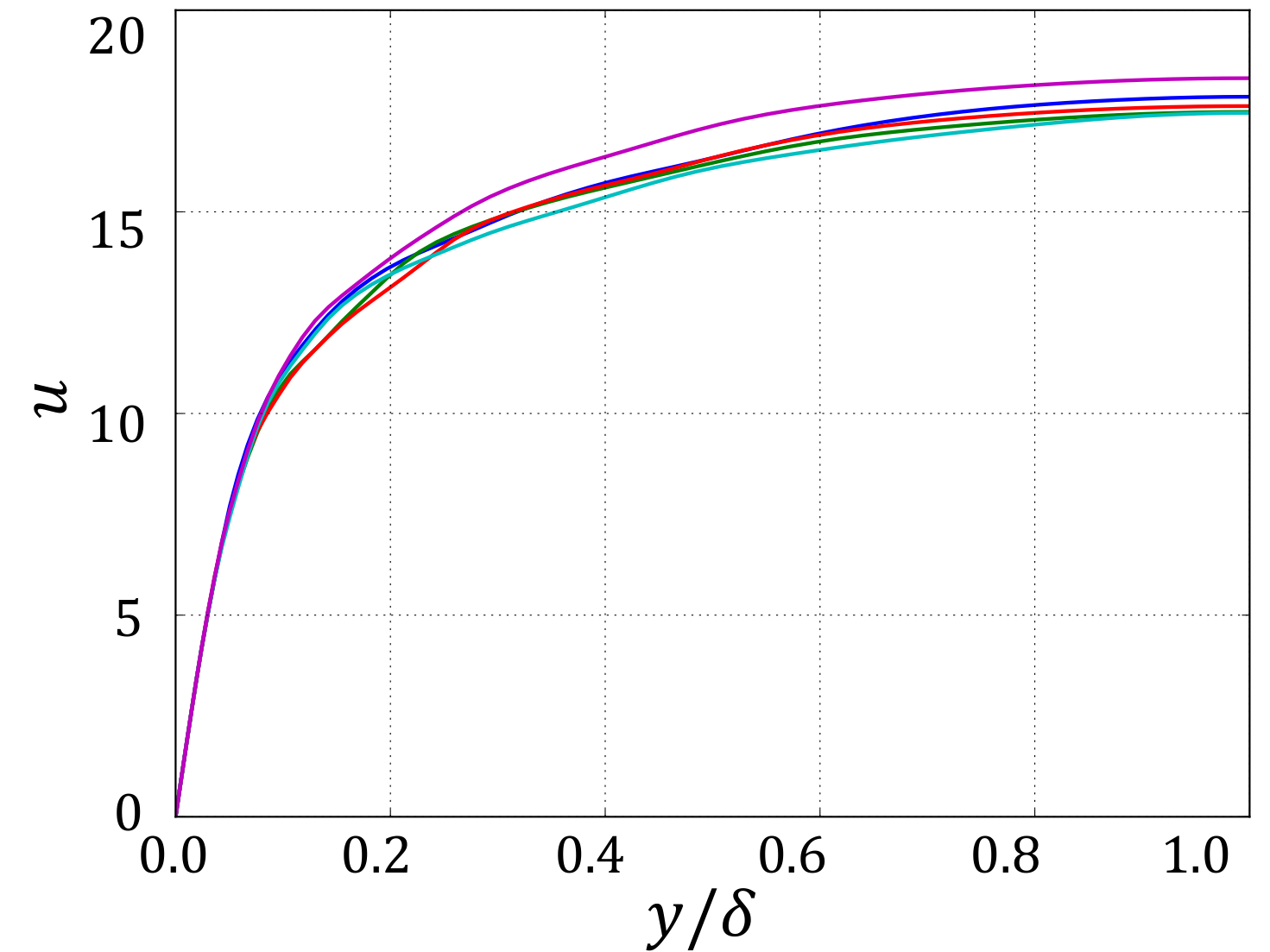}}\\
    \caption{(a) Realizations of true eddy viscosity field with samples drawn from the Gaussian processes for the discrepancy for plane channel flow at frictional Reynolds number $Re_\tau = 180$. The logarithmic discrepancy $ \delta_{\log \nu} \equiv \log(\nu_t^{\text{opt}}/\nu_t^{\text{rans}})$ of the $k$--$\omega$ model is inferred by minimizing velocity discrepancies with the DNS data. (b) Velocities propagated from the sampled eddy viscosity, indicating the uncertainties in the predicted velocities.  Horizontal axis is the wall-normal distance normalized by half channel width $\delta$. Figure reproduced with permission from \citet{dow2011quantification}.}
  \label{fig:qiqi}
\end{figure}

 Specifically, ~\citet{dow2011quantification} first used DNS data from plane channel flows to infer an optimal eddy viscosity field $\nu_t^{\text{opt}}$.  They further constructed a zero-mean Gaussian process for the logarithmic discrepancy $\delta_{\log \nu} = \log(\nu_t/\nu_t^{\text{rans}}) $. Equivalently, the field of true eddy viscosity $\nu_t(\bm{x})$ is modeled as a random field as follows:
\begin{equation}
 \label{eq:nu-discrepancy}
\log \nu_t  = \log \nu_t^{\text{rans}}  + \delta_{\log \nu}
\qquad \text{with} \qquad \delta_{\log \nu}    \sim \mathcal{GP}(0, K(\bm{x}, \bm{x}'))
\end{equation}
where the covariance kernel $K$ was chosen as a squared exponential function with its hyperparameters including variance $\sigma$ and length scale $l$ (see Equation~\ref{eq:sq-exp}) determined by using maximum likelihood estimation by using the inferred optimal eddy viscosity field as data.  After the hyperparameters were determined, they sampled the Gaussian processes to obtain realizations of possible eddy viscosity fields (Figure~\ref{fig:qiqi}a) in similar yet slightly different geometries, e.g., plane channel with wavy walls.  This slight extrapolation is based on the assumption that the eddy viscosity discrepancies $\delta_{\log \nu}(\bm{x})$ in a class of similar flows conform to the same statistical model. Such realizations of the eddy viscosity obtained from the Gaussian process were used to solve the RANS equations and to obtain an ensemble of velocity predictions as shown in Figure~\ref{fig:qiqi}b.  The obtained ensemble represents the uncertainties in RANS-predicted velocities, which can be further processed to obtain uncertainties for other quantities of interests. This methodology has recently been extended to more complex flows in a U-bend channel~\citep{hayek2018adjoint-based}.  As with the UQ approach based on transport equations~\cite{singh2016using}, all the predictions in the ensemble are still constrained by the Boussinesq assumption originating from the baseline model.

\subsection{Introducing uncertainties in Reynolds stresses}
\label{sec:nonpara-tau}

Reynolds stress plays a unique and particularly important role in RANS modeling -- it is the term through which most turbulence models enter the RANS momentum equations as can be seen in Figure~\ref{fig:hierarchy}.  In the derivation of the RANS equations there is a closure problem. That is, the Reynolds stress term in the obtained averaged-equations needs to be modeled. While a Reynolds stress transport equation (RSTE) can be derived from the NS equations, the RSTE itself contains even more unclosed terms. On the other hand, if the true Reynolds stress field is supplied to the RANS equation, theoretically the true velocity and all other quantities can also be obtained, provided that the numerical uncertainties are negligible and that the RANS equations are well-conditioned.  
As the Reynolds stress is the only modeled term in the RANS equations, inaccuracy in its modeling is the source of model-form uncertainty for RANS simulations, at least for single phase, fully turbulent flows without transition~\cite{pope2000turbulent}.  Transition modeling is an important topic~\cite{zhang2018efficient} but it is beyond the scope of this review. The insight on the importance of Reynolds stress in turbulence modeling was abstracted as composite model theory~\cite{oliver2015validating}, which is detailed in \ref{app:composite}.

Based on the observations above, it is natural to introduce uncertainties to the Reynolds stresses. So far, two distinct approaches have been proposed to characterize the uncertainties in the Reynolds stresses:
\begin{itemize}
  \item formulating a stochastic differential equation (SDE) for the Reynolds stress discrepancy tensor~$\delta_\tau$ driven by a Wiener process (random walk forcing model)~\citep{oliver2013representing}, and
  \item using realizability constraints to guide the perturbations of single-point Reynolds stresses.
  \end{itemize}
  In both approaches the Reynolds stress discrepancy is considered a random tensor field characterized by physical constraints (e.g., conservation laws or realizability). Both approaches are introduced below.

\subsubsection{Stochastic differential equation of Reynolds stress discrepancy}
In the first approach, several forms of SDEs were explored for the Reynolds discrepancy field~$ \delta_\tau$ in a plane channel flow, a typical one of which reads as follows~\citep{oliver2013representing}:
\begin{equation}
  - C_{pr} \delta_\tau \frac{d {U}}{d x_2}
  - \frac{d}{d x_2}\left[
    (\nu + C_\nu \nu_t) \frac{d \delta_\tau}{d x_2} \right]
    =
    C_\sigma (\nu + C_\nu \nu_t)^{5/4} \left( \frac{d U }{d x_2} \right)^{7/4} \frac{dW}{d x_2}
  \end{equation}
  with the three terms indicating production, diffusion, and residual; $x_2$ is the wall-normal coordinate and $ U$ is the horizontal velocity; $W$ indicates a Wiener process; $C_{pr}$, $C_\nu$, and $C_\sigma$ are coefficients to be calibrated from data. The SDE has a form that is similar to, but simpler than, the Reynolds stress transport equations. Specifically, the SDE shares the same convection-diffusion-production form\footnote{The convection term disappears in mean equations of the plane channel flow.} as the RSTE, but the SDE has a stochastic residual term on the right hand side of the SDE in place of the unclosed terms (e.g., triple correlation and pressure--rate-of-strain) in the exact RSTE. The solution to the SDE provides an indication of the uncertainties in the Reynolds stresses, which can be propagated to the velocities and other quantities of interests.  The SDE-based approach yields uncertainties for the entire field~$\delta_\tau(\bm{x})$, which is in contrast to the single-point realizability constraints examined in Section~\ref{sec:realizability}. If one considers the discrepancy $\delta_\tau(\bm{x})$ a tensorial random field, the cross-component and spatial correlations are both accounted for through the SDE.  Unfortunately, the construction of the SDE heavily relies on physical insights and modeling heuristics. Consequently, it is not straightforward to extend the formulation above to more complex flows beyond plane channel flows.

  \subsubsection{Estimating uncertainty bounds guided by realizability maps}
  \label{sec:realizability}
  In the second approach, perturbations are introduced directly to the modeled Reynolds stresses, based on which uncertainty propagation and statistical inferences are performed. A common starting point of these methods is the following decomposition of the Reynolds stress tensor:
\begin{equation}
  \label{eq:tau-decomp}
   -\boldsymbol{\tau} = 2 k \left( \frac{1}{3} \mathbf{I} +  \mathbf{a} \right)
  = 2 k \left( \frac{1}{3} \mathbf{I} + \mathbf{V} \Lambda \mathbf{V}^\top \right)
\end{equation}
where $k$ is the turbulent kinetic energy, which indicates the \emph{magnitude} of $ \boldsymbol{\tau}$; $\mathbf{I}$ is the second-order identity tensor; $\mathbf{a}$ is the anisotropy tensor; $\mathbf{V} = [\mathbf{v}_1, \mathbf{v}_2, \mathbf{v}_3]$ and $\Lambda = \textrm{diag}[{\lambda}_1, {\lambda}_2, {\lambda}_3]$ where ${\lambda}_1+ {\lambda}_2 + {\lambda}_3=0$ are the orthonormal eigenvectors and eigenvalues of $\mathbf{a}$, respectively, indicating the \emph{shape} (aspect ratio) and \emph{orientation} of $\bm{\tau}$, if the latter is visualized as an ellipsoid~\cite{simonsen2005turbulent}.

Transformation of the eigenvalues leads to invariants that can be mapped to the well-known Lumley triangle~\citep{lumley1978computational} or the recently proposed barycentric triangle~\citep{banerjee2007presentation}, both of which provide a map for all realizable states of turbulence.  Any realizable turbulence state can be mapped to a point within or on the edge of the triangles after the respective transformations.  In the case of the barycentric map, the following linear transformation from eigenvalues $(\lambda_1, \lambda_2, \lambda_3)$ of the anisotropy $\mathbf{a}$ to the barycentric coordinates $(c_1, c_2, c_3)$ is adopted:
\begin{subequations}
  \label{eq:lambda2c}
\begin{align}
  c_1 & = \lambda_1 - \lambda_2 \\
  c_2 & = 2(\lambda_2 - \lambda_3) \\
  c_3 & = 3 \lambda_3 + 1 \ .
\end{align}
\end{subequations}
The barycentric triangle and the mapping above are similar to the Lumley triangle but overcomes several shortcomings of the latter, including (i) the tendency to cluster towards the isotropic state and (ii) the nonlinearity in the mapping from the eigenvalues to tensor invariants. Like the Lumley triangle, the barycentric triangle has clear physical interpretation in that it indicates the \emph{componentality} of the turbulence~\cite{emory2014componentality-based,emory2014estimating}.  For example, the upper corner ($c_3 = 1$) corresponds to  three-component isotropic turbulence while the lower left corner ($c_2 = 1$) corresponds to two-component axisymmetric turbulence, which occurs in flows close to a solid wall (e.g., point  $P_2$ in Figure~\ref{fig:wing-body-schematic}).

\begin{figure}[!htbp]
  \centering
   \subfloat[Perturbation in Barycentric coordinates] {\includegraphics[width=0.42\textwidth]{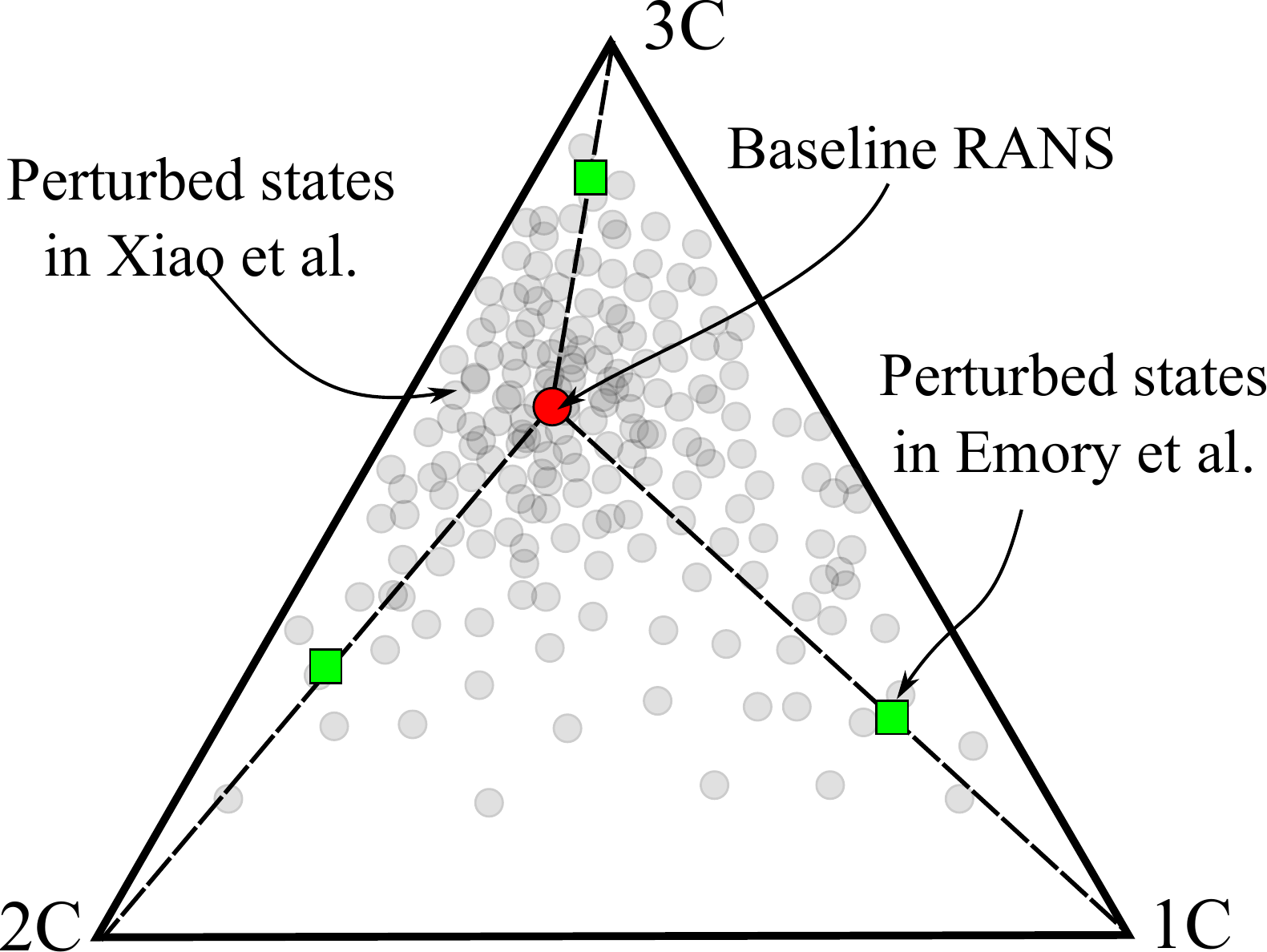}}
   \subfloat[Perturbation with random matrix sampling] {\includegraphics[width=0.43\textwidth]{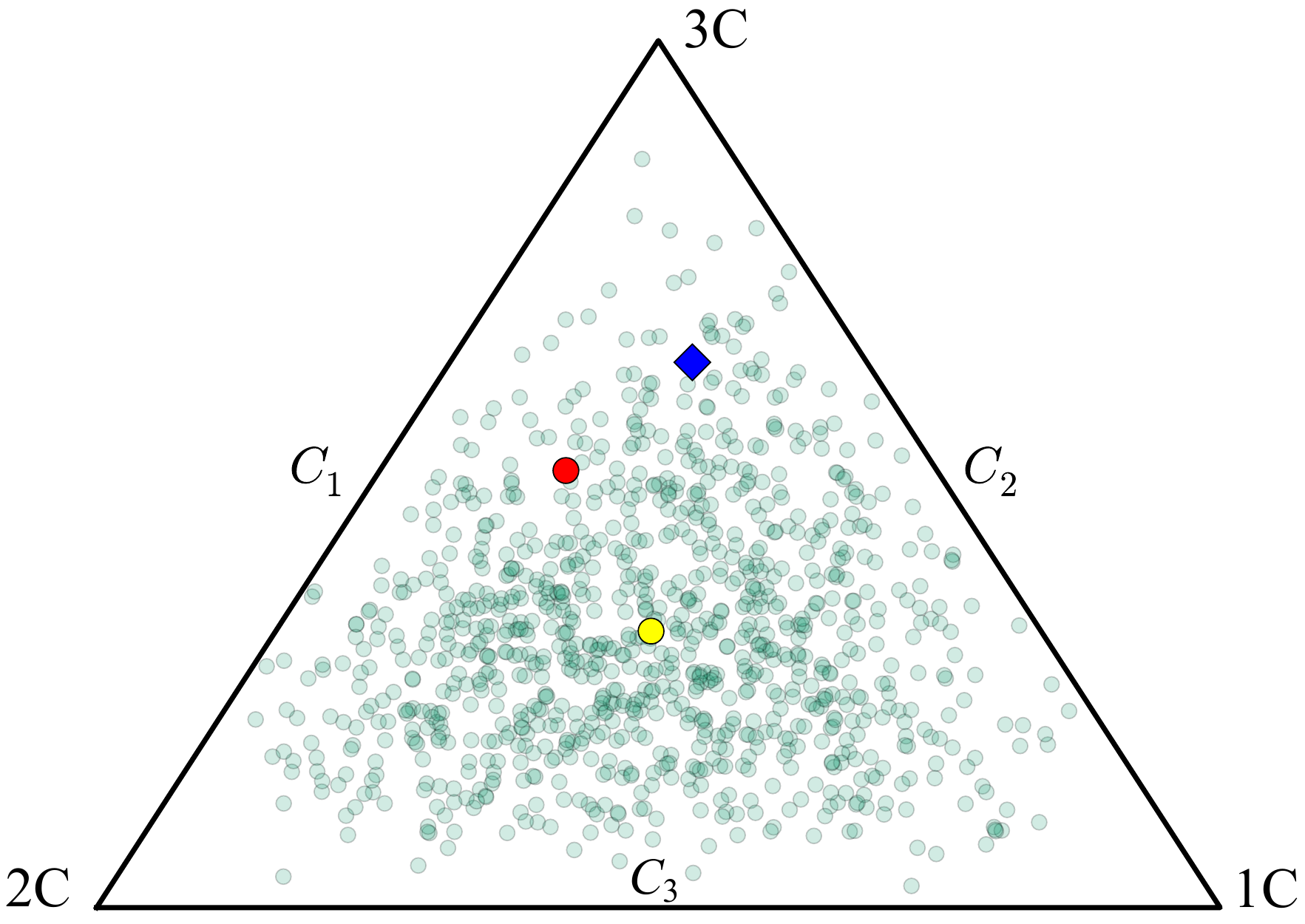}}
   \caption[]{Model-form uncertainty quantification through perturbation of Reynolds stresses within the physically realizable limit enclosed by the Barycentric triangle.   (a) Comparison between the perturbation schemes of \citet{emory2013modeling} and \citet{xiao2016quantifying}.  (b) Perturbation scheme based on random matrix theory~\cite{xiao2017random}, which is compared with the physics-based perturbations~\citet{xiao2016quantifying} shown in (a).
Legend in Panel (b):
baseline RANS prediction \tikz\draw[black,fill=red] (0,0) circle (.7ex); ;
DNS data \tikz\draw[black,fill=blue,rotate around={45:(0,0)}] (0,0) rectangle (1.1ex, 1.1ex);  ;
sample mean \tikz\draw[black,fill=yellow] (0,0) circle (.7ex);   ;
 perturbed states (samples) \tikz\draw[black,fill=lightgreen] (0,0) circle (.7ex); . 
  Figure reproduced with permission from \citet{xiao2017random}.}
  \label{fig:emory-xiao}
\end{figure}

 The realizability requirements on Reynolds stresses have been studies extensively in the early years of turbulence model development.  Efforts from \citet{schumann1977realizability}, \citet{lumley1978computational}, and \citet{pope1985pdf}, among others, have led to a class of realizable Reynolds stress models~\cite{speziale1994on}.  However, in the context of quantifying model-form uncertainties in RANS simulations, \citet{emory2011modeling,emory2013modeling} pioneered the use of realizability maps to guide the exploration of Reynolds stress uncertainties. They proposed introducing separate perturbations to $k$, $\Lambda$, and $\mathbf{V}$ resulting from the decomposition above to obtain a few representative limiting states:
\begin{equation}
  \label{eq:tau-decomp-star}
   -\boldsymbol{\tau}^\star
  = 2 k^\star \left( \frac{1}{3} \mathbf{I} + \mathbf{V}^\star \Lambda^\star \mathbf{V^\star}^\top \right)
\end{equation}
where $\star$ indicates perturbed states from the RANS-predicted baseline, e.g., $ \Lambda^\star = \Lambda^{\text{rans}} + \delta \Lambda$. The initial focus was placed on the eigenvalues, as the realizability map provides a straightforward and rigorous bound on how they can be perturbed. One possibility of perturbing the anisotropy is to perturb it towards one-component (1C), two-component (2C), and three-component (3C) limiting states of realizable turbulence, represented by the three corresponding vertices of the barycentric triangle (see Figure~\ref{fig:emory-xiao}a).

Nevertheless, the realizability map does not provide a direct bound on the magnitude $k$ and the eigenvectors $\mathbf{V}$.  In order to utilize the realizability map to bound $k$ and $\mathbf{V}$, it is important to recognize that $k$, $\Lambda$, and $\mathbf{V}$ are not independent but intimately coupled.  They are different characteristics of the same Reynolds stress tensor, which is governed by a coupled Reynolds stress transport equation (RSTE). In fact, with some algebra the RSTE can be transformed to three individual transport equations for the turbulent kinetic energy(TKE) $k$, eigenvalues $\Lambda$, and eigenvectors $\mathbf{V}$ as well as their discrepancies~\cite{pope2000turbulent,thompson2016strategy}, although only the TKE transport equation~\eqref{eq:k} is commonly used in turbulence modeling.  The coupling among the three variables can be utilized in many ways. For example, the anisotropy bounds obtained from the realizability map~\cite{emory2013modeling} can be used to estimate the bounds on the TKE production $\mathcal{P}_k = \bm{\tau} \cddots \mathbf{S}$, which is further substituted into transport equation (\ref{eq:k}) to obtain the TKE corresponding to the limiting states~\cite{gorle2012epistemic}. The obtained TKE fields can be used to estimate their uncertainties. Similarly, \citet{thompson2016strategy} exploited the coupling to estimate the uncertainties in the eigenvectors by using the realizability bounds of the eigenvalues.  Unfortunately, a large number of unclosed terms in the Reynolds stress transport equation makes it much more difficult than estimating the uncertainties in the TKE.  Recently, the eigenvectors perturbation has also been investigated~\citep{iaccarino2017eigenspace,mishra2017uncertainty}. Two extreme bounding cases of the perturbation for Reynolds stress eigenvectors are considered. In one case the semi-major axis of the Reynolds stress ellipsoid is aligned with the stretching eigen-direction of the mean rate of strain tensor; in another case the semi-major axis of the Reynolds stress ellipsoid is aligned with the compressive eigen-direction of the mean rate-of-strain tensor. The two limiting states are chosen to explore the possible extreme scenarios of turbulent production.

In summary, the above-mentioned studies by Iaccarino and co-workers~\cite{emory2011modeling,emory2013modeling,gorle2012epistemic,thompson2016strategy,iaccarino2017eigenspace} used barycentric triangle as guide to comprehensively explore the limiting states of Reynolds stresses.  They form an efficient, physics-based scheme to estimate RANS model uncertainty by using only five simulations. Moreover, the parameterization scheme of Reynolds stress perturbations becomes the foundation of more sophisticated methods that use statistical inference and machine learning to quantify and reduce the RANS model uncertainties~\cite{xiao2016quantifying,wang2017physics-informed,wu2018data-driven}.

\subsubsection{Systematic sampling of Reynolds stress uncertainty within realizability constraints}

In statistical inference for quantifying and reducing model-form uncertainties, it is insufficient to merely perturb Reynolds stresses towards limiting states~\cite{xiao2016quantifying}). Rather, statistical sampling and inference requires a systematic scheme for parameterizing the perturbations to the TKE, the eigenvalues, and particularly the eigenvectors. Perturbations on $k$ and $\Lambda$ can be represented as random fields, albeit with realizability constraints. To this end, the uncertainties associated with RANS modeled TKE $k^{\text{rans}}$ can be represented in terms of its logarithmic discrepancy, in a similar way to that for the eddy viscosity~\cite{wang2016quantification} in Equation~\eqref{eq:nu-discrepancy}. However, parameterizing the perturbations on the eigenvectors is more challenging due to the need to maintain their \emph{orthonormal} property, which is necessary to ensure that the perturbed Reynolds stresses remain symmetric positive semidefinite tensors. To this end, it is most convenient to represent the perturbation from $\mathbf{V}$ to $\mathbf{V}^\star$ as a rigid-body rotation, i.e., $\mathbf{V}^\star = Q^\delta \mathbf{V}$ with $Q^\delta$ being an orthonormal rotation matrix representing the perturbation. In fact, a rotation can be represented more compactly by using a set of Euler angles ($\varphi_1, \varphi_2, \varphi_3$). That is, any rigid-body rotation in a three-dimensional space (with a few rare exceptions) can be achieved by the following three consecutive intrinsic rotations about the axes of the local coordinate system ($x$--$y$--$z$) of the rigid body~\citep{goldstein1980euler}: (i) a rotation about the $z$ axis by angle $\varphi_1$, (ii) a rotation about the $x$ axis by $\varphi_2$, followed by (iii) another rotation about its $z$ axis by $\varphi_3$.  The Euler-angle based representation has been used for quantifying RANS model-form uncertainties~\citep{wang2016quantification}. Alternatively, the same transformation can be represented as a unit quaternion.  Euler's rotation theorem states there exists a unique axis of unit vector $\mathbf{n} \equiv [n_1, n_2, n_3]$ passing through the origin and an angle $\vartheta$ such that $\mathbf{V}^\star$ can be obtained via rotating $\mathbf{V}$ by $\vartheta$ about an axis $\mathbf{n}$, and thus the rigid-body rotation can be represented by a unit quaternion~\cite{horn1987closed-form}:
\begin{equation}
  \label{eq:quaternion}
  \mathbf{h} = \left[\cos{\frac{\vartheta}{2}}, \; n_{1}\sin{\frac{\vartheta}{2}}, \; n_{2}\sin{\frac{\vartheta}{2}}, \;  n_{3}\sin{\frac{\vartheta}{2}}\right]^\top
\end{equation}
where $\|\mathbf{h}\|$ = 1.  In uncertainty quantification and machine learning for RANS modeling, the two representations of Reynolds stress perturbation based on Euler angle and unit quaternion have been compared, and the latter was found to be superior~\citep{wu2017representation}.

\subsubsection{Random matrix approach for quantifying Reynolds stress uncertainty}

The realizability constraint of Reynolds stresses plays a critical role in all the RANS model-form uncertainty quantification methods outlined above. However, physics-based decomposition as in Equation~(\ref{eq:tau-decomp}) is only one of the possible ways to guarantee realizability.  \citet{xiao2017random} proposed an alternative approach where the Reynolds stress tensor $\boldsymbol{\tau}$ is modeled as a $3 \times 3$ random matrix that conforms to a maximum entropy distribution defined on the set of positive semi-definite matrices.  Reynolds stress uncertainty can thus be estimated by directly sampling from the defined distribution, with the realizability of all samples guaranteed without using the realizability maps.  The validity of the random matrix approach can be clearly seen from the equivalence among the following three interpretations of the Reynolds stress realizability.  That is, a Reynolds stress tensor is physically realizable if and only if it satisfies one of the following conditions:
\begin{enumerate}[(1)]
\item it is the covariance matrix of a real-valued vector (i.e., the velocity),
\item it resides within or on the edge of the barycentric triangle (or Lumley triangle) after transformations (e.g., Equation~(\ref{eq:lambda2c}) for the former), or
\item it is symmetric and positive semi-definite.
\end{enumerate}
The three conditions are, respectively, (i) the origin of the realizability constraint,  (ii) the foundation for the physics-based approach, and (iii) the basis for the random matrix approach.  All three conditions above are equivalent~\cite{xiao2017random}.

The random matrix approach and the physics-based approach are compared in detail in \citet{wang2016quantification}. It was concluded that both approaches yield qualitatively similar results, particularly when the perturbations are small and far away from the limiting states (edges of the barycentric triangle). When the perturbations are large, physics-based perturbations may result in samples falling outside the barycentric triangle, which must be capped and result in a slightly distorted distribution. In contrast, the random matrix approach does not have this issue. Another important difference is that the physics-based approach perturbs the three components (magnitude, shape, and orientation) separately, while the random matrix approach perturbs all three components simultaneously, with $k$, $\Lambda$, and $\mathbf{V}$ implicitly constrained by the maximum entropy principle.

\subsubsection{Quantifying and reducing Reynolds stress uncertainties with data}

\begin{figure}[!htbp]
\centering
\includegraphics[width=0.5\textwidth]{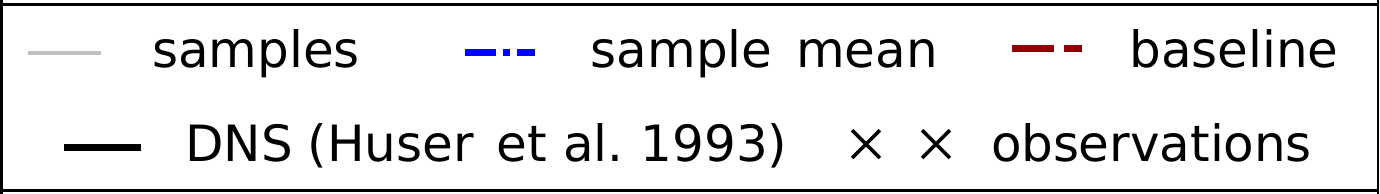}\\
\subfloat[Prior ensemble]{\includegraphics[width=0.45\textwidth]{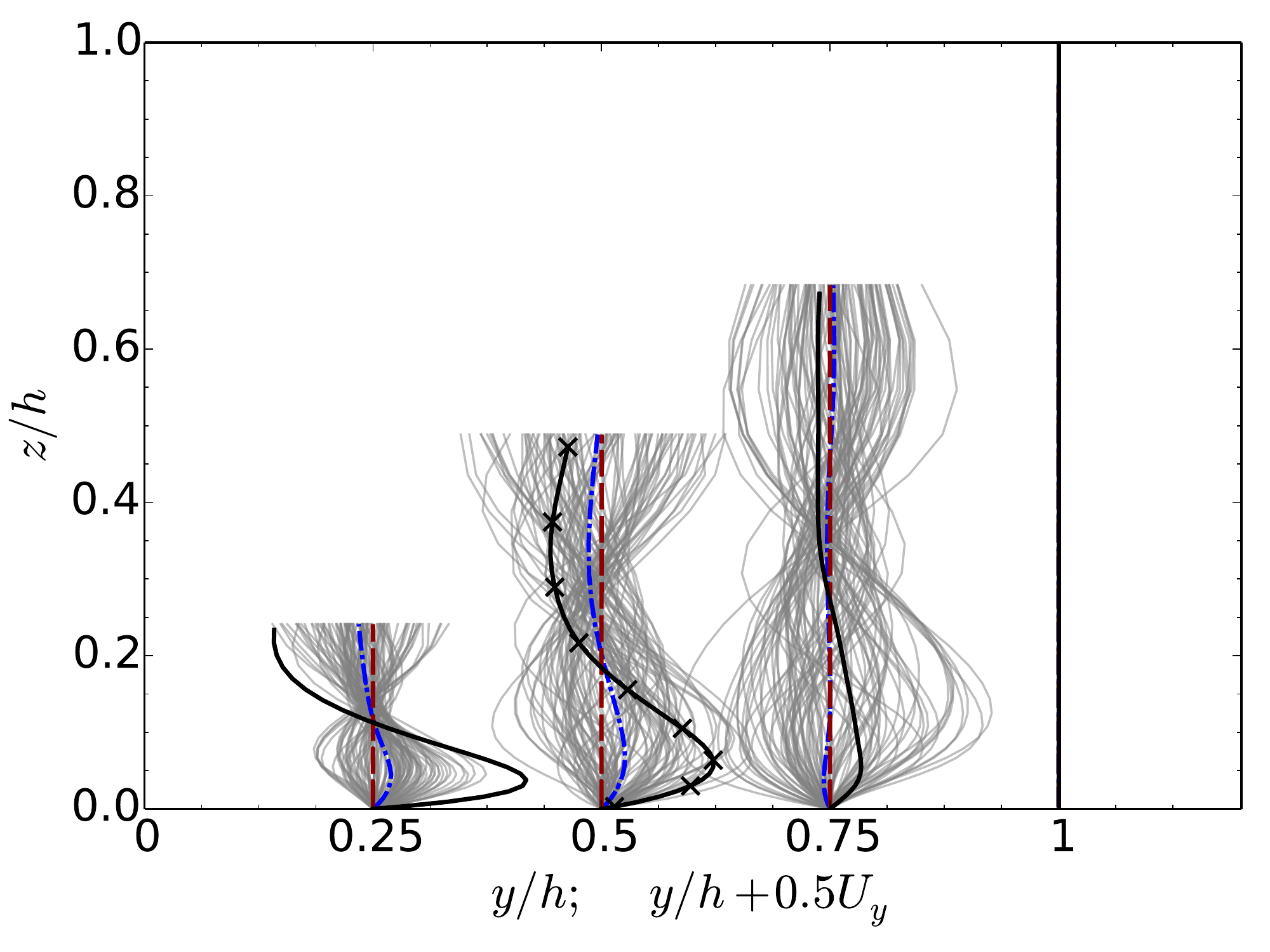}}\hspace{0.05em}
\subfloat[Posterior ensemble]{\includegraphics[width=0.45\textwidth]{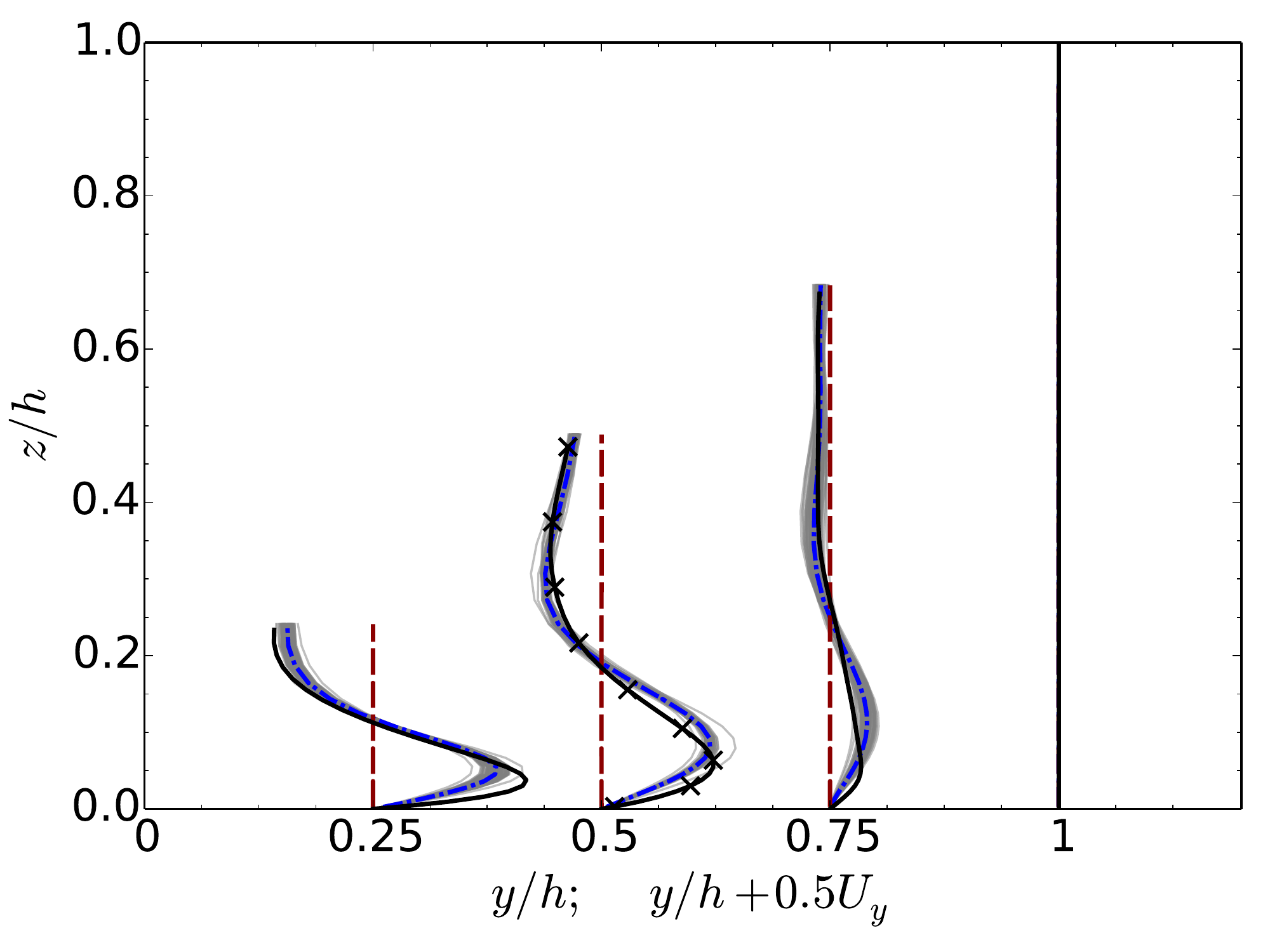}}\hspace{0.05em}
\caption{Inference of full-field in-plane mean velocity of the fully developed turbulent flow in a square duct, showing the lower left quadrant. (a) Prior velocity ensemble and (b) posterior velocity ensemble at four spanwise locations with comparison to baseline and benchmark results.  The velocity profiles in the prior ensemble are scaled by a factor of 0.3 for clarity. The upper half of the domain is omitted due to diagonal symmetry. Figure reprinted with permission from \citet{xiao2016quantifying}.}
\label{fig:duct}
\end{figure}

The works reviewed above all involved forward analysis, i.e., propagation of uncertainties introduced in the Reynolds stresses to velocities or derived quantities of interest. As with the parametric approaches and other non-parametric approaches introducing uncertainties in viscosity (Section~\ref{sec:nonpara-nut}) and turbulent transport equations (Section~\ref{sec:nonpara-beta}), backward analysis (statistical inference) can also be performed on Reynolds stresses to quantify and reduce uncertainties in RANS model predictions.  The objective is to find a Reynolds stress field that yields the best agreement with the data (e.g., sparse observations of velocities) accounting for the state covariance and the error covariance. Both variational methods and ensemble methods introduced in Section~\ref{sec:prob-theory-da} can be used. \citet{xiao2016quantifying} used the ensemble Kalman method to infer the Reynolds stress and full-field velocities from sparse velocity data. Constraints and empirical prior knowledge about the Reynolds stress field and its discrepancies are built into the inference, specifically including:
\begin{enumerate}[(1)]
\item realizability at any point,
\item smoothness of the Reynolds stress field and its discrepancy for incompressible flows, and
\item empirical knowledge on the regions where Reynolds stress discrepancies are large.
\end{enumerate}
Utilizing these constraints and prior knowledge greatly reduces the dimension of the inverse problem, which has the same effects as the regularization terms in the full-field inversion and optimization problems as in Equations~\eqref{eq:beta-opt} and \eqref{eq:nu-opt-mod}. The realizability is ensured by parameterizing the Reynolds stress in terms of the physics-based decomposition as in Equation~\eqref{eq:tau-decomp}. The smoothness is ensured by representing the random fields in terms of its Karhunen--Loeve expansion, truncated to the first $n$ terms:
\begin{equation}
  \label{eq:KL}
  \bm{\tau}(\bm{x}) = \sum_{\alpha=1}^{n} w_\alpha  \phi_\alpha(\bm{x})
\end{equation}
where $w_\alpha$ are random variables to be inferred, and $ \{ \phi_\alpha(\bm{x}) \}_{\alpha=1}^n$ are a set of orthogonal basis functions corresponding to the covariance kernel of the random field $\bm{\tau}$. The basis functions can be computed from the Fredholm integral equation by solving an eigenvalue problem for the kernel and embody the empirical knowledge on Reynolds stress discrepancy as mentioned above. Figure~\ref{fig:duct} shows representative results from a fully developed square duct flow, presenting the in-plane velocities before and after incorporating the data, i.e., prior and posterior of velocity distributions. It clearly shows that the velocity predictions at all cross-sections are markedly improved, even in locations where velocity observations are not available. The calibrated discrepancy can even be used to correct square duct flows at a higher Reynolds number or flows in different geometries such as a rectangular duct~\cite{wu2016bayesian}.  Furthermore, it was demonstrated that incorporating empirical knowledge is clearly valuable and has similar effects as increasing the amount of observation data~\cite{wang2016incorporating}.

Among the prior knowledge used for the statistical inference, the regions where Reynolds stress discrepancies are large have primarily depended on input from users based on their empirical knowledge. However, the combination of physical and modeling insights with modern data science has opened new opportunities.  \citet{gorle2014deviation} proposed an analytical marker function based on the deviation from parallel shear flow and used it to predict discrepancies in RANS-modeled Reynolds stress. Their ideas are based on the insightful observation that commonly used eddy viscosity models were developed and tuned for parallel shear flows (boundary layers). A departure from such flows typically leads to violations of assumptions in these models. Moreover, emerging machine learning techniques have made it possible to provide more accurate maps of where large discrepancies exist.  \citet{ling2015evaluation} developed a machine learning method to evaluate potential inadequacy of RANS models by using DNS databases.  This approach has been recently applied to more complex flows (e.g., jet in crossflow~\cite{ling2017uncertainty}). The results include several fields of binary labels (whether the specified model assumption is violated), which could be further processed to obtain a variance of Reynolds stress discrepancy that can be incorporated into the covariance kernel field.

The same decomposition scheme in Equation~(\ref{eq:tau-decomp}), which has been used for Reynolds-stress-based uncertainty quantification and statistical inferences, can be used as a parameterization scheme for correcting RANS-predicted Reynolds stresses by using machine learning and training data.  \citet{tracey2013application} represented discrepancies in barycentric coordinates as a function of local mean-flow variables and leveraged machine learning to train the function.  \citet{wang2017physics-informed} and \citet{wu2018data-driven} developed a more systematic strategy to predict discrepancies in the magnitude, anisotropy, and orientation of the Reynolds stress tensor in terms of an invariant feature set for a set of tensor variables of the mean flow (e.g., $\mathbf{S}$, $\bm{\Omega}$, $\nabla p$, $\nabla k$), referred to as the integrity basis~\cite{ling2016machine}. They showed improved results in the prediction of Reynolds stresses and mean velocities~\citep{wu2018data-driven} for two canonical flows, i.e., separated flows over periodic hills and secondary flows in a square duct.  As an alternative, \citet{ling2016reynolds} proposed a neural network architecture with embedded invariance properties to learn and predict the coefficients of an objective formulation for the tensorial function $\bm{\tau} = f(\mathbf{S}, \bm{\Omega})$. These works illustrated how physical constraints can be embedded in machine learning. Nevertheless, a unique challenge for directly correcting or predicting the Reynolds stress tensors with data-driven models is the possible ill-conditioning of the RANS equations. For example, small errors in the machine-learning-predicted Reynolds stresses can lead to large errors in the propagated velocities~\cite{wu2018rans}. In order to overcome this difficulty, \citet{wu2018data-driven} proposed learning the linear and nonlinear parts of the Reynolds stress separately, with the linear part treated implicitly to improve model conditioning. Numerous other approaches have been proposed for augmenting and improving turbulence models based on machine learning~\cite{duraisamy2019turbulence}.
Finally, data-driven, machine-learning based methods have also been used in improving CFD models of thermal fluids flow with focus on boiling flows in nuclear reactor thermo hydraulics~\citep[e.g.,][]{liu2018data,chang2018reynolds,chang2018classification,hanna2017coarse} and in high-Mach number flows~\citep{wang2018predicting}. 

\subsection{Spatial correlations in Reynolds stress  discrepancy}

Most of the approaches for RANS uncertainty quantification reviewed above have focused on the uncertainty bounds of the Reynolds stress at a single point.  This is partly because the mathematical rigor of such a bound can only be maintained by the realizability constraint.  However, an equally important source of uncertainty comes from the spatial variation of the Reynolds stress discrepancy. After all, it is the \emph{divergence} of the Reynolds stress field that appears in the RANS momentum equations.  When quantifying uncertainties in RANS-based predictions, \citet{emory2013modeling} specified a spatial field for the eigenvalue perturbations based on the empirical understanding on the performance of the RANS model for the particular problem. \citet{xiao2016quantifying,xiao2017random} used the same argument to define a non-stationary Gaussian process model for the Reynolds stress discrepancies. They used Karhunen--Loeve (KL) expansions~\citep{le-matre2010spectral} to approximately represent the perturbation field with leading modes in KL expansion (see Equation~(\ref{eq:KL})).  Specification of such spatial distribution is probably the weakest link in the entire process of Reynolds stress-based uncertainty quantification.

Since the true Reynolds stress and the RANS modeled counterpart are described by their respective transport equations, the model discrepancies should conform to a transport equation of the same structure. Building upon such insight, \citet{edeling2017data-free} proposed a ``return-to-eddy-viscosity'' model, which is a transport equation with a source term describing the deviation of the turbulence state from equilibrium state assumed by the linear eddy viscosity models:
\begin{equation}
  \label{eq:trans-bary}
\frac{D c_1}{Dt}=a_{1c}\frac{\varepsilon}{k}\left(c_1^{\text{rans}} - c_1\right)+\frac{\partial}{\partial x_i}
\left[\left(\nu+\frac{\nu_t}{\sigma_{1c}}\right)\frac{\partial c_1}{\partial x_i}\right]\\
\end{equation}
where $c_1^{\text{rans}}$ is the barycentric coordinates corresponding to the baseline RANS modeled Reynolds stress tensor; $a_{1c}$ and $\sigma_{1c}$ are model coefficients to be calibrated. A similar PDE is formulated for $c_2$.  These heuristically justified, physics-inspired PDEs provide a bound for the Reynolds stress field.  Moreover, the model coefficients in the PDEs above can be calibrated by using data and Bayesian inference, and the calibrated equations are further used for predictions~\citep{edeling2017data-free}.

\begin{figure}[!htbp]
\centering
\subfloat[Modes from statistical covariance kernel]{\includegraphics[width=0.95\textwidth]{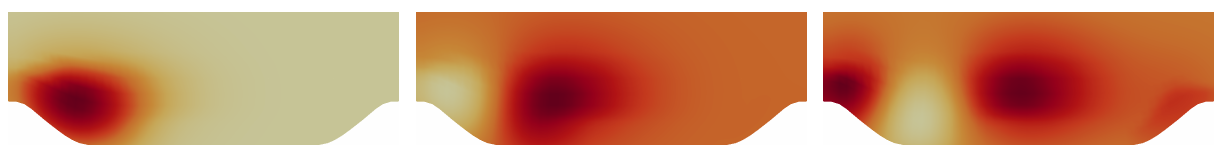}}\hspace{0.05em}
\subfloat[Modes from PDE-informed covariance kernel]{\includegraphics[width=0.95\textwidth]{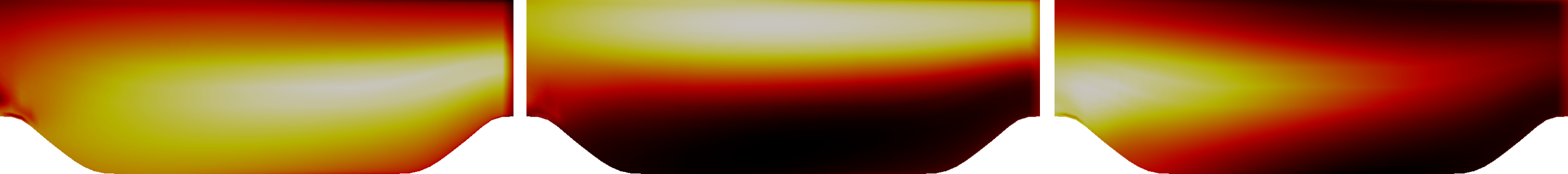}}
\caption{Comparison of modes (eigen-functions) obtained from a purely statistical kernel (squared exponential function) and a PDE-informed kernel as in Equation~(\ref{eq:prop-sigma}). Panels (a) and (b) reproduced from \citet{xiao2016quantifying} and ~\citet{wu2018pde-informed}, respectively.}
\label{fig:pde-cov}
\end{figure}

More recently, \citet{wu2018pde-informed} utilized the fundamental connection between PDEs and covariance to provide a physically anchored covariance structure, which has a clear advantages over purely statistical covariance structures previously used for model discrepancies~\citep{xiao2016quantifying,xiao2017random}. Specifically, they constructed an approximate, linearized PDE for the model discrepancy:
\begin{equation}
\label{eq:trans-err}
\frac{D \delta}{Dt}-\frac{\partial}{\partial x_i}
\left[\left(\nu+ \frac{\nu_T}{\sigma_\phi}\right)\frac{\partial \delta}{\partial x_i}\right]=\mathcal{S}
\qquad \text{or more compactly} \qquad \mathcal{L}(\delta) = \mathcal{S}
\end{equation}
where $\delta$ denotes the field of model discrepancy such as the discrepancy in the RANS-modeled Reynolds stresses, and $\mathcal{S}$ indicates the unclosed source terms. Equation~\eqref{eq:trans-err} can be generalized as $\mathcal{L}\left( \delta \right)=\mathcal{S}$, where $\mathcal{L}$ corresponds to the linearized differential operator  on the left-hand side. In previous works~\citep{xiao2016quantifying,xiao2017random}, purely statistical covariance structures such as the squared exponential kernel in Gaussian processes were specified for the model discrepancy $\delta$.  The physics-inspired transport equation (\ref{eq:trans-err}) requires the specification of a kernel of the source term $\mathcal{S}$ and provides a physical covariance structure of the error term $\delta$ by transforming the covariance with the differential operator as follows:
\begin{equation}
\label{eq:prop-sigma}
\Sigma_{\delta}=\mathcal{L}^{-1}\Sigma_{S}\left(\mathcal{L}^{-1}\right)^\top
\end{equation}
where $\mathcal{L}^{-1}$ is the inverse operator of the linearized PDE~(\ref{eq:trans-err}), and $\Sigma$ denotes covariance. \citet{wu2018pde-informed} showed that such a physics-inspired covariance structure better accounts for the spatial correlation of the discrepancy term $\delta$ than the squared exponential covariance kernel. Sample results for flow over periodic hills are presented in Figure~\ref{fig:pde-cov}, which shows the first three modes (i.e., $\{ \phi_\alpha(x) \}_{\alpha = 1}^3$ as in Equation~\eqref{eq:KL}) obtained by using a squared exponential kernel (Figure~\ref{fig:pde-cov}a) and a physics-informed kernel (Figure~\ref{fig:pde-cov}b), e.g., from Equation~\eqref{eq:prop-sigma}.  In this geometry, the general flow direction is from left to right. The streamline-aligned covariance structure endowed by the convection is evident, while the modes obtained from the squared exponential kernel exhibits nonphysical, spatially isotropic structures.

\section{Uncertainties in large eddy simulations}
\label{sec:les}

As recalled in Section~\ref{sec:introduction}, large eddy simulation (LES) is a turbulence simulation method that resolves larger scale turbulence and models sub-grid scales~\citep{sagaut2006large}. For LES performed on an adequate mesh, most of the important turbulence scales are resolved except in near-wall regions of wall-bounded flows. Consequently, uncertainties associated with the subgrid scale (SGS) model no longer  dominate. Instead, LES are influenced by uncertainties of a number of sources that are of comparable order of magnitude, including:
\begin{enumerate}[(1)]
\item uncertainties due to SGS models, including their parameters,
\item uncertainties associated with initial and boundary conditions,
\item uncertainties in the numerical discretization (mesh and numerical scheme).
\end{enumerate}
Note that items 2--3 are not model uncertainties but input uncertainties and numerical uncertainties, respectively.  This is in stark contrast to RANS simulations, where the model uncertainty clearly dominates other uncertainties.  The literature on prediction accuracies of LES is vast and is mostly from deterministic perspectives. In such frameworks, the problem of concern should be more precisely referred to as errors and not uncertainties. Nevertheless, some studies have tackled the problem from an uncertainty quantification point of view, which are shortly reviewed here. Due to the much higher computational cost of LES as compared to RANS simulations, most studies are limited to uncertainty propagation and sensitivity analysis, i.e., propagation of assumed probability distributions through an LES solver to investigate the sensitivity of the output quantities with respect to the input.  In order to reduce the number of samples and overall computational costs for uncertainty propagation, many of the studies reviewed below built surrogate models by using different methods, e.g., polynomial chaos expansion~\cite{le-matre2010spectral,najm2009uncertainty}, probabilistic collocation method~\cite{tatang1995direct}, or sparse grid method~\citep{bungartz2004sparse}.

A review of recent work about quantification and reduction of uncertainties arising in LES is presented in the following of this section.
We point out here that high-fidelity simulations such as LES and DNS generally have smaller uncertainties than RANS simulations.  
However, even DNS have their own uncertainties, e.g., due to statistical averaging and numerical methods, which must be considered in many situations when using DNS data for RANS model development and calibration. 
While an in-depth discussion of DNS uncertainties is beyond the scope of the current review, in \ref{app:dns} we survey a few aspects that are most relevant for RANS modeling.

\subsection{Uncertainties in SGS models}
Traditional LES computations rely on the explicit introduction of a closure model for the subgrid-scale terms arising from the filtering of the Navier--Stokes equations. A large number of SGS models have been developed over years, almost all of which require specification of model constants, although some (e.g. dynamic Smagorinsky model) allow for a dynamic computation of the parameters from a test filter~\citep{germano1991dynamic}.  The most widely used SGS  model is the algebraic Smagorinsky model, which models the SGS viscosity as:
\begin{equation}
  \label{eq:sgs}
\nu_\mathrm{\text{sgs}}
= (C_s \Delta_g)^2\left| \overline{\mathbf{S}} \right|
\qquad
\text{with} \qquad \left| \overline{\mathbf{S}} \right| \equiv \sqrt{2 \overline{S}_{ij} \overline{S}_{ij}}
\end{equation}
where $\overline{\mathbf{S}}$ is the rate-of-strain based on the filtered velocity field, $\Delta_g$ is the grid size, and $C_s$ is a coefficient that needs to be specified (referred to as Smagorinsky constant, usually chosen to be in the range from 0.1 to 0.2) and has the effect of determining the strength of SGS dissipation.  \citet{meyers2006on} derived the exact expression of the Smagorinsky constant $C_s$ by using Pope's formulation for the turbulent kinetic energy spectrum. The  derivation demonstrate that $C_s$ depends both on the specific flow and on the filter, indicating that it should be treated as an uncertain quantity. The dependence of $C_s$ on the filter size deserves special attention for LES with implicit filtering, where the filter size is not explicitly specified but determined by the local grid size (see further discussions on numerical uncertainties in Section~\ref{sec:les-num}).  \citet{lucor2007sensitivity} performed LES for decaying homogeneous isotropic turbulence and propagated the uncertainties associated with the Smagorinsky constant.  Specifically, they reconstructed accurately the solution statistics with a typical number of 22 samples. They carried out uncertainty propagation corresponding to different grid resolutions and found that an optimal value of the constant can be found for each level of grid refinement~\citep{meyers2006on}. This finding confirmed the close interactions between the SGS model and the numerical discretization.

\citet{meldi2011is} investigated the sensitivity of the $C_s$ constant to the algebraic function and its parameters used to describe the initial energy spectrum.
 \citet{khalil2015uncertainty} performed LES for turbulent bluff-body stabilized flame and studied the uncertainties associated with Smagorinsky constant, Prandtl number, and Schmidt number.
 \citet{safta2017uncertainty} investigated LES of channel flow and studied uncertainties associated with model coefficients $C_{\mu_{\epsilon}}$ and $C_{\epsilon}$ in the $k^{sgs}$ model~\citep{yoshizawa1993bridging}.
Unlike previous uncertainty propagation studies for LES, \citet{templeton2015calibration} first used Bayesian inference to calibrate the model coefficients $C_{\mu_{\epsilon}}$ and $C_{\epsilon}$ based on a DNS database of forced isotropic turbulence in a periodic box~\citep{li2008public}. The quantified uncertainties in the calibrated model coefficients were then propagated to predictions in LES of turbulent channel flows.
\citet{tran2016sparse} also used Bayesian inference to quantify the uncertainties associated with the Smagorinsky constant as well as filter length and the exponent in van Driest damping function from synthetic data (corresponding to a reference LES) for the flow around a cylinder.



While the above-mentioned studies used parametric approaches to address uncertainties associated with model coefficients,  non-parametric UQ approaches for LES recently started drawing attention. \citet{jofre2018framework} estimated the structural uncertainties in the SGS stress model in LES of the canonical plane channel flows. They perturbed the SGS stresses obtained from  baseline model in a similar way as in the RANS simulations~\cite{emory2011modeling,emory2013modeling,gorle2013framework}. By directly introducing perturbation into the SGS stresses, the explored uncertainty space is no longer constrained by the baseline SGS model.

\subsection{Uncertainties in the boundary conditions for LES}
Boundary conditions are a crucial ingredient of the overall model in LES, as they may influence the development of shear and boundary layers and transition to turbulence. \citet{congedo2013numerical} investigated the sensitivity of LES to uncertainties in the numerical inlet conditions by studying the turbulent flow in a pipe with an axisymmetric expansion. The study focused on the effect of the inlet bulk velocity, swirl ratio, and turbulent intensity on the resulting mean flow fields.  The results were compared with experimental data, for which an estimate of the observational uncertainty was available.
On the other hand, \citet{carnevale2013uncertainty} studied the heat transfer in a channel with pins and used uncertainty propagation to investigate the flow sensitivity to the Reynolds number, which is representative of the uncertainties associated with inlet mean velocity, fluid density, or geometrical variations. They compared the results to those of RANS simulations, showing that the epistemic uncertainty due to the modeling, i.e., RANS model versus LES model, dominates the aleatoric uncertainties such as the solution sensitivity to the Reynolds number. However, compared to RANS simulations, the LES results are more sensitive to the inlet Reynolds number.

\subsection{Uncertainties due to the numerical discretization}
\label{sec:les-num}
As has been pointed out above, LES suffer from strong interactions between modeling and numerical errors. This is particularly true for  LES with implicit filtering, which is dominant in practical LES, where the mesh is part of the model in that it provides the local filtering bandwidth, as is evident from Equation~(\ref{eq:sgs}).
In theory, only the dynamics of the large scales is computed and the smaller scales are modeled. In practice, scale separation in LES is difficult to establish, since the low-pass filtering arises from a complex combination of implicit filtering by the grid and the discretization schemes. Even when explicit filters are applied, the approximations introduced by the
discretization methods modify the actual shape of the filter function. The intricate interactions
between SGS modeling errors and numerical errors (and the ill-defined filter resulted therefrom) have attracted attention (e.g.,~\cite{ghosal1996analysis,vreman1996comparision,kravchenko1997on}).
Here we mention a few studies that analyzied the numerical parameters from a probabilistic perspective.
For example, \citet{meldi2012quantification} performed LES for a spatially evolving mixing layer and studied the uncertainty propagation for grid stretching ratio in the turbulent and transitional regions. \citet{mariotti2016stochastic} studied the flow around a 5:1 rectangular cylinder and propagated the uncertainties associated with grid resolution in the spanwise direction and the weight of the explicit low-pass filter.

With the increasing availability of computational resources and the increasing use of LES in industrial simulations, uncertainty quantification in LES is expected follow a similar development path as for RANS but with equal emphasis on all the above-mentioned uncertainties sources. It will evolve from the current data-free, parametric approaches to more sophisticated, data-driven, non-parametric approaches, and from the current proof-of-concept studies to gradual deployment in industrial simulations.

\section{Conclusions and future research}
\label{sec:conclusion}


This review summarized techniques for quantifying uncertainties associated with turbulence models in computational fluid dynamics simulations. We focused on uncertainty quantification in RANS models, because they are expected to remain the workhorse tool for industrial CFD simulations in decades to come, thanks to their lower computational costs and better robustness than scale-resolving methods.  Quantifying uncertainties in RANS predictions are of strategic importance towards the goal of certified numerical simulations of fluid flows.

The literature survey shows that RANS uncertainty quantification has been a rapidly evolving field in the past decade. Most of the recent research focused on statistical approaches to estimate prediction uncertainties due to turbulence models and on data-driven methods to reduce such uncertainties. Development of such statistically rigorous techniques for quantifying and reducing RANS model uncertainties has been fostered by:
\begin{enumerate}[(1)]
\item the considerable increase of computer resources,
\item the ever-increasing mass of high fidelity experimental and numerical data, and
\item the development of statistical sampling and inference methods guided by physical constraints and prior knowledge in turbulence modeling.
\end{enumerate}

This article classifies existing literature of model uncertainty quantification into parametric and non-parametric approaches, which are reviewed separately. In the parametric approaches, uncertainties are introduced into the coefficients in RANS closure models. That is, the coefficients are modeled as random variables, whose prior distributions are then propagated to the predictions through RANS simulations or updated by incorporating observation data within the Bayesian inference framework. Extensions of the parametric approaches are multi-model approaches such as Bayesian model averaging and Bayesian model--scenario averaging methods. In these methods, predictions of new flow configurations (scenarios) are formulated as an average of predictions from an ensemble of competing models, weighted by their respective posterior probabilities and the similarity of respective calibration scenarios to the prediction scenario.  Parametric and multi-model ensemble methods are robust, non-intrusive, and relatively mature.  When combined with surrogate models that replace RANS models to allow for efficient sampling, they can be used in uncertainty quantification involving complex, three-dimensional engineering flows.

A drawback of parametric approaches is that  any calibration and inference of the parameters are inevitably based on, and will influence, the entire flow field. However, a turbulent flow may simultaneously contain regions ranging from equilibrium regions that are well predicted by simple models to highly non-equilibrium regions (e.g., separation, shock waves, streamline curvature, rotation) where even advanced models would fail.  Non-parametric approaches provide an attractive alternative to tackle turbulence modeling uncertainties while accounting for locality of turbulent flows.  These approaches rely on random fields to represent the RANS model discrepancy, which are estimated from physical bounds and further propagated to predictions or inferred from observation data. However, the unique challenge here is that the uncertainty propagation and statistical inference involve random fields of much higher dimensions.  Sampling and inference in such a high-dimensional space remain an active field of research with many open challenges and opportunities.

Another thrilling subject for future research is the application of non-parametric approaches for predictions. Data assimilation and machine learning algorithms have been recently applied to extrapolating estimated discrepancy fields to configurations that are relatively close to the ones contained in the training flows~\cite{wu2018data-driven}. However, using the estimated uncertainties to drastically different configurations remains a delicate and possibly dangerous task.  Introducing sound physical constraints in the representation of the discrepancy and using physics-based transport equations for describing its spatial correlation structure seems to be a promising approach~\cite{wu2018pde-informed}.  Bridging uncertainty quantification and data-driven modeling, such a physics-informed approach has the potential of yielding RANS models that can predict turbulent flows with quantified uncertainties, paving the way toward certified CFD simulations~\cite{duraisamy2019turbulence}.

\section*{Acknowledgment}

PC acknowledges support from French Agence Nationale de la Recherche (ANR). 
HX acknowledges support and mentoring from the Department of Aerospace and Ocean Engineering at Virginia Tech and particularly Prof. C.J. Roy and Prof. E.G. Paterson. 
The authors would like to thank Dr. J.-L. Wu, Dr. J.-X. Wang, Dr. R. P. Dwight, Dr. W. N. Edeling, and M. Schmeltzer for their collaborations and Mr. C. Michel\'en-Str\"ofer for useful suggestions on the manuscript. Finally, the authors gratefully acknowledge Dr. J.-L. Wu for providing figures, materials, and suggestions for this review. Finally, the authors thank the three anonymous reviewers for their constructive and valuable comments, which helped improving the quality and clarity of this review article.

\section*{References}


\begin{thebibliography}{187}
\providecommand{\natexlab}[1]{#1}
\providecommand{\url}[1]{\texttt{#1}}
\expandafter\ifx\csname urlstyle\endcsname\relax
  \providecommand{\doi}[1]{doi: #1}\else
  \providecommand{\doi}{doi: \begingroup \urlstyle{rm}\Url}\fi

\bibitem[Sagaut et~al.(2013)Sagaut, Deck, and Terracol]{sagaut2006multiscale}
Sagaut P, Deck S, Terracol M.
\newblock \emph{Multiscale and Multiresolution Approaches in Turbulence}.
\newblock Imperial College Press, second edition, 2013.

\bibitem[Sagaut(2006)]{sagaut2006large}
Sagaut P.
\newblock \emph{Large Eddy Simulations for Incompressible Flows: An
  Introduction}.
\newblock Springer, 2006.

\bibitem[Spalart(2009)]{spalart2009detached-eddy}
Spalart PR.
\newblock Detached-eddy simulation.
\newblock \emph{Annual Review of Fluid Mechanics}, 41:\penalty0 181--202, 2009.

\bibitem[Fr{\"o}hlich and von Terzi(2008)]{frohlich2008hybrid}
Fr{\"o}hlich J, von Terzi D.
\newblock Hybrid {LES/RANS} methods for the simulation of turbulent flows.
\newblock \emph{Progress in Aerospace Sciences}, 44\penalty0 (5):\penalty0
  349--377, 2008.

\bibitem[Chaouat(2017)]{chaouat2017state}
Chaouat B.
\newblock The state of the art of hybrid rans/les modeling for the simulation
  of turbulent flows.
\newblock \emph{Flow, Turbulence and Combustion}, 99:\penalty0 279--327, 2017.

\bibitem[Cabot and Moin(2000)]{cabot2000approximate}
Cabot W, Moin P.
\newblock Approximate wall boundary conditions in the large-eddy simulation of
  high {R}eynolds number flow.
\newblock \emph{Flow, Turbulence and Combustion}, 63\penalty0 (1-4):\penalty0
  269--291, 2000.

\bibitem[Piomelli and Balaras(2002)]{piomelli2002wall}
Piomelli U, Balaras E.
\newblock Wall-layer models for large-eddy simulations.
\newblock \emph{Ann. Rev. Fluid Mech.}, 34\penalty0 (1):\penalty0 349--374,
  2002.

\bibitem[Kawai and Larsson(2012)]{kawai2012wall}
Kawai S, Larsson J.
\newblock Wall-modeling in large eddy simulation: Length scales, grid
  resolution, and accuracy.
\newblock \emph{Phys. Fluids}, 24\penalty0 (1):\penalty0 015105, 2012.

\bibitem[Yang et~al.(2015)Yang, Sadique, Mittal, and
  Meneveau]{yang2015integral}
Yang XIA, Sadique J, Mittal R, Meneveau C.
\newblock Integral wall model for large eddy simulations of wall-bounded
  turbulent flows.
\newblock \emph{Phys. Fluids}, 27\penalty0 (2):\penalty0 025112, 2015.

\bibitem[Wilcox(2006)]{wilcox2006turbulence}
Wilcox DC.
\newblock \emph{Turbulence modeling for CFD}.
\newblock DCW Industries, third edition, 2006.

\bibitem[{Durbin}(2018)]{durbin2018some}
{Durbin} PA.
\newblock Some recent developments in turbulence closure modeling.
\newblock \emph{Annual Review of Fluid Mechanics}, 50:\penalty0 77--103, 2018.

\bibitem[Girimaji(2006)]{girimaji2006partially-averaged}
Girimaji SS.
\newblock Partially-averaged {N}avier-{S}tokes model for turbulence: A
  {R}eynolds-averaged {N}avier--{S}tokes to direct numerical simulation
  bridging method.
\newblock \emph{Journal of Applied Mechanics}, 73\penalty0 (3):\penalty0
  413--421, 2006.

\bibitem[Xiao and Jenny(2012)]{xiao2012consistent}
Xiao H, Jenny P.
\newblock A consistent dual-mesh framework for hybrid {LES/RANS} modeling.
\newblock \emph{Journal of Computational Physics}, 231\penalty0 (4):\penalty0
  1848--1865, 2012.

\bibitem[Johnson et~al.(2005)Johnson, Tinoco, and Yu]{johnson2005thirty}
Johnson FT, Tinoco EN, Yu~NJ.
\newblock Thirty years of development and application of {CFD} at {Boeing
  Commercial Airplanes}, {Seattle}.
\newblock \emph{Computers \& Fluids}, 34\penalty0 (10):\penalty0 1115--1151,
  2005.

\bibitem[Spalart(2015)]{spalart2015philosophies}
Spalart PR.
\newblock Philosophies and fallacies in turbulence modeling.
\newblock \emph{Progress in Aerospace Sciences}, 74:\penalty0 1--15, 2015.

\bibitem[Duraisamy et~al.(2019)Duraisamy, Iaccarino, and
  Xiao]{duraisamy2019turbulence}
Duraisamy K, Iaccarino G, Xiao H.
\newblock Turbulence modeling in the age of data.
\newblock \emph{Annual Review of Fluid Mechanics}, 51:\penalty0 357--377, 2019.

\bibitem[Perot and Moin(1996)]{perot1996new}
Perot B, Moin P.
\newblock A new approach to turbulence modeling.
\newblock Technical report, Proceedings of Summer Research Program, Center of
  Turbulence Research, Stanford University, Stanford, CA, USA, 1996.

\bibitem[Draper(1995)]{draper1995assessment}
Draper D.
\newblock Assessment and propagation of model uncertainty.
\newblock \emph{Journal of the Royal Statistical Society. Series B
  (Methodological)}, pages 45--97, 1995.

\bibitem[Kennedy and O'Hagan(2001)]{kennedy2001bayesian}
Kennedy MC, O'Hagan A.
\newblock Bayesian calibration of computer models.
\newblock \emph{Journal of the Royal Statistical Society: Series B (Statistical
  Methodology)}, 63\penalty0 (3):\penalty0 425--464, 2001.

\bibitem[Speziale(1987)]{speziale1987on}
Speziale CG.
\newblock On nonlinear $k$-$l$ and $k$-$\varepsilon$ models of turbulence.
\newblock \emph{Journal of Fluid Mechanics}, 178:\penalty0 459--475, 1987.

\bibitem[Gatski and Speziale(1993)]{gatski1993on}
Gatski T, Speziale C.
\newblock On explicit algebraic stress models for complex turbulent flows.
\newblock \emph{Journal of Fluid Mechanics}, 254:\penalty0 59--79, 1993.

\bibitem[Launder et~al.(1975)Launder, Reece, and Rodi]{launder1975progress}
Launder BE, Reece GJ, Rodi W.
\newblock Progress in the development of a {Reynolds-stress} turbulence
  closure.
\newblock \emph{Journal of Fluid Mechanics}, 68\penalty0 (03):\penalty0
  537--566, 1975.

\bibitem[Ray et~al.(2018{\natexlab{a}})Ray, Lefantzi, Arunajatesan, and
  Dechant]{ray2018learning}
Ray J, Lefantzi S, Arunajatesan S, Dechant L.
\newblock Learning an eddy viscosity model using shrinkage and {Bayesian}
  calibration: A jet-in-crossflow case study.
\newblock \emph{{ASCE-ASME} Journal of Risk and Uncertainty in Engineering
  Systems, {Part B}: Mechanical Engineering}, 4\penalty0 (1):\penalty0 011001,
  2018{\natexlab{a}}.

\bibitem[Tinoco et~al.(2018)Tinoco, Brodersen, Keye, Laflin, Feltrop, Vassberg,
  Mani, Rider, Wahls, Morrison, Hue, Roy, Mavriplis, and
  Murayama]{tinoco2018summary}
Tinoco EN, Brodersen OP, Keye S, Laflin KR, Feltrop E, Vassberg JC, Mani M,
  Rider B, Wahls RA, Morrison JH, Hue D, Roy CJ, Mavriplis DJ, Murayama M.
\newblock Summary data from the sixth {AIAA} {CFD} drag prediction workshop:
  {CRM} cases.
\newblock \emph{Journal of Aircraft}, 55\penalty0 (4):\penalty0 1352--1379,
  2018.

\bibitem[Baldwin and Lomax(1978)]{baldwin1978thin}
Baldwin B, Lomax H.
\newblock Thin layer approximation and algebraic model for separated turbulent
  flows.
\newblock AIAA Paper 78-257, 1978.

\bibitem[Cinnella et~al.(2016)Cinnella, Dwight, and
  Edeling]{cinnella2016review}
Cinnella P, Dwight R, Edeling WN.
\newblock Review of uncertainty quantification in turbulence modelling to date.
\newblock Minisymposium ``UQ in turbulence modelling'', SIAM Uncertainty
  Quantification conference, Lausanne, Switzerland. 5-8 April, 2016.
\newblock doi: 10.13140/RG.2.1.4512.5523.

\bibitem[Parussini and Pediroda(2007)]{parussini2007fictitious}
Parussini L, Pediroda V.
\newblock Fictitious domain with least-squares spectral element method to
  explore geometric uncertainties by non-intrusive polynomial chaos method.
\newblock \emph{Computer modeling in engineering and science}, 22\penalty0
  (1):\penalty0 41--63, 2007.

\bibitem[Liu et~al.(2017)Liu, Litvinenko, Schillings, and
  Schulz]{liu2017quantification}
Liu D, Litvinenko A, Schillings C, Schulz V.
\newblock Quantification of airfoil geometry-induced aerodynamic
  uncertainties---comparison of approaches.
\newblock \emph{SIAM/ASA Journal on Uncertainty Quantification}, 5\penalty0
  (1):\penalty0 334--352, 2017.

\bibitem[Avdonin and Polifke(2018)]{avdonin2018quantification}
Avdonin A, Polifke W.
\newblock Quantification of the impact of uncertainties in operating conditions
  on the flame transfer function with non-intrusive polynomial chaos expansion.
\newblock In \emph{ASME Turbo Expo: Power for Land, Sea, and Air}, volume 4A:
  Combustion, Fuels, and Emissions. ASME, 2018.
\newblock Paper GT2018-75476.

\bibitem[Gorl\'e et~al.(2015)Gorl\'e, Garcia-Sanchez, and
  Iaccarino]{gorle2015quantifying}
Gorl\'e C, Garcia-Sanchez C, Iaccarino G.
\newblock Quantifying inflow and {RANS} turbulence model form uncertainties for
  wind engineering flows.
\newblock \emph{Journal of Wind Engineering and Industrial Aerodynamics},
  144:\penalty0 202--212, 2015.

\bibitem[Mariotti et~al.(2016{\natexlab{a}})Mariotti, Salvetti, Omrani, and
  Witteween]{mariotti2016freestream}
Mariotti A, Salvetti M, Omrani S, Witteween J.
\newblock Stochastic analysis of the impact of freestream conditions on the
  aerodynamics of a rectangular 5:1 cylinder.
\newblock \emph{Computers and Fluids}, 136:\penalty0 170--192,
  2016{\natexlab{a}}.

\bibitem[James et~al.(2013)James, Witten, Hastie, and
  Tibshirani]{james2013introduction}
James G, Witten D, Hastie T, Tibshirani R.
\newblock \emph{An introduction to statistical learning}, volume 112.
\newblock Springer, 2013.

\bibitem[Klir(2006)]{klir2006uncertainty}
Klir GJ.
\newblock \emph{Uncertainty and Information: Foundations of Generalized
  Information Theory}.
\newblock Wiley Interscience, 2006.

\bibitem[Ferson(1996)]{ferson1996what}
Ferson S.
\newblock What {Monte Carlo} methods cannot do.
\newblock \emph{Human and Ecological Risk Assessment}, 2\penalty0 (4):\penalty0
  990--1007, 1996.

\bibitem[Ferson and Ginzburg(1996)]{ferson1996different}
Ferson S, Ginzburg LR.
\newblock Different methods are needed to propagate ignorance and variability.
\newblock \emph{Reliability Engineering and System Safety}, 54\penalty0
  (2-3):\penalty0 133--144, 1996.

\bibitem[Liu et~al.(2004)Liu, Chen, Arendt, and Huang]{liu2004arithmetic}
Liu Y, Chen W, Arendt P, Huang HZ.
\newblock Arithmetic with uncertain numbers: Rigorous and (often) best possible
  answers.
\newblock \emph{Reliability Engineering and System Safety}, 85:\penalty0
  135--152, 2004.

\bibitem[Shafer(1976)]{shafer1976mathematical}
Shafer G.
\newblock \emph{A Mathematical Theory of Evidence}.
\newblock Princeton University Press, Princeton, NJ, 1976.

\bibitem[Zadeh(1996)]{zadeh1996fuzzy}
Zadeh LA.
\newblock Fuzzy sets.
\newblock In \emph{Fuzzy Sets, Fuzzy Logic, And Fuzzy Systems: Selected Papers
  by Lotfi A Zadeh}, pages 394--432. World Scientific, 1996.

\bibitem[Bernardini and Tonon(2010)]{bernardini2010bounding}
Bernardini A, Tonon F.
\newblock \emph{Bounding Uncertainty in Civil Engineering}.
\newblock Springer-Verlag, 2010.

\bibitem[Oberkampf and Roy(2010)]{oberkampf2010verification}
Oberkampf WL, Roy CJ.
\newblock \emph{Verification and Validation In Scientific Computing}.
\newblock Cambridge University Press, 2010.

\bibitem[Roy and Oberkampf(2011)]{roy2011comprehensive}
Roy CJ, Oberkampf WL.
\newblock A comprehensive framework for verification, validation, and
  uncertainty quantification in scientific computing.
\newblock \emph{Computer Methods in Applied Mechanics and Engineering},
  200:\penalty0 2131--2144, 2011.

\bibitem[Dow and Wang(2011)]{dow2011quantification}
Dow E, Wang Q.
\newblock Quantification of structural uncertainties in the $k$--$\omega$
  turbulence model.
\newblock In \emph{52nd AIAA/ASME/ASCE/AHS/ASC Structures, Structural Dynamics
  and Materials Conference}, Denver, Colorado, April 2011. AIAA.
\newblock Paper 2011-1762.

\bibitem[Singh and Duraisamy(2016)]{singh2016using}
Singh AP, Duraisamy K.
\newblock Using field inversion to quantify functional errors in turbulence
  closures.
\newblock \emph{Physics of Fluids}, 28:\penalty0 045110, 2016.

\bibitem[Xiao et~al.(2016)Xiao, Wu, Wang, Sun, and Roy]{xiao2016quantifying}
Xiao H, Wu~JL, Wang JX, Sun R, Roy C.
\newblock Quantifying and reducing model-form uncertainties in
  {Reynolds-averaged} {Navier--Stokes} simulations: A data-driven,
  physics-informed {Bayesian} approach.
\newblock \emph{Journal of Computational Physics}, 324:\penalty0 115--136,
  2016.

\bibitem[Ling et~al.(2016{\natexlab{a}})Ling, Kurzawski, and
  Templeton]{ling2016reynolds}
Ling J, Kurzawski A, Templeton J.
\newblock Reynolds averaged turbulence modelling using deep neural networks
  with embedded invariance.
\newblock \emph{Journal of Fluid Mechanics}, 807:\penalty0 155--166,
  2016{\natexlab{a}}.

\bibitem[Weatheritt and Sandberg(2016)]{weatheritt2016novel}
Weatheritt J, Sandberg R.
\newblock A novel evolutionary algorithm applied to algebraic modifications of
  the {RANS} stress--strain relationship.
\newblock \emph{Journal of Computational Physics}, 325:\penalty0 22--37, 2016.

\bibitem[Weatheritt and Sandberg(2017)]{weatheritt2017development}
Weatheritt J, Sandberg RD.
\newblock The development of algebraic stress models using a novel evolutionary
  algorithm.
\newblock \emph{International Journal of Heat and Fluid Flow}, 68:\penalty0
  298--318, 2017.

\bibitem[Turgeon et~al.(2001)Turgeon, Pelletier, and
  Borggaard]{turgeon2001application}
Turgeon {\'E}, Pelletier D, Borggaard J.
\newblock Application of a sensitivity equation method to the $k$--$\epsilon$
  model of turbulence.
\newblock In \emph{15th AIAA computational fluid dynamics conference}, pages
  25--34, 2001.

\bibitem[Dunn et~al.(2011)Dunn, Shotorban, and Frendi]{dunn2011uncertainty}
Dunn MC, Shotorban B, Frendi A.
\newblock Uncertainty quantification of turbulence model coefficients via
  {Latin} hypercube sampling method.
\newblock \emph{Journal of Fluids Engineering}, 133\penalty0 (4):\penalty0
  041402, 2011.

\bibitem[Platteeuw et~al.(2008)Platteeuw, Loeven, and
  Bijl]{platteeuw2008uncertainty}
Platteeuw PDA, Loeven GJA, Bijl H.
\newblock Uncertainty quantification applied to the $k$--$\epsilon$ model of
  turbulence using the probabilistic collocation method.
\newblock In \emph{10th AIAA Non-Deterministic Approaches Conference}, 2008.
\newblock Paper no.: 2008-2150.

\bibitem[Margheri et~al.(2014)Margheri, Meldi, Salvetti, and
  Sagaut]{margheri2014epistemic}
Margheri L, Meldi M, Salvetti M, Sagaut P.
\newblock Epistemic uncertainties in {RANS} model free coefficients.
\newblock \emph{Computers \& Fluids}, 102:\penalty0 315--335, 2014.

\bibitem[Schaefer et~al.(2016)Schaefer, Hosder, West, Rumsey, Carlson, and
  Kleb]{schaefer2016uncertainty}
Schaefer J, Hosder S, West T, Rumsey C, Carlson JR, Kleb W.
\newblock Uncertainty quantification of turbulence model closure coefficients
  for transonic wall-bounded flows.
\newblock \emph{AIAA Journal}, 55\penalty0 (1):\penalty0 195--213, 2016.

\bibitem[Emory et~al.(2011)Emory, Pecnik, and Iaccarino]{emory2011modeling}
Emory M, Pecnik R, Iaccarino G.
\newblock Modeling structural uncertainties in {Reynolds}-averaged computations
  of shock/boundary layer interactions.
\newblock In \emph{49th AIAA Aerospace Sciences Meeting including the New
  Horizons Forum and Aerospace Exposition}, 2011.
\newblock Paper 2011-479.

\bibitem[Emory et~al.(2013)Emory, Larsson, and Iaccarino]{emory2013modeling}
Emory M, Larsson J, Iaccarino G.
\newblock Modeling of structural uncertainties in {Reynolds}-averaged
  {Navier}-{Stokes} closures.
\newblock \emph{Physics of Fluids}, 25\penalty0 (11):\penalty0 110822, 2013.

\bibitem[Iaccarino et~al.(2017)Iaccarino, Mishra, and
  Ghili]{iaccarino2017eigenspace}
Iaccarino G, Mishra AA, Ghili S.
\newblock Eigenspace perturbations for uncertainty estimation of single-point
  turbulence closures.
\newblock \emph{Physical Review Fluids}, 2\penalty0 (2):\penalty0 024605, 2017.

\bibitem[Mishra and Iaccarino(2017)]{mishra2017uncertainty}
Mishra AA, Iaccarino G.
\newblock Uncertainty estimation for {Reynolds-Averaged Navier--Stokes}
  predictions of high-speed aircraft nozzle jets.
\newblock \emph{AIAA Journal}, 55:\penalty0 1--6, 2017.

\bibitem[Edeling et~al.(2017)Edeling, Iaccarino, and
  Cinnella]{edeling2017data-free}
Edeling WN, Iaccarino G, Cinnella P.
\newblock Data-free and data-driven {RANS} predictions with quantified
  uncertainty.
\newblock \emph{Flow, Turbulence and Combustion}, 100:\penalty0 1--24, 2017.

\bibitem[Xiao et~al.(2017)Xiao, Wang, and Gahnem]{xiao2017random}
Xiao H, Wang JX, Gahnem RG.
\newblock A random matrix approach for quantifying model-form uncertainties in
  turbulence modeling.
\newblock \emph{Computer Methods in Applied Mechanics and Engineering},
  313:\penalty0 941--965, 2017.

\bibitem[Poroseva et~al.(2006)Poroseva, Hussaini, and
  Woodruff]{poroseva2006improving}
Poroseva SV, Hussaini MY, Woodruff SL.
\newblock Improving the predictive capability of turbulence models using
  evidence theory.
\newblock \emph{AIAA Journal}, 44\penalty0 (6):\penalty0 1220--1228, 2006.

\bibitem[Edeling et~al.(2014{\natexlab{a}})Edeling, Cinnella, and
  Dwight]{edeling2014predictive}
Edeling W, Cinnella P, Dwight RP.
\newblock Predictive {RANS} simulations via {Bayesian} model-scenario
  averaging.
\newblock \emph{Journal of Computational Physics}, 275:\penalty0 65--91,
  2014{\natexlab{a}}.

\bibitem[Edeling et~al.(2018)Edeling, Schmelzer, Cinnella, and
  Dwight]{edeling2018bayesian}
Edeling WN, Schmelzer M, Cinnella P, Dwight RP.
\newblock {Bayesian} predictions of {Reynolds}-averaged {Navier}--{Stokes}
  uncertainties using maximum a posteriori estimates.
\newblock \emph{{AIAA} Journal}, 5\penalty0 (6):\penalty0 2018--2029, 2018.

\bibitem[Cheung et~al.(2011)Cheung, Oliver, Prudencio, Prudhomme, and
  Moser]{cheung2011bayesian}
Cheung SH, Oliver TA, Prudencio EE, Prudhomme S, Moser RD.
\newblock Bayesian uncertainty analysis with applications to turbulence
  modeling.
\newblock \emph{Reliability Engineering \& System Safety}, 96\penalty0
  (9):\penalty0 1137--1149, 2011.

\bibitem[Kato and Obayashi(2013)]{kato2013approach}
Kato H, Obayashi S.
\newblock Approach for uncertainty of turbulence modeling based on data
  assimilation technique.
\newblock \emph{Computers \& Fluids}, 85:\penalty0 2--7, 2013.

\bibitem[Kato et~al.(2015)Kato, Yoshizawa, Ueno, and Obayashi]{kato2015data}
Kato H, Yoshizawa A, Ueno G, Obayashi S.
\newblock A data assimilation methodology for reconstructing turbulent flows
  around aircraft.
\newblock \emph{Journal of Computational Physics}, 283:\penalty0 559--581,
  2015.

\bibitem[Ray et~al.(2016)Ray, Lefantzi, Arunajatesan, and
  Dechant]{ray2016bayesian}
Ray J, Lefantzi S, Arunajatesan S, Dechant L.
\newblock Bayesian parameter estimation of a $k$--$\varepsilon$ model for
  accurate jet-in-crossflow simulations.
\newblock \emph{AIAA Journal}, 54\penalty0 (8):\penalty0 2432--2448, 2016.

\bibitem[Edeling et~al.(2014{\natexlab{b}})Edeling, Cinnella, Dwight, and
  Bijl]{edeling2014bayesian}
Edeling WN, Cinnella P, Dwight RP, Bijl H.
\newblock Bayesian estimates of parameter variability in the $k$--$\varepsilon$
  turbulence model.
\newblock \emph{Journal of Computational Physics}, 258:\penalty0 73--94,
  2014{\natexlab{b}}.

\bibitem[Papadimitriou and Papadimitriou(2015)]{papadimitriou2015bayesian}
Papadimitriou DI, Papadimitriou C.
\newblock Bayesian uncertainty quantification of turbulence models based on
  high-order adjoint.
\newblock \emph{Computers \& Fluids}, 120:\penalty0 82--97, 2015.

\bibitem[Wu et~al.(2016{\natexlab{a}})Wu, Wang, and Xiao]{wu2016bayesian}
Wu~JL, Wang JX, Xiao H.
\newblock A {Bayesian} calibration--prediction method for reducing model-form
  uncertainties with application in {RANS} simulations.
\newblock \emph{Flow, Turbulence and Combustion}, 97\penalty0 (3):\penalty0
  761--786, 2016{\natexlab{a}}.

\bibitem[Wang et~al.(2016{\natexlab{a}})Wang, Wu, and
  Xiao]{wang2016incorporating}
Wang J, Wu~JL, Xiao H.
\newblock Incorporating prior knowledge for quantifying and reducing model-form
  uncertainty in {RANS} simulations.
\newblock \emph{International Journal for Uncertainty Quantification},
  6\penalty0 (2), 2016{\natexlab{a}}.

\bibitem[Parish and Duraisamy(2016)]{parish2016paradigm}
Parish EJ, Duraisamy K.
\newblock A paradigm for data-driven predictive modeling using field inversion
  and machine learning.
\newblock \emph{Journal of Computational Physics}, 305:\penalty0 758--774,
  2016.

\bibitem[Le~Ma{\^\i}tre and Knio(2010)]{le-matre2010spectral}
Le~Ma{\^\i}tre OP, Knio OM.
\newblock \emph{Spectral methods for uncertainty quantification: with
  applications to computational fluid dynamics}.
\newblock Springer, 2010.

\bibitem[Ghanem and Spanos(2003)]{ghanem2003stochastic}
Ghanem RG, Spanos PD.
\newblock \emph{Stochastic Finite Elements: A Spectral Approach}.
\newblock Dover Publications, revised edition, 2003.

\bibitem[Glasserman(2004)]{glasserman2004monte-carlo}
Glasserman P.
\newblock \emph{{Monte Carlo} Methods in Financial Engineering}.
\newblock Springer, 2004.

\bibitem[Helton and Davis(2003)]{helton2003latin}
Helton JC, Davis FJ.
\newblock Latin hypercube sampling and the propagation of uncertainty in
  analyses of complex systems.
\newblock \emph{Reliability Engineering {\&} System Safety}, 81\penalty0
  (1):\penalty0 23--69, 2003.

\bibitem[Giles(2008)]{giles2008multilevel}
Giles MB.
\newblock Multilevel {Monte Carlo} path simulation.
\newblock \emph{Operations Research}, 56\penalty0 (3):\penalty0 607--617, 2008.

\bibitem[Barth et~al.(2011)Barth, Schwab, and Zollinger]{barth2011multi-level}
Barth A, Schwab C, Zollinger N.
\newblock Multi-level {Monte Carlo} finite element method for elliptic {PDEs}
  with stochastic coefficients.
\newblock \emph{Numerische Mathematik}, 119\penalty0 (1):\penalty0 123--161,
  2011.

\bibitem[Mishra et~al.(2012)Mishra, Schwab, and {\v
  S}ukys]{mishra2012multi-level}
Mishra S, Schwab C, {\v S}ukys J.
\newblock Multi-level {Monte Carlo} finite volume methods for nonlinear systems
  of conservation laws in multi-dimensions.
\newblock \emph{Journal of Computational Physics}, 231\penalty0 (8):\penalty0
  3365--3388, 2012.

\bibitem[M{\"u}ller et~al.(2013)M{\"u}ller, Jenny, and
  Meyer]{muller2013multilevel}
M{\"u}ller F, Jenny P, Meyer DW.
\newblock Multilevel {Monte Carlo} for two phase flow and buckley--leverett
  transport in random heterogeneous porous media.
\newblock \emph{Journal of Computational Physics}, 250:\penalty0 685--702,
  2013.

\bibitem[M{\"u}ller et~al.(2014)M{\"u}ller, Meyer, and
  Jenny]{muller2014solver-based}
M{\"u}ller F, Meyer DW, Jenny P.
\newblock Solver-based vs. grid-based multilevel {Monte Carlo} for two phase
  flow and transport in random heterogeneous porous media.
\newblock \emph{Journal of Computational Physics}, 268:\penalty0 39--50, 2014.

\bibitem[Peherstorfer et~al.(2018)Peherstorfer, Willcox, and
  Gunzburger]{peherstorfer2018survey}
Peherstorfer B, Willcox K, Gunzburger M.
\newblock Survey of multifidelity methods in uncertainty propagation,
  inference, and optimization.
\newblock \emph{SIAM Review}, 60\penalty0 (3):\penalty0 550--591, 2018.

\bibitem[{Moselhy} and {Marzouk}(2012)]{moselhy2012bayesian}
{Moselhy} TAE, {Marzouk} YM.
\newblock Bayesian inference with optimal maps.
\newblock \emph{Journal of Computational Physics}, 231\penalty0 (23):\penalty0
  7815--7850, 2012.

\bibitem[Parno and Marzouk(2018)]{parno2018transport}
Parno MD, Marzouk YM.
\newblock Transport map accelerated {Markov} chain {Monte Carlo}.
\newblock \emph{SIAM/ASA Journal on Uncertainty Quantification}, 6\penalty0
  (2):\penalty0 645--682, 2018.

\bibitem[Guillas et~al.(2014)Guillas, Glover, and
  Malki-Epshtein]{guillas2014bayesian}
Guillas S, Glover N, Malki-Epshtein L.
\newblock Bayesian calibration of the constants of the k--$\varepsilon$
  turbulence model for a {CFD} model of street canyon flow.
\newblock \emph{Computer methods in applied mechanics and engineering},
  279:\penalty0 536--553, 2014.

\bibitem[Mons et~al.(2016)Mons, Chassaing, Gomez, and
  Sagaut]{mons2016reconstruction}
Mons V, Chassaing JC, Gomez T, Sagaut P.
\newblock Reconstruction of unsteady viscous flows using data assimilation
  schemes.
\newblock \emph{Journal of Computational Physics}, 316:\penalty0 255--280,
  2016.

\bibitem[Iglesias et~al.(2013)Iglesias, Law, and
  Stuart]{iglesias2013ensemble-kalman}
Iglesias MA, Law KJH, Stuart AM.
\newblock {Ensemble Kalman} methods for inverse problems.
\newblock \emph{Inverse Problems}, 29\penalty0 (4):\penalty0 045001 (20pp),
  2013.

\bibitem[Evensen(2003)]{evensen2003ensemble}
Evensen G.
\newblock The ensemble {Kalman} filter: theoretical formulation and practical
  implementation.
\newblock \emph{Ocean Dynamics}, 53\penalty0 (4):\penalty0 343--367, 2003.

\bibitem[Evensen(2009)]{evensen2009data}
Evensen G.
\newblock \emph{Data Assimilation:The Ensemble {Kalman} Filter}.
\newblock Springer, 2009.

\bibitem[Ernst et~al.(2015)Ernst, Sprungk, and Starkloff]{ernst2015analysis}
Ernst OG, Sprungk B, Starkloff HJ.
\newblock Analysis of the ensemble and polynomial chaos {Kalman} filters in
  {Bayesian} inverse problems.
\newblock \emph{SIAM/ASA Journal on Uncertainty Quantification}, 3\penalty0
  (1):\penalty0 823--851, 2015.

\bibitem[Schillings and Stuart(2017)]{schillings2017analysis}
Schillings C, Stuart AM.
\newblock Analysis of the ensemble {Kalman} filter for inverse problems.
\newblock \emph{SIAM Journal on Numerical Analysis}, 55\penalty0 (3):\penalty0
  1264--1290, 2017.

\bibitem[Jones and Launder(1972)]{jones1972prediction}
Jones WP, Launder BE.
\newblock The prediction of laminarization with a two-equation model of
  turbulence.
\newblock \emph{International journal of heat and mass transfer}, 15\penalty0
  (2):\penalty0 301--314, 1972.

\bibitem[Speziale et~al.(1992)Speziale, Abid, and
  Anderson]{speziale1992critical}
Speziale CG, Abid R, Anderson EC.
\newblock Critical evaluation of two-equation models for near-wall turbulence.
\newblock \emph{AIAA Journal}, 30\penalty0 (2):\penalty0 324--331, 1992.

\bibitem[Mohamed and Larue(1990)]{mohamed1990decay}
Mohamed MS, Larue JC.
\newblock The decay power law in grid-generated turbulence.
\newblock \emph{Journal of Fluid Mechanics}, 219:\penalty0 195--214, 1990.

\bibitem[Kim et~al.(1987)Kim, Moin, and Moser]{kim1987turbulence}
Kim J, Moin P, Moser R.
\newblock Turbulence statistics in fully developed channel flow at low
  {Reynolds} number.
\newblock \emph{Journal of Fluid Mechanics}, 177:\penalty0 133--166, 1987.

\bibitem[Pope(2000)]{pope2000turbulent}
Pope SB.
\newblock \emph{Turbulent Flows}.
\newblock Cambridge University Press, Cambridge, 2000.

\bibitem[Zanoun et~al.(2003)Zanoun, Durst, and Nagib]{zanoun2003evaluating}
Zanoun ES, Durst F, Nagib H.
\newblock Evaluating the law of the wall in two-dimensional fully developed
  turbulent channel flows.
\newblock \emph{Physics of Fluids}, 15\penalty0 (10):\penalty0 3079--3089,
  2003.

\bibitem[Tavoularis and Karnik(1989)]{tavoularis1989further}
Tavoularis S, Karnik U.
\newblock Further experiments on the evolution of turbulent stresses and scales
  in uniformly sheared turbulence.
\newblock \emph{Journal of Fluid Mechanics}, 204:\penalty0 457--478, 1989.

\bibitem[Durbin(2017)]{durbin2017personal}
Durbin P.
\newblock Personal communication, 2017.

\bibitem[Menter(1994)]{menter1994two-equation}
Menter FR.
\newblock Two-equation eddy-viscosity turbulence models for engineering
  applications.
\newblock \emph{AIAA Journal}, 32\penalty0 (8):\penalty0 1598--1605, 1994.

\bibitem[Schaefer et~al.(2017)Schaefer, Cary, Mani, and
  Spalart]{schaefer2017uncertainty}
Schaefer JA, Cary AW, Mani M, Spalart PR.
\newblock {Uncertainty Quantification and Sensitivity Analysis of SA Turbulence
  Model Coefficients in Two and Three Dimensions}.
\newblock AIAA Paper 2017-1710, 2017.

\bibitem[Spalart and Allmaras(1992)]{spalart1992one-equation}
Spalart PR, Allmaras SR.
\newblock A one-equation turbulence model for aerodynamic flows.
\newblock AIAA Paper 1992-439, 1992.

\bibitem[Turgeon et~al.(2004)Turgeon, Pelletier, and
  Borggaard]{turgeon2004general}
Turgeon {\'E}, Pelletier D, Borggaard J.
\newblock A general continuous sensitivity equation formulation for the
  $k$--$\varepsilon$ model of turbulence.
\newblock \emph{International Journal of Computational Fluid Dynamics},
  18\penalty0 (1):\penalty0 29--46, 2004.

\bibitem[Launder and Sharma(1974)]{launder1974application}
Launder BE, Sharma BI.
\newblock Application of the energy-dissipation model of turbulence to the
  calculation of flow near a spinning disc.
\newblock \emph{Letters in Heat and Mass Transfer}, 1\penalty0 (2):\penalty0
  131--138, 1974.

\bibitem[Xiu and Karniadakis(2002)]{xiu2002wiener-askey}
Xiu D, Karniadakis GE.
\newblock The {W}iener--{A}skey polynomial chaos for stochastic differential
  equations.
\newblock \emph{SIAM journal on scientific computing}, 24\penalty0
  (2):\penalty0 619--644, 2002.

\bibitem[Lefantzi et~al.(2015)Lefantzi, Ray, Arunajatesan, and
  Dechant]{lefantzi2015estimation}
Lefantzi S, Ray J, Arunajatesan S, Dechant L.
\newblock Estimation of $k$--$\varepsilon$ parameters using surrogate models
  and jet-in-crossflow data.
\newblock Technical report, Sandia National Laboratories, Livermore, CA, USA,
  2015.

\bibitem[Ray et~al.(2018{\natexlab{b}})Ray, Dechant, Lefantzi, Ling, and
  Arunajatesan]{ray2018robust}
Ray J, Dechant L, Lefantzi S, Ling J, Arunajatesan S.
\newblock Robust {B}ayesian calibration of k--$\varepsilon$ model for
  compressible jet-in-crossflow simulations.
\newblock \emph{AIAA Journal}, 2018{\natexlab{b}}.
\newblock In press.

\bibitem[Evensen(1994)]{evensen1994sequential}
Evensen G.
\newblock Sequential data assimilation with a nonlinear quasi-geostrophic model
  using {Monte Carlo} methods to forecast error statistics.
\newblock \emph{Journal of Geophysical Research: Oceans}, 99\penalty0
  (C5):\penalty0 10143--10162, 1994.

\bibitem[Kline et~al.(1969)Kline, Coles, and Hirst]{kline1969computation}
Kline SJ, Coles DE, Hirst E.
\newblock \emph{Computation of turbulent boundary layers -- 1968
  {AFOSR-IFP-Stanford} Conference: proceedings held at Stanford University,
  August 18-25, 1968}.
\newblock Thermosciences Division, Stanford University, 1969.

\bibitem[Brynjarsd{\'o}ttir and O'Hagan(2014)]{brynjarsdottir2013learning}
Brynjarsd{\'o}ttir J, O'Hagan A.
\newblock Learning about physical parameters: The importance of model
  discrepancy.
\newblock \emph{Inverse Problems}, 30:\penalty0 114007, 2014.

\bibitem[Nitschke et~al.(2017)Nitschke, Cinnella, Lucor, and
  Chassaing]{nitschke2018model}
Nitschke C, Cinnella P, Lucor D, Chassaing JC.
\newblock Model-form and predictive uncertainty quantification in linear
  aeroelasticity.
\newblock \emph{Journal of Fluids and Structures}, 73:\penalty0 137--161, 2017.

\bibitem[Oliver and Moser(2011)]{oliver2011bayesian}
Oliver TA, Moser RD.
\newblock Bayesian uncertainty quantification applied to {RANS} turbulence
  models.
\newblock \emph{Journal of Physics: Conference Series}, 318:\penalty0 042032,
  2011.

\bibitem[Prudencio and Cheung(2012)]{prudencio2012parallel}
Prudencio E, Cheung SH.
\newblock Parallel adaptive multilevel sampling algorithms for the bayesian
  analysis of mathematical models.
\newblock \emph{International Journal for Uncertainty Quantification},
  2\penalty0 (3):\penalty0 215--237, 2012.

\bibitem[Chien(1982)]{chien1982predictions}
Chien KY.
\newblock Predictions of channel and boundary-layer flows with a
  low-{Reynolds}-number turbulence model.
\newblock \emph{AIAA Journal}, 20\penalty0 (1):\penalty0 33--38, 1982.

\bibitem[Diomede et~al.(2008)Diomede, Davolio, Marsigli, Miglietta, Moscatello,
  Papetti, Paccagnella, Buzzi, and Malguzzi]{diomede2008discharge}
Diomede T, Davolio S, Marsigli C, Miglietta M, Moscatello A, Papetti P,
  Paccagnella T, Buzzi A, Malguzzi P.
\newblock Discharge prediction based on multi-model precipitation forecasts.
\newblock \emph{Meteorology and Atmospheric Physics}, 101\penalty0
  (3-4):\penalty0 245--265, 2008.

\bibitem[Duan et~al.(2007)Duan, Ajami, Gao, and
  Sorooshian]{duan2007multi-model}
Duan Q, Ajami NK, Gao X, Sorooshian S.
\newblock Multi-model ensemble hydrologic prediction using {Bayesian} model
  averaging.
\newblock \emph{Advances in Water Resources}, 30\penalty0 (5):\penalty0
  1371--1386, 2007.

\bibitem[Tebaldi and Knutti(2007)]{tebaldi2007use}
Tebaldi C, Knutti R.
\newblock The use of the multi-model ensemble in probabilistic climate
  projections.
\newblock \emph{Philosophical Transactions of the Royal Society of London A:
  Mathematical, Physical and Engineering Sciences}, 365\penalty0
  (1857):\penalty0 2053--2075, 2007.

\bibitem[Hoeting et~al.(1999)Hoeting, Madigan, Raftery, and
  Volinsky]{hoeting1999bayesian}
Hoeting JA, Madigan D, Raftery AE, Volinsky CT.
\newblock Bayesian model averaging: a tutorial.
\newblock \emph{Statistical Science}, 14\penalty0 (4):\penalty0 382--401, 1999.

\bibitem[Soize(2005)]{soize2005comprehensive}
Soize C.
\newblock A comprehensive overview of a non-parametric probabilistic approach
  of model uncertainties for predictive models in structural dynamics.
\newblock \emph{Journal of Sound and Vibration}, 288\penalty0 (3):\penalty0
  623--652, 2005.

\bibitem[Wu et~al.(2016{\natexlab{b}})Wu, Wang, and Xiao]{wu2016quantifying}
Wu~JL, Wang JX, Xiao H.
\newblock Quantifying model form uncertainty in {RANS} simulation of wing-body
  junction flow.
\newblock Unpublished manuscript, arXiv:1605.05962, 2016{\natexlab{b}}.

\bibitem[Devenport and Simpson(1990)]{devenport1990time-dependent}
Devenport WJ, Simpson RL.
\newblock Time-dependent and time-averaged turbulence structure near the nose
  of a wing-body junction.
\newblock \emph{Journal of Fluid Mechanics}, 210:\penalty0 23--55, 1990.

\bibitem[Ling and Templeton(2015)]{ling2015evaluation}
Ling J, Templeton J.
\newblock Evaluation of machine learning algorithms for prediction of regions
  of high {Reynolds averaged Navier Stokes} uncertainty.
\newblock \emph{Physics of Fluids (1994-present)}, 27\penalty0 (8):\penalty0
  085103, 2015.

\bibitem[Singh et~al.(2017)Singh, Medida, and
  Duraisamy]{singh2017machine-learning-augmented}
Singh AP, Medida S, Duraisamy K.
\newblock Machine-learning-augmented predictive modeling of turbulent separated
  flows over airfoils.
\newblock \emph{AIAA Journal}, 55\penalty0 (7):\penalty0 2215--2227, 2017.

\bibitem[Wang et~al.(2017{\natexlab{a}})Wang, Wu, and
  Xiao]{wang2017physics-informed}
Wang JX, Wu~JL, Xiao H.
\newblock Physics-informed machine learning approach for reconstructing
  {Reynolds} stress modeling discrepancies based on {DNS} data.
\newblock \emph{Physical Review Fluids}, 2\penalty0 (3):\penalty0 034603,
  2017{\natexlab{a}}.

\bibitem[Wu et~al.(2018{\natexlab{a}})Wu, Xiao, and
  Paterson]{wu2018data-driven}
Wu~JL, Xiao H, Paterson E.
\newblock Physics-informed machine learning approach for augmenting turbulence
  models: A comprehensive framework.
\newblock \emph{Physical Review Fluids}, 3\penalty0 (7):\penalty0 074602,
  2018{\natexlab{a}}.

\bibitem[{Hayek} et~al.(2018){Hayek}, {Wang}, and
  {Laskowski}]{hayek2018adjoint-based}
{Hayek} ME, {Wang} Q, {Laskowski} GM.
\newblock Adjoint-based optimization of {RANS} eddy viscosity model for
  {U}-bend channel flow.
\newblock In \emph{2018 AIAA Aerospace Sciences Meeting}, 2018.

\bibitem[Zhang and Fu(2018)]{zhang2018efficient}
Zhang J, Fu~S.
\newblock An efficient {Bayesian} uncertainty quantification approach with
  application to k-$\omega$-$\gamma$ transition modeling.
\newblock \emph{Computers \& Fluids}, 161:\penalty0 211--224, 2018.

\bibitem[Oliver et~al.(2015)Oliver, Terejanu, Simmons, and
  Moser]{oliver2015validating}
Oliver TA, Terejanu G, Simmons CS, Moser RD.
\newblock Validating predictions of unobserved quantities.
\newblock \emph{Computer Methods in Applied Mechanics and Engineering},
  283:\penalty0 1310--1335, 2015.

\bibitem[Oliver and Moser(2013)]{oliver2013representing}
Oliver TA, Moser RD.
\newblock Representing uncertainty due to inaccurate models.
\newblock Unpublished presentation, 2013.

\bibitem[Simonsen and Krogstad(2005)]{simonsen2005turbulent}
Simonsen A, Krogstad P{\AA}.
\newblock Turbulent stress invariant analysis: {Clarification} of existing
  terminology.
\newblock \emph{Physics of Fluids}, 17\penalty0 (8):\penalty0 088103, 2005.

\bibitem[Lumley(1978)]{lumley1978computational}
Lumley JL.
\newblock Computational modeling of turbulent flows.
\newblock \emph{Advances in applied mechanics}, 18\penalty0 (123):\penalty0
  213, 1978.

\bibitem[Banerjee et~al.(2007)Banerjee, Krahl, Durst, and
  Zenger]{banerjee2007presentation}
Banerjee S, Krahl R, Durst F, Zenger C.
\newblock Presentation of anisotropy properties of turbulence, invariants
  versus eigenvalue approaches.
\newblock \emph{Journal of Turbulence}, 8\penalty0 (32):\penalty0 N32, 2007.

\bibitem[Emory and Iaccarino(2014)]{emory2014componentality-based}
Emory M, Iaccarino G.
\newblock Componentality-based wall-blocking for {RANS} models.
\newblock Technical report, Proceedings of Summer Research Program, Center of
  Turbulence Research, Stanford University, Stanford, CA, USA, 2014.

\bibitem[Emory(2014)]{emory2014estimating}
Emory MA.
\newblock \emph{Estimating model-form uncertainty in {Reynolds}-averaged
  {Navier}-{Stokes} closures}.
\newblock PhD thesis, Stanford University, 2014.

\bibitem[Schumann(1977)]{schumann1977realizability}
Schumann U.
\newblock Realizability of {Reynolds}-stress turbulence models.
\newblock \emph{Physics of Fluids (1958-1988)}, 20\penalty0 (5):\penalty0
  721--725, 1977.

\bibitem[Pope(1985)]{pope1985pdf}
Pope S.
\newblock {PDF} methods for turbulent reactive flows.
\newblock \emph{Progress in Energy and Combustion Science}, 11\penalty0
  (2):\penalty0 119--192, 1985.

\bibitem[Speziale et~al.(1994)Speziale, Abid, and Durbin]{speziale1994on}
Speziale CG, Abid R, Durbin PA.
\newblock On the realizability of {Reynolds} stress turbulence closures.
\newblock \emph{Journal of Scientific Computing}, 9\penalty0 (4):\penalty0
  369--403, 1994.

\bibitem[Thompson et~al.(2016{\natexlab{a}})Thompson, Sampaio, Edeling, Mishra,
  and Iaccarino]{thompson2016strategy}
Thompson R, Sampaio L, Edeling W, Mishra AA, Iaccarino G.
\newblock A strategy for the eigenvector perturbations of the {Reynolds} stress
  tensor in the context of uncertainty quantification.
\newblock In \emph{Proceedings of the Summer Program, Center for Turbulence
  Research}, page~10, 2016{\natexlab{a}}.

\bibitem[Gorl{\'e} et~al.(2012)Gorl{\'e}, Emory, Larsson, and
  Iaccarino]{gorle2012epistemic}
Gorl{\'e} C, Emory M, Larsson J, Iaccarino G.
\newblock Epistemic uncertainty quantification for {RANS} modeling of the flow
  over a wavy wall.
\newblock \emph{Center for Turbulence Research, Annual Research Briefs}, 2012.

\bibitem[Wang et~al.(2016{\natexlab{b}})Wang, Sun, and
  Xiao]{wang2016quantification}
Wang JX, Sun R, Xiao H.
\newblock Quantification of uncertainties in turbulence modeling: A comparison
  of physics-based and random matrix theoretic approaches.
\newblock \emph{International Journal of Heat and Fluid Flows}, 62:\penalty0
  577--592, 2016{\natexlab{b}}.

\bibitem[Goldstein(1980)]{goldstein1980euler}
Goldstein H.
\newblock The {Euler} angles.
\newblock \emph{Classical Mechanics}, pages 143--148, 1980.

\bibitem[Horn(1987)]{horn1987closed-form}
Horn BKP.
\newblock Closed-form solution of absolute orientation using unit quaternions.
\newblock \emph{Journal of Optical Society of America}, 4\penalty0
  (4):\penalty0 629--642, 1987.

\bibitem[Wu et~al.(2018{\natexlab{b}})Wu, Sun, Laizet, and
  Xiao]{wu2017representation}
Wu~JL, Sun R, Laizet S, Xiao H.
\newblock Representation of {Reynolds} stress perturbations with application in
  machine-learning-assisted turbulence modeling.
\newblock \emph{Computer Methods in Applied Mechanics and Engineering},
  2018{\natexlab{b}}.
\newblock In Press. arXiv:1709.05683.

\bibitem[Gorl{\'e} et~al.(2014)Gorl{\'e}, Larsson, Emory, and
  Iaccarino]{gorle2014deviation}
Gorl{\'e} C, Larsson J, Emory M, Iaccarino G.
\newblock The deviation from parallel shear flow as an indicator of linear
  eddy-viscosity model inaccuracy.
\newblock \emph{Physics of Fluids (1994-present)}, 26\penalty0 (5):\penalty0
  051702, 2014.

\bibitem[Ling et~al.(2017)Ling, Ruiz, Lacaze, and
  Oefelein]{ling2017uncertainty}
Ling J, Ruiz A, Lacaze G, Oefelein J.
\newblock Uncertainty analysis and data-driven model advances for a
  jet-in-crossflow.
\newblock \emph{Journal of Turbomachinery}, 139\penalty0 (2):\penalty0 021008,
  2017.

\bibitem[Tracey et~al.(2013)Tracey, Duraisamy, and
  Alonso]{tracey2013application}
Tracey B, Duraisamy K, Alonso J.
\newblock Application of supervised learning to quantify uncertainties in
  turbulence and combustion modeling.
\newblock In \emph{51st AIAA Aerospace Sciences Meeting}, 2013.
\newblock {Dallas}, TX, paper 2013-0259.

\bibitem[Ling et~al.(2016{\natexlab{b}})Ling, Jones, and
  Templeton]{ling2016machine}
Ling J, Jones R, Templeton J.
\newblock Machine learning strategies for systems with invariance properties.
\newblock \emph{Journal of Computational Physics}, 318:\penalty0 22--35,
  2016{\natexlab{b}}.

\bibitem[Wu et~al.(2018{\natexlab{c}})Wu, Xiao, Sun, and Wang]{wu2018rans}
Wu~JL, Xiao H, Sun R, Wang Q.
\newblock {RANS} equations with {Reynolds} stress closure can be
  ill-conditioned.
\newblock arXiv:1803.05581, 2018{\natexlab{c}}.

\bibitem[Liu et~al.(2018)Liu, Dinh, Sato, and Niceno]{liu2018data}
Liu Y, Dinh N, Sato Y, Niceno B.
\newblock Data-driven modeling for boiling heat transfer: using deep neural
  networks and high-fidelity simulation results.
\newblock \emph{Applied Thermal Engineering}, 144:\penalty0 305--320, 2018.

\bibitem[Chang and Dinh(2018{\natexlab{a}})]{chang2018reynolds}
Chang CW, Dinh NT.
\newblock Reynolds-averaged turbulence modeling using type {I} and type {II}
  machine learning frameworks with deep learning.
\newblock \emph{arXiv:1804.01065}, 2018{\natexlab{a}}.

\bibitem[Chang and Dinh(2018{\natexlab{b}})]{chang2018classification}
Chang CW, Dinh NT.
\newblock Classification of machine learning frameworks for data-driven thermal
  fluid models.
\newblock \emph{arXiv:1801.06621}, 2018{\natexlab{b}}.

\bibitem[Hanna et~al.(2017)Hanna, Dinh, Youngblood, and
  Bolotnov]{hanna2017coarse}
Hanna BN, Dinh NT, Youngblood RW, Bolotnov IA.
\newblock Coarse-grid computational fluid dynamic ({CG-CFD}) error prediction
  using machine learning.
\newblock \emph{arXiv:1710.09105}, 2017.

\bibitem[Wang et~al.(2018)Wang, Huang, Duan, and Xiao]{wang2018predicting}
Wang JX, Huang J, Duan L, Xiao H.
\newblock Predicting {R}eynolds stresses in high-{M}ach-number turbulent
  boundary layers with physics-informed machine learning.
\newblock \emph{arXiv:1808.07752}, 2018.

\bibitem[Wu et~al.(2018{\natexlab{d}})Wu, Str{\"o}fer, and
  Xiao]{wu2018pde-informed}
Wu~JL, Str{\"o}fer CM, Xiao H.
\newblock {PDE}-informed construction of covariance kernel in uncertainty
  quantification of random fields.
\newblock In preparation, 2018{\natexlab{d}}.

\bibitem[Najm(2009)]{najm2009uncertainty}
Najm HN.
\newblock Uncertainty quantification and polynomial chaos techniques in
  computational fluid dynamics.
\newblock \emph{Annual Review of Fluid Mechanics}, 41:\penalty0 35--52, 2009.

\bibitem[Tatang(1995)]{tatang1995direct}
Tatang MA.
\newblock \emph{Direct incorporation of uncertainty in chemical and
  environmental engineering systems}.
\newblock PhD thesis, Massachusetts Institute of Technology, 1995.

\bibitem[Bungartz and Griebel(2004)]{bungartz2004sparse}
Bungartz HJ, Griebel M.
\newblock Sparse grids.
\newblock \emph{Acta numerica}, 13:\penalty0 147--269, 2004.

\bibitem[Germano et~al.(1991)Germano, Piomelli, Moin, and
  Cabot]{germano1991dynamic}
Germano M, Piomelli U, Moin P, Cabot WH.
\newblock A dynamic subgrid‐scale eddy viscosity model.
\newblock \emph{Physics of Fluids A: Fluid Dynamics}, 3\penalty0 (7):\penalty0
  1760--1765, 1991.
\newblock \doi{10.1063/1.857955}.

\bibitem[Meyers and Sagaut(2006)]{meyers2006on}
Meyers J, Sagaut P.
\newblock On the model coefficients for the standard and the variational
  multi-scale {Smagorinsky} model.
\newblock \emph{Journal of Fluid Mechanics}, 569:\penalty0 287--319, 2006.

\bibitem[Lucor et~al.(2007)Lucor, Meyers, and Sagaut]{lucor2007sensitivity}
Lucor D, Meyers J, Sagaut P.
\newblock Sensitivity analysis of large-eddy simulations to subgrid-scale-model
  parametric uncertainty using polynomial chaos.
\newblock \emph{Journal of Fluid Mechanics}, 585:\penalty0 255--279, 2007.

\bibitem[Meldi et~al.(2011)Meldi, Lucor, and Sagaut]{meldi2011is}
Meldi M, Lucor D, Sagaut P.
\newblock Is the {Smagorinsky} coefficient sensitive to uncertainty in the form
  of the energy spectrum?
\newblock \emph{Physics of Fluids}, 23\penalty0 (12):\penalty0 125109, 2011.

\bibitem[Khalil et~al.(2015)Khalil, Lacaze, Oefelein, and
  Najm]{khalil2015uncertainty}
Khalil M, Lacaze G, Oefelein JC, Najm HN.
\newblock Uncertainty quantification in {LES} of a turbulent bluff-body
  stabilized flame.
\newblock \emph{Proceedings of the Combustion Institute}, 35\penalty0
  (2):\penalty0 1147--1156, 2015.

\bibitem[Safta et~al.(2017)Safta, Blaylock, Templeton, Domino, Sargsyan, and
  Najm]{safta2017uncertainty}
Safta C, Blaylock M, Templeton J, Domino S, Sargsyan K, Najm H.
\newblock Uncertainty quantification in {LES} of channel flow.
\newblock \emph{International Journal for Numerical Methods in Fluids},
  83\penalty0 (4):\penalty0 376--401, 2017.

\bibitem[Yoshizawa(1993)]{yoshizawa1993bridging}
Yoshizawa A.
\newblock Bridging between eddy-viscosity-type and second-order turbulence
  models through a two-scale turbulence theory.
\newblock \emph{Physical Review E}, 48\penalty0 (1):\penalty0 273, 1993.

\bibitem[Templeton et~al.(2015)Templeton, Blaylock, Domino, Hewson, Kumar,
  Ling, Najm, Ruiz, Safta, Sargsyan, et~al.]{templeton2015calibration}
Templeton JA, Blaylock ML, Domino SP, Hewson JC, Kumar PR, Ling J, Najm HN,
  Ruiz A, Safta C, Sargsyan K, others .
\newblock Calibration and forward uncertainty propagation for large-eddy
  simulations of engineering flows.
\newblock Technical report, Sandia National Laboratories (SNL-CA), Livermore,
  CA, 2015.

\bibitem[Li et~al.(2008)Li, Perlman, Wan, Yang, Meneveau, Burns, Chen, Szalay,
  and Eyink]{li2008public}
Li~Y, Perlman E, Wan M, Yang Y, Meneveau C, Burns R, Chen S, Szalay A, Eyink G.
\newblock A public turbulence database cluster and applications to study
  {Lagrangian} evolution of velocity increments in turbulence.
\newblock \emph{Journal of Turbulence}, 9:\penalty0 N31, 2008.

\bibitem[Tran et~al.(2016)Tran, Webster, and Zhang]{tran2016sparse}
Tran H, Webster CG, Zhang G.
\newblock A sparse grid method for {Bayesian} uncertainty quantification with
  application to large eddy simulation turbulence models.
\newblock In \emph{Sparse Grids and Applications-Stuttgart 2014}, pages
  291--313. Springer, 2016.

\bibitem[Jofre et~al.(2018)Jofre, Domino, and Iaccarino]{jofre2018framework}
Jofre L, Domino SP, Iaccarino G.
\newblock A framework for characterizing structural uncertainty in large-eddy
  simulation closures.
\newblock \emph{Flow, Turbulence and Combustion}, 100\penalty0 (2):\penalty0
  341--363, 2018.

\bibitem[Gorl{\'e} and Iaccarino(2013)]{gorle2013framework}
Gorl{\'e} C, Iaccarino G.
\newblock A framework for epistemic uncertainty quantification of turbulent
  scalar flux models for {Reynolds}-averaged {Navier}-{Stokes} simulations.
\newblock \emph{Physics of Fluids}, 25\penalty0 (5):\penalty0 055105, 2013.

\bibitem[Congedo et~al.(2013)Congedo, Duprat, Balarac, and
  Corre]{congedo2013numerical}
Congedo PM, Duprat C, Balarac G, Corre C.
\newblock Numerical prediction of turbulent flows using {Reynolds-averaged
  Navier--Stokes} and large-eddy simulation with uncertain inflow conditions.
\newblock \emph{International Journal for Numerical Methods in Fluids},
  72\penalty0 (3):\penalty0 341--358, 2013.

\bibitem[Carnevale et~al.(2013)Carnevale, Montomoli, D’Ammaro, Salvadori, and
  Martelli]{carnevale2013uncertainty}
Carnevale M, Montomoli F, D’Ammaro A, Salvadori S, Martelli F.
\newblock Uncertainty quantification: A stochastic method for heat transfer
  prediction using {LES}.
\newblock \emph{Journal of Turbomachinery}, 135\penalty0 (5):\penalty0 051021,
  2013.

\bibitem[Ghosal(1996)]{ghosal1996analysis}
Ghosal S.
\newblock An analysis of numerical errors in large-eddy simulations of
  turbulence.
\newblock \emph{Journal of Computational Physics}, 125\penalty0 (1):\penalty0
  187--206, 1996.

\bibitem[Vreman et~al.(1996)Vreman, Geurts, and Kuerten]{vreman1996comparision}
Vreman B, Geurts B, Kuerten H.
\newblock Comparision of numerical schemes in large-eddy simulation of the
  temporal mixing layer.
\newblock \emph{International Journal for Numerical Methods in Fluids},
  22\penalty0 (4):\penalty0 297--311, 1996.

\bibitem[Kravchenko and Moin(1997)]{kravchenko1997on}
Kravchenko A, Moin P.
\newblock On the effect of numerical errors in large eddy simulations of
  turbulent flows.
\newblock \emph{Journal of Computational Physics}, 131\penalty0 (2):\penalty0
  310--322, 1997.

\bibitem[Meldi et~al.(2012)Meldi, Salvetti, and
  Sagaut]{meldi2012quantification}
Meldi M, Salvetti MV, Sagaut P.
\newblock Quantification of errors in large-eddy simulations of a spatially
  evolving mixing layer using polynomial chaos.
\newblock \emph{Physics of Fluids}, 24\penalty0 (3):\penalty0 035101, 2012.

\bibitem[Mariotti et~al.(2016{\natexlab{b}})Mariotti, Siconolfi, and
  Salvetti]{mariotti2016stochastic}
Mariotti A, Siconolfi L, Salvetti M.
\newblock Stochastic sensitivity analysis of large-eddy simulation predictions
  of the flow around a 5:1 rectangular cylinder.
\newblock \emph{European Journal of Mechanics-B/Fluids}, 62:\penalty0 149--165,
  2016{\natexlab{b}}.

\bibitem[Ma et~al.(2015)Ma, Lu, and Tryggvason]{ma2015using}
Ma~M, Lu~J, Tryggvason G.
\newblock Using statistical learning to close two-fluid multiphase flow
  equations for a simple bubbly system.
\newblock \emph{Physics of Fluids}, 27\penalty0 (9):\penalty0 092101, 2015.

\bibitem[Ma et~al.(2016)Ma, Lu, and Tryggvason]{ma2016using}
Ma~M, Lu~J, Tryggvason G.
\newblock Using statistical learning to close two-fluid multiphase flow
  equations for bubbly flows in vertical channels.
\newblock \emph{International Journal of Multiphase Flow}, 85:\penalty0
  336--347, 2016.

\bibitem[Randall et~al.(2003)Randall, Khairoutdinov, Arakawa, and
  Grabowski]{randall2003breaking}
Randall D, Khairoutdinov M, Arakawa A, Grabowski W.
\newblock Breaking the cloud parameterization deadlock.
\newblock \emph{Bulletin of the American Meteorological Society}, 84\penalty0
  (11):\penalty0 1547--1564, 2003.

\bibitem[Tiedtke(1993)]{tiedtke1993representation}
Tiedtke M.
\newblock Representation of clouds in large-scale models.
\newblock \emph{Monthly Weather Review}, 121\penalty0 (11):\penalty0
  3040--3061, 1993.

\bibitem[Chevallier et~al.(1998)Chevallier, Ch{\'e}ruy, Scott, and
  Ch{\'e}din]{chevallier1998neural}
Chevallier F, Ch{\'e}ruy F, Scott N, Ch{\'e}din A.
\newblock A neural network approach for a fast and accurate computation of a
  longwave radiative budget.
\newblock \emph{Journal of Applied Meteorology}, 37\penalty0 (11):\penalty0
  1385--1397, 1998.

\bibitem[Mansour et~al.(1988)Mansour, Kim, and
  Moin]{mansour1988reynolds-stress}
Mansour NN, Kim J, Moin P.
\newblock Reynolds-stress and dissipation-rate budgets in a turbulent channel
  flow.
\newblock \emph{Journal of Fluid Mechanics}, 194:\penalty0 15--44, 1988.

\bibitem[Jeyapaul et~al.(2014)Jeyapaul, Coleman, and
  Rumsey]{jeyapaul2014assessment}
Jeyapaul E, Coleman GN, Rumsey CL.
\newblock Assessment of higher-order {RANS} closures in a decelerated planar
  wall-bounded turbulent flow.
\newblock In \emph{44th AIAA Fluid Dynamics Conference}, page 2088, 2014.

\bibitem[Hoyas and Jim{\'e}nez(2008)]{hoyas2008reynolds}
Hoyas S, Jim{\'e}nez J.
\newblock Reynolds number effects on the {Reynolds}-stress budgets in turbulent
  channels.
\newblock \emph{Physics of Fluids}, 20\penalty0 (10):\penalty0 101511, 2008.

\bibitem[Oliver et~al.(2014)Oliver, Malaya, Ulerich, and
  Moser]{oliver2014estimating}
Oliver TA, Malaya N, Ulerich R, Moser RD.
\newblock Estimating uncertainties in statistics computed from direct numerical
  simulation.
\newblock \emph{Physics of Fluids}, 26\penalty0 (3):\penalty0 035101, 2014.

\bibitem[Thompson et~al.(2016{\natexlab{b}})Thompson, Sampaio,
  de~Bragan{\c{c}}a~Alves, Thais, and Mompean]{thompson2016methodology}
Thompson RL, Sampaio LEB, de~Bragan{\c{c}}a~Alves FA, Thais L, Mompean G.
\newblock A methodology to evaluate statistical errors in {DNS} data of plane
  channel flows.
\newblock \emph{Computers \& Fluids}, 130:\penalty0 1--7, 2016{\natexlab{b}}.

\bibitem[Poroseva et~al.(2016)Poroseva, Colmenares~F, and
  Murman]{poroseva2016on}
Poroseva SV, Colmenares~F JD, Murman SM.
\newblock On the accuracy of {RANS} simulations with {DNS} data.
\newblock \emph{Physics of Fluids}, 28\penalty0 (11):\penalty0 115102, 2016.

\bibitem[Wang et~al.(2017{\natexlab{b}})Wang, Wu, Ling, Iaccarino, and
  Xiao]{wang2017comprehensive}
Wang JX, Wu~JL, Ling J, Iaccarino G, Xiao H.
\newblock A comprehensive physics-informed machine learning framework for
  predictive turbulence modeling.
\newblock arXiv:1701.07102, 2017{\natexlab{b}}.

\bibitem[Ko et~al.(2008)Ko, Lucor, and Sagaut]{ko2008sensitivity}
Ko~J, Lucor D, Sagaut P.
\newblock Sensitivity of two-dimensional spatially developing mixing layers
  with respect to uncertain inflow conditions.
\newblock \emph{Physics of Fluids}, 20\penalty0 (7):\penalty0 077102, 2008.

\end{thebibliography}

\appendix

\section{Algorithms in uncertainty quantification}

\subsection{Plain Monte Carlo sampling}
\label{app:mc}

The algorithms for plain Monte Carlo sampling is rather straightforward.  Given the probability distribution $p(\bm{\theta})$ of the model parameters, Monte Carlo simulations can be used to obtain the distributions of the output. Specifically, the procedure of uncertainty propagation based on plain Monte Carlo simulation is as follows:
\begin{enumerate}[(1)]
    \item \textbf{Sampling.} Draw a number of samples~$\{\theta_1, \theta_2, \cdots, \theta_n \}$ from the specified prior probability distribution~$p(\bm{\theta})$.
    \item \textbf{Propagation.} For each of the sample, the model is evaluated to obtain the outputs~$\{y_1, y_2, \cdots, y_N\}$.
    \item \textbf{Aggregation.} The distribution of the QoI is estimated from the propagated samples.
    \end{enumerate}
This procedure is illustrated pictorially in Fig.~\ref{fig:forward-backward}a.

\subsection{Exact Bayesian inference with Markov chain Monte Carlo sampling}
\label{app:mcmc}

Much like the ergodicity assumption for the ensemble averaging to obtain the RANS equations, the MCMC sampling requires the ergodicity assumption. That is, any set within the state space can be reached from any other set with nonzero probability within finite steps. The MCMC procedure with Metropolis--Hastings sampling algorithm is as follows:
\begin{enumerate} [(1)]
\item Initialize the state $\bm{\theta}^{(0)}$.
\item Based on the current state $\bm{z}^{(i)}$, make a proposal of next state (e.g., a random walk), i.e., sample $\bm{z}^\star \sim q(\bm{z}^\star | \bm{z}^{(i)})$.
\item Evaluate the posterior density $p(\bm{z}^\star)$ and the ratio $\chi = p(\bm{z}^\star)/p(\bm{z}^{(i)})$.
\item Accept the proposal (i.e., move to $z^\star$) if $\chi \ge 1$; otherwise accept the proposal with probability~$\chi$.
\item Repeat  steps 2--4
\end{enumerate}
This procedure is illustrated in Fig.~\ref{fig:mcmc}.  Intuitively, the sampler always accepts to go to a more likely state, which increases samples in high posterior probability regions.  On the other hand, it also allows for the possibility of going to less likely states which allows for exploring the tails (rare events regions in the state space) and increase mixing (traveling back and forth in different regions).

\subsection{Approximate Bayesian inference with iterative Ensemble Kalman method}
\label{app:enkf}

In the example below, we assume the velocity at some locations is the observed physical state for notation simplicity. The augmented system state $\bm{z}(\bm{x})$ is written as a vector formed by stacking the unknown parameters and the physical states $\boldsymbol{\xi}(\bm{x})$:
\begin{equation}
\bm{z} = [\xi_1, \cdots, \xi_n;  \bm{\theta}]^\top,
\end{equation}
in which $\top$ indicates vector transpose, and  $\bm{\theta} = [\theta_1, \theta_2, \cdots, \theta_r] $ is a vector of $r$ parameters.

Given the prior distributions for parameters ($\bm{\theta}$) to be inferred and the covariance matrix~$\mathsf{R}$ of the  observations $y^{obs}$, the EnKF based inversion algorithm proceeds as follows:
\begin{enumerate}
\item \textbf{Sampling of prior distribution.}

  From the prior distributions of the parameters, $M$ samples are drawn. Each sample consists of a combination of values for $\bm{\theta}$.

      \item \textbf{Propagation.}

        The output $\hat{y}_{i}$ are
	computed  by using the updated parameters $\bm{\theta}$ from the previous
	analysis step (or from the initial sampling if this is the first propagation step). The
	propagation is performed until next converged results are obtained.  The
	$\hat{\cdot}$ indicates predicted quantities that will be corrected in the analysis step below.
	The propagation is performed for each sample in the ensemble, leading to the propagated ensemble
	$\{\hat{\bm{z}}_j\}_{j = 1}^M$.  Each sample $\hat{\bm{z}}_j$ is a vector containing a
	realization of the velocity field and the parameters $\bm{\theta}$ (see
	Equation~(\ref{eq:aug-state})).  The mean $\bar{\bm{z}}$ and covariance $\mathsf{P}$ of the
	propagated ensemble are estimated from the samples.

      \item \textbf{Analysis/Correction.}

        The computed physical fields (velocities) $\hat{\xi}_{i}$ in the whole field are compared and sampled to compare with observations $\xi_{i}^{obs}$.  The ensemble covariance $\mathsf{P}$ and the error covariance $\mathsf{R}$ are used to compute the Kalman gain matrix $\mathsf{K}$ as follows:
        \begin{equation}
          \label{eq:kalman-gain}
          \mathsf{K}^{(n+1)} = \mathsf{P}^{(n+1)} \, \mathsf{H}^\top \,
          \left(\mathsf{H} \mathsf{P}^{(n+1)} \, \mathsf{H}^\top + \mathsf{R}\right)^{-1}
        \end{equation}
        Each sample is then corrected as follows by using the Kalman gain matrix:
	\begin{equation}
	\label{eq:analysis}
	\bm{z}_j = \hat{\bm{z}}_j + \mathsf{K} ({\boldsymbol{\xi}_j} - \mathsf{H}
	\hat{\bm{z}}_j)
	\end{equation}
	where superscript $\bm{z}_j$ is the corrected system state; $\boldsymbol{\xi} = [\xi_1,
	\cdots, \xi_n]'$ are the velocity, the part of the system state vector that can be
	observed; $\mathsf{H}$ is the observation matrix. After the correction, the analyzed state contains
	updated velocities and parameters.

      \item Repeat propagation and analysis Steps 2--3 for next iteration step until convergence is achieved.
\end{enumerate}

The corrected state obtained in Step 3 is a linear combination of the prediction and observations, with the Kalman gain matrix $\mathsf{K}$ being the weight of the observations.

\begin{figure}[htbp]
  \centering
  \includegraphics[width=0.55\textwidth]{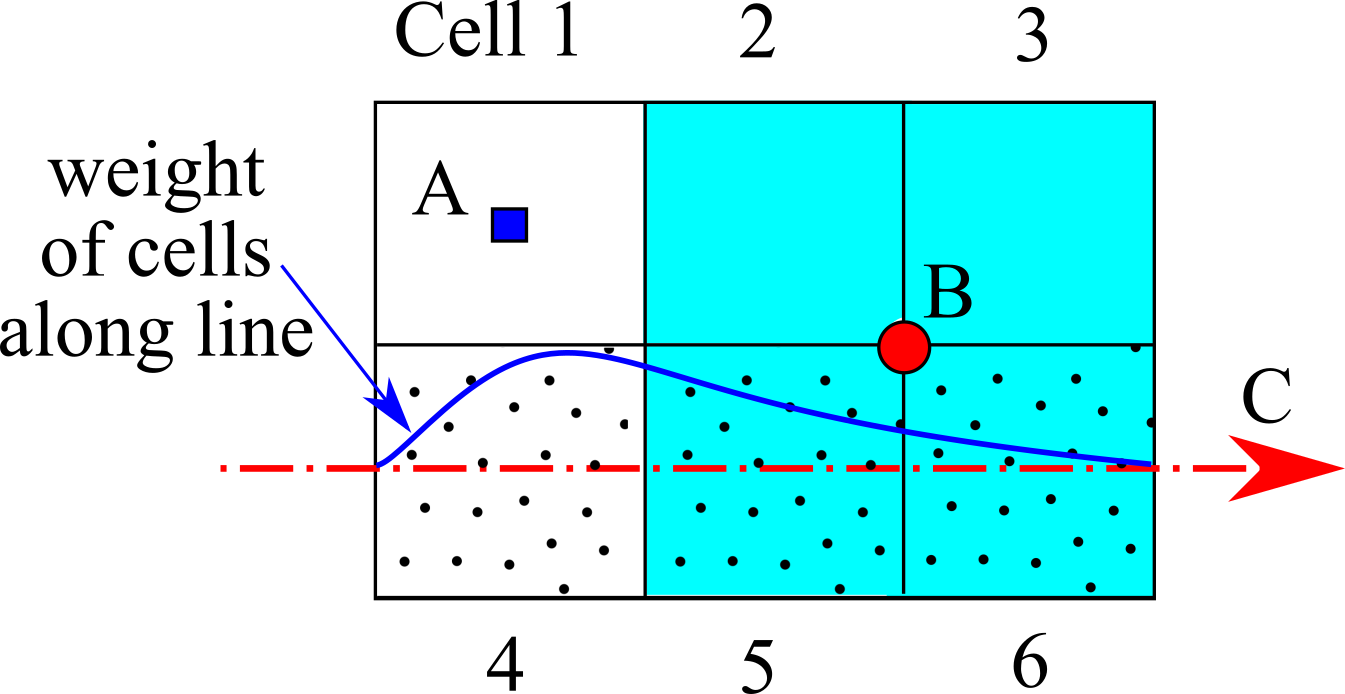}
  \caption{A simple domain with 6 CFD cells and three observations used to illustrate the observation matrix, which defines the mapping from the system state to the observations.}
  \label{fig:obsH}
\end{figure}

The observation matrix $\mathsf{H}: \mathbb{R}^{m+r} \mapsto \mathbb{R}^n$ has a size of $n \times (m+r)$, which maps a vector in the $m$ dimensional state space to a vector in the $n$ dimensional observation space. While point measurements of velocities are used as observations, other derived quantities such as lift, drag, pressure coefficients, surface coefficients, or velocities along a line of sight can be also used by choosing appropriate observation operators.  For all forms of experimental data, the observation matrix $\mathsf{H}$ in the filtering techniques relates the simulated system states to the observed quantities, i.e., $\mathbf{y} = \mathsf{H} \bm{z}$. It is a mapping from system state space to the observation space.  Example of observation operator is shown here. Consider the simple system shown in Fig.~\ref{fig:obsH} to illustrate the principle.  The simulation domain is discretized with 6 cells and the quantity of concern is the horizontal velocity only. Hence, the state vector has a dimension of $6$ by $1$. Three quantities are observed, a volumetric measurement of the velocity at cell 1, a velocity measurement at point B (which is the average of cells 2, 3, 5, 6), and an integrated measurement of the velocity along the line C, with weight factors of $1/2$, $1/3$, $1/6$ for cells 4, 5, 6, respectively. The mapping $\mathbf{y} = \mathsf{H} \bm{z}$ can be written as:
\begin{equation}
  \label{eq:H}
\begin{bmatrix}
y_a \\
y_b\\
y_c
\end{bmatrix}
=
\begin{pmatrix}
  1 & 0 & 0 & 0 & 0 & 0 \\
  0 & 1/4 & 1/4 & 0 & 1/4 & 1/4 \\
  0 & 0 & 0 & 1/2 & 1/3 & 1/6
\end{pmatrix}
\;
\begin{bmatrix}
z_1 \\
 \vdots \\
z_6
\end{bmatrix} .
\end{equation}

\section{Composite model theory and openbox treatment of model inadequacy}
\label{app:composite}

\begin{figure}[!htbp]
  \centering
  \subfloat[Black-box, physics neutral approach] {\includegraphics[width=0.48\textwidth, height=0.48\textwidth]{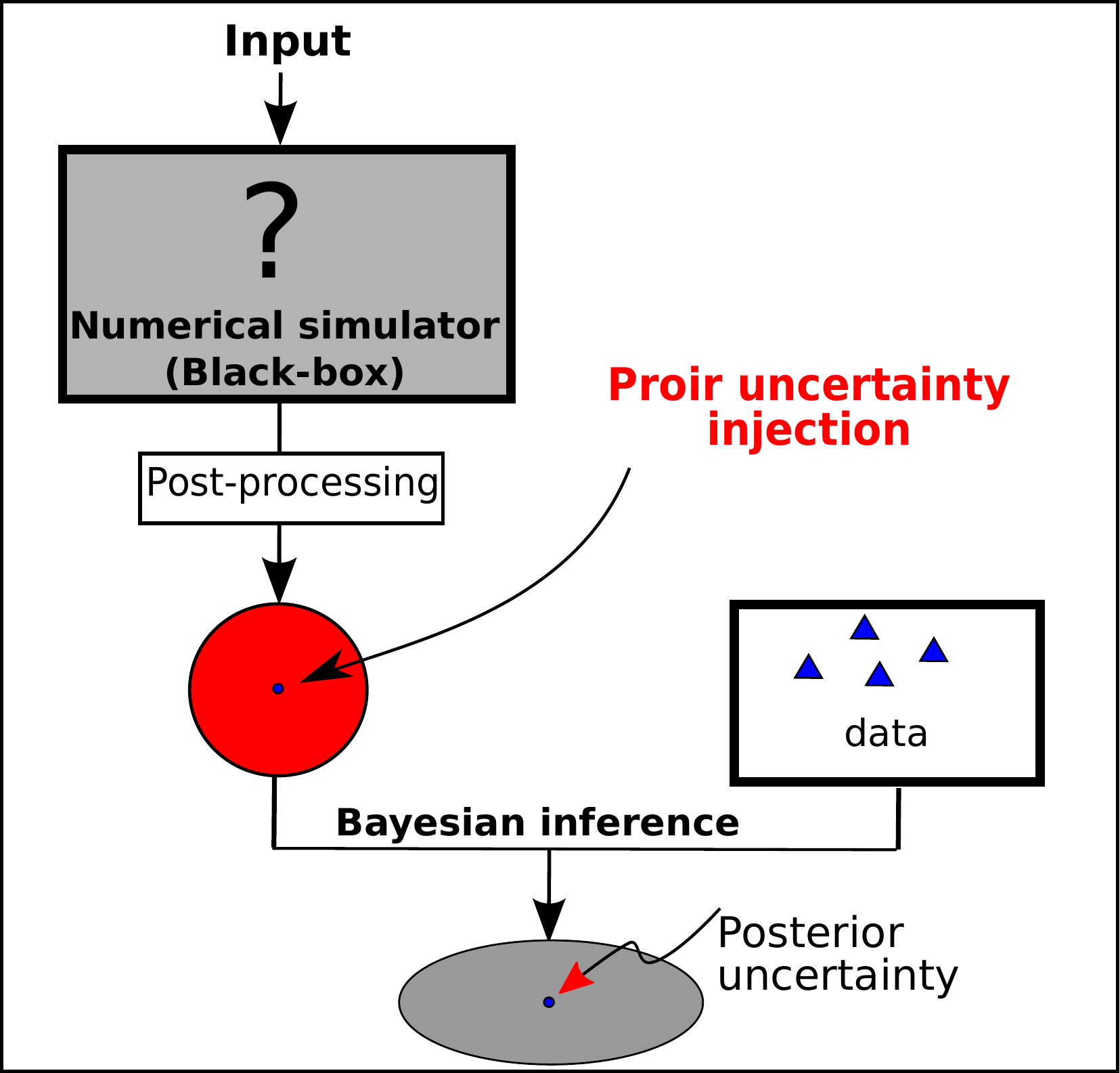}} \hspace{2mm}
   \subfloat[Open-box, physics-informed approach] {\includegraphics[width=0.47\textwidth, height=0.48\textwidth]{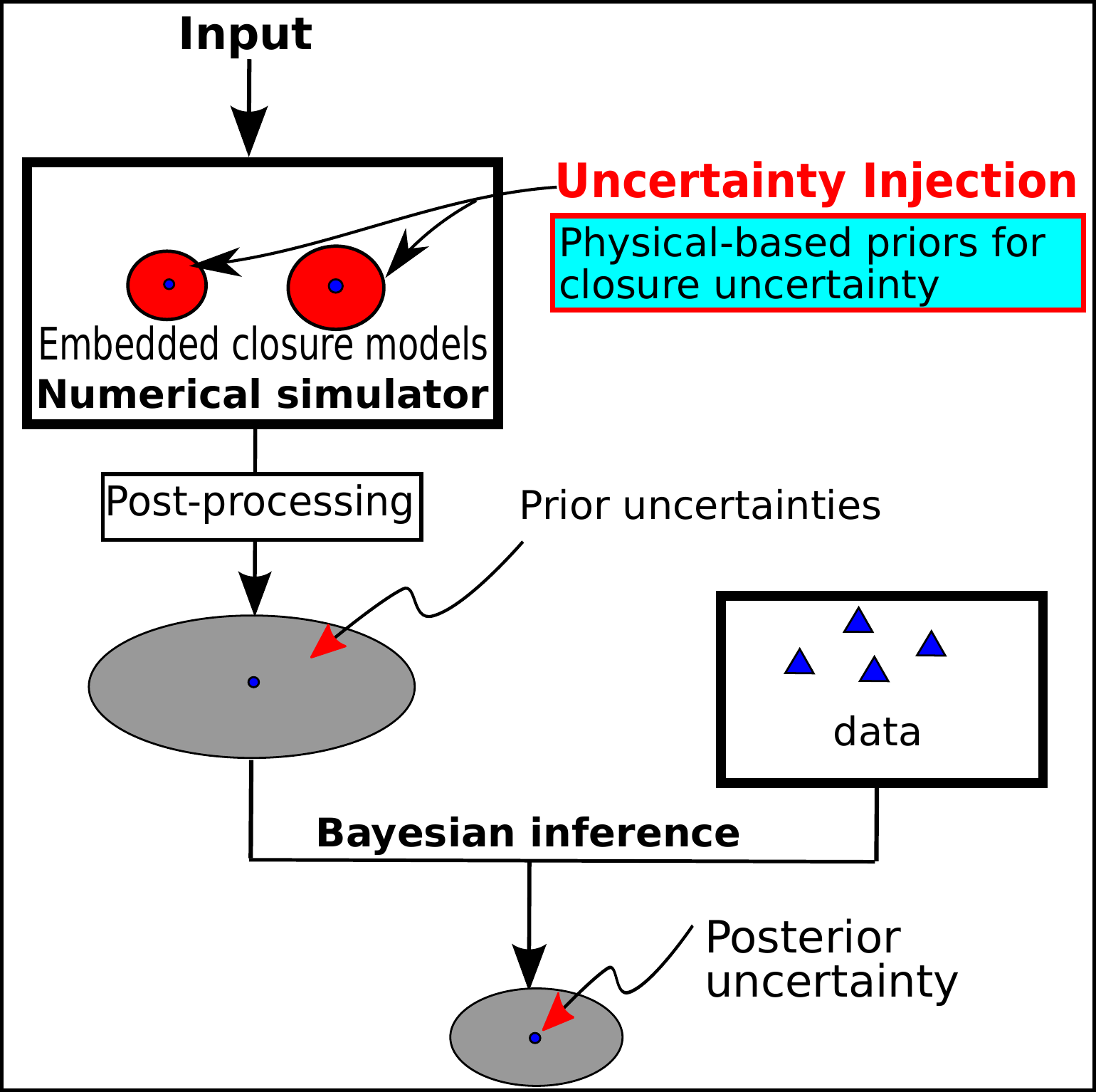}}
   \caption{Schematic illustration of the difference between (a) the traditional physics-neutral approach and (b) the recently developed open-box, physics-informed approach for uncertainty quantification and model calibration.}
  \label{fig:openbox}
\end{figure}

Introducing uncertainties into Reynolds stresses, in both parametric and non-parametric approaches, is motivated by a key consensus in the turbulence modeling community: Reynolds stress is the source of uncertainty in the RANS equations. This consensus is formulated officially as ``composite model theory'' by \citet{oliver2015validating}.  That is, RANS simulations are based on reliable theories describing the conservation laws of mass, momentum, and energy, but contain approximate embedded models to account for the unresolved or unknown physics, i.e., the Reynolds stresses, leading to a composite model.  This theory clearly separates the numerical model (simulator) into two components, i.e., rigorous equations and approximate closure models, and states that uncertainties should be introduced where they originate physically.  This insight resulted in open-box approaches for uncertainty quantification and statistical inference. This is a major advance in model uncertainty quantification in RANS simulations compared to the earlier framework of Kennedy and O'Hagan~\cite{kennedy2001bayesian}, where model inadequacy are introduced directly to the quantities of interest or the observed quantities and the numerical model (simulator) is treated as a blackbox. The open-box and blackbox approaches are compared schematically in Fig.~\ref{fig:openbox}.

Composite models are ubiquitous in various disciplines of science and engineering.  For example, in multiphase flow simulations, models are used to describe interphase mass and momentum exchanges in averaged equations~\citep{ma2015using,ma2016using}; in climate and weather modeling, parameterization are used to account for unresolved or unknown physics including radiation, cloud, and boundary layer processes~\citep{randall2003breaking,tiedtke1993representation,chevallier1998neural}. In all these examples, the conservation laws are all expressed in well-grounded PDEs, albeit containing unclosed terms.

\section{Uncertainties in DNS and their impact on RANS modeling}
\label{app:dns}

DNS data has long been considered the golden standard for evaluating the merits of turbulence models~\citep{mansour1988reynolds-stress,jeyapaul2014assessment}. Evaluation of turbulence models can be done either \textit{a priori} by comparing RANS-predicted Reynolds stresses with DNS data or \textit{a posteriori} by comparing the fields solved by using the model of concern with mean fields from DNS.
However, DNS are, like experimental observations, affected by more or less large uncertainties that may affect the comparisons: these mainly consist of sampling errors and discretization errors.

Although it is the instantaneous quantities such as velocities and pressure that are solved for in DNS, their statistical moments (e.g., means and covariance) are usually the quantities of interest, obtained by averaging a sufficiently large number of temporally uncorrelated  samples of instantaneous fields.  Sampling errors are caused by the fact that the samples may be correlated, and that the number of samples may not be sufficient to achieve statistical convergence, see  \citet{hoyas2008reynolds} for a discussion.

For a properly performed DNS, the mesh must resolve all relevant flow scales and the sampling error is generally considered dominant. However, the mesh usually has to be chosen based on empirical judgment and sampling and discretization errors may be coupled.  A Bayesian approach to account for sampling errors when estimating discretization errors is proposed in \citet{oliver2014estimating}.

Recently, evaluations of emerging data-driven turbulence models motivated an exercise that involves solving the RANS equations with specified Reynolds stresses, which is referred to as \emph{propagation}.  It has long been assumed that propagating accurate Reynolds stresses would lead to accurate velocities.  However, various authors
\citet{thompson2016strategy,thompson2016methodology,poroseva2016on,wang2017comprehensive,wu2018data-driven} found significant discrepancies between the propagated velocities and the DNS velocities.  On the other hand, \citet{wang2017comprehensive} performed the same propagation for fully developed turbulent flows in square ducts at various Reynolds numbers and found that the propagated velocities agree with DNS data satisfactorily. Such apparently conflicting findings were explained by different model conditioning in various flows, i.e., different sensitivity levels of the mean velocities to Reynolds stresses~\cite{wu2018rans}.

In addition to the preceding sources of uncertainty, DNS also suffers from uncertainties associated with the specification of the boundary conditions in a similar way as LES does. Among studies based on probabilistic approaches we mention \citet{ko2008sensitivity}, who examined the sensitivity in the DNS of two-dimensional  plane mixing layers to uncertainties in the inflow boundary conditions.

\end{document}